\documentstyle[12pt,fvtijmp1]{article}
\def\SECTION{\section}
\def\SUBSECTION{\subsection}
\def\asop{$As$-operation}
\def\r*op{$R^{*}$-operation}
\def\rop{$R$-operation}
\def\asy#1#2{\ \displaystyle\mathop\simeq_{
\mathstrut #1\rightarrow#2}\ } 

\def\.{\raise0.3ex\lbox{\scriptscriptstyle\circ}}
\def\lbox#1{\hbox{\mathsurround0pt$#1$}}
\begin{document}

\thispagestyle{empty}
\vbox to1.2in{\vfill}         % VERTICAL SKIP
\hfill \hbox{FERMILAB-PUB-91/347-T}
\begin{center}

{\large
    EUCLIDEAN ASYMPTOTIC EXPANSIONS\\
[2mm]
    OF GREEN FUNCTIONS OF QUANTUM FIELDS.\\
[2mm]
    (I) {\sc Expansions of Products of Singular Functions}\\
[6mm]}  
{ F.~V.~Tkachov}\\
[5mm]
Fermi National Accelerator Laboratory \\
P.O.Box 500, Batavia, Illinois 60510 USA\\
[2mm]
and\\
[2mm]
Institute for Nuclear Research\\ 
of the Russian Academy of Sciences\\
Moscow 117312, Russia.%
\footnote{
Permanent address.
}\\
[15mm]

\end{center}

{
\centerline{\sc ABSTRACT}
\vskip2mm
\noindent
The problem of asymptotic expansions of Green functions in 
perturbative QFT is studied for the class of Euclidean asymptotic 
regimes. Phenomenological applications are analyzed to obtain a 
meaningful mathematical formulation of the problem. 
It is shown that the problem reduces to studying asymptotic expansions 
of products of a class of singular functions in the sense of the 
distribution theory. Existence, uniqueness and explicit expressions 
for such expansions ({\sl \asop\ for products of singular functions}) 
in dimensionally regularized form are obtained using the so-called 
{\sl extension principle}.
}

\vbox to1cm{\vfill}         % VERTICAL SKIP

\hfill\hbox{{\sl In memory of S.G.Gorishny (1958-1988)}}

\hfill

\setcounter{footnote}{0}
\newpage

\thispagestyle{myheadings}\markright{}

\SECTION{Introduction}
\label{introduction}

\begin{flushright}

\it... the perturbative version of the Wilson short distance\\
expansion is completely clear since more than ten years.\\
 ... A new theory is not needed.\\[2mm]
\sl A referee for NUCLEAR PHYSICS B.

\end{flushright}

\hfill

\SUBSECTION{The problem of asymptotic expansions in perturbative QFT}
\label{theproblem}

Approximations and asymptotic methods pervade applications of 
mathematics in natural sciences: nothing simplifies a problem more than 
reduction of the number of its independent parameters---and 
the idea of asymptotic expansion is one of the two most useful in this respect.

In applied quantum field theory, one deals with amplitudes---or, 
more generally, with Green functions of elementary or composite operators, 
which depend on momentum and mass parameters.
Detailed investigation of dynamics of physical processes is rarely possible 
with more than just a few independent parameters. Therefore, one 
normally considers asymptotic regimes in which almost all the momenta 
and masses are much larger (or smaller) than the chosen few.  
Since the most informative dynamical framework is currently provided by 
the perturbation theory, Green functions are represented as infinite sums 
over a hierarchy of Feynman diagrams, and the initial problem 
falls into two parts. 

First, one has to expand individual Feynman diagrams 
with respect to external parameters---masses and momenta.
This is the analytical part of the problem.
Empirical evidence indicated long ago that diagrams can be expanded 
in powers and logarithms of masses and momenta. 
For a wide class of the asymptotic regimes a 
formal proof of this fact was given by D.~Slavnov~\cite{dslav:73}. 
The reasoning of \cite{dslav:73} was based on a modification of the 
techniques used in the proof of the Weinberg theorem 
(which established the power-and-log nature of leading 
asymptotics in high-energy regimes; for a 
detailed discussion of the theorem see \cite{manoukian:book}).
That result was extended to other asymptotic regimes using various 
techniques \cite{french}--\cite{kosins-masl}. 

The main difficulty with asymptotic expansions of multiloop diagrams
is that formal Taylor expansions of the integrand result in 
non-integrable singularities. This indicates that the integrals 
depend on the expansion parameter non-analytically. 
The papers \cite{dslav:73}--\cite{kosins-masl} followed, with variations,
the old idea of splitting integration space---whether in momentum 
or parametric representations---into regions in such a way as 
to allow one to extract the non-analytic (usually logarithmic) 
contribution from each region, eventually, by explicit integration. 
However, complexity of multiloop diagrams 
exacerbated by ultraviolet renormalization 
made obtaining convenient explicit expressions for coefficients of 
such expansions unfeasible within a framework of elementary integral calculus.
Nevertheless, expansions that involve simplest functions of the expansion 
parameter (e.g.\ powers and logarithms) are exactly what is needed 
in the final respect in applications, and even such limited information 
on the analytical form of expansions can be useful.%
\footnote{
cf.\ an early attempt to develop calculational algorithms for 
coefficient functions of operator product expansions \cite{gluing:82}.
}

The second part of the perturbative expansion problem is combinatorial. 
The observation that asymptotic expansions of non-perturbatively 
defined objects should have a non-perturbative form 
is due to K.Wilson \cite{wilson:69} who also discovered that such operator 
product expansions (OPE) can be highly useful in phenomenology. 
W.Zimmermann \cite{zimm}, using what became known as the 
Bogoliubov-Parasiuk-Hepp-Zimmermann (BPHZ) techniques, 
was first to show that it is indeed possible to construct an OPE 
for a class of short-distance asymptotic regimes.%
\footnote{
It was also realized that the short-distance OPE 
is closely related to other expansion problems e.g.\ the problem of 
decoupling of heavy particles and low-energy effective Lagrangians---%
see a discussion and references below in subsect.~\ref{ss1.3};
in fact,  short-distance OPE and related problems 
constitute a subclass of Euclidean regimes
studied in the present paper.}

The main achievements of \cite{zimm} were, first, 
the demonstration of how the expansion in a ``global" OPE form is 
combinatorially restored from terms corresponding to separate Feynman 
diagrams; second, the required smallness of the remainder 
of the expansion was proved in presence of UV renormalization. 
However, unlike the expansions obtained in \cite{dslav:73}, the 
coefficient functions of the OPE of \cite{zimm} were not pure powers 
and logarithms of the expansion parameter (i.e.\ short distance or 
large momentum transfer) but also contained a non-trivial dependence 
on masses of the particles (apparently because masses were needed as 
regulators for infrared divergences). Therefore, 
although the results of \cite{zimm} are firmly established theorems, they 
fell short of providing an adequate basis for applications 
(for a more detailed discussion see subsect.~\ref{ss1.2} below): 
first, the expressions for coefficient functions were 
unmanageable from purely calculational point of view%
\footnote{
This explains why the first 
large-scale calculations of OPE beyond tree level \cite{bardeen:78} were
performed by ``brute force": the coefficient functions of an OPE were found by
straightforward calculation of asymptotics of the relevant non-expanded 
amplitudes and then by explicit verification
of the fact that the asymptotics have the form which agrees with the 
OPE ansatz.  The more sophisticated methods of \cite{ fvt:ope83}, \cite{alg:83} were 
discovered outside the BPHZ framework---using the 
ideas described in the present paper.
}; 
second, infinite 
expansions could not be obtained in models with massless particles, e.g.
QCD; third, the logarithms of masses contained in coefficient 
functions are---as became clear later 
\cite{fvt:mass}---non-perturbative within QCD in the sense that they 
cannot be reliably evaluated within perturbation theory using 
asymptotic freedom. All the above drawbacks have the same origin: 
a lack of the property that is now known as {\em perfect factorization\/}
(we will discuss this point later on in subsect.~\ref{ss1.2}).

Lastly, although the BPHZ techniques may be fine as an instrument of 
verification of the results discovered  by other methods, 
its heuristic potential turned out to be inadequate: 
it proved to be of little help in finding {\em new\/} results. 
Indeed, both the formula of the Bogoliubov $R$-operation \cite{bog-shir} 
and the OPE were  discovered using heuristics that are foreign to the 
BPHZ method, while the attempts to obtain full-fledged OPE-like results 
for a wider class of asymptotic regimes (non-Euclidean, or Minkowskian regimes) 
have so far largely failed.

A fully satisfactory solution of the expansion problem---which it is 
the aim of the present and companion publications to describe---has been found 
only for the class of the so-called {\em Euclidean asymptotic regimes\/}
(for precise definitions see sect.~\ref{s2}) in a series of publications  
\cite{fvt:q82}--\cite{IV}, including efficient calculational formulae for
coefficient functions of OPE.   
The derivation in \cite{fvt:q82}--\cite{IV} employed 
a new mathematical techniques based on a novel concept of 
asymptotic expansions in the sense of distributions. Algorithmically, 
the key technical instrument here is the so-called {\em asymptotic operation}
($As$-operation). As has become clear \cite{paradigm}, the techniques 
based on the \asop\ offers a comprehensive and powerful alternative 
to what is known as the BPHZ theory.

\SUBSECTION{The lesson of the $R$-operation}
\label{heuristics}

Bogoliubov's 1952 derivation of the fundamental formula for subtraction 
of UV divergences (the $R$-operation) \cite{bog-shir} provides a fine example 
of a highly non-trivial reasoning which led to a highly non-trivial result---%
a reasoning that has nothing to do with how the $R$-operation is treated (i.e.\ 
``proved") within the framework of the BPHZ method.
By examining Bogoliubov's reasoning one can exhibit 
the central dilemma of the theory of multiloop diagrams, and the lessons 
learnt thereby have a direct bearing on the theory of asymptotic 
expansions as well. 

The dilemma is as follows. On the one hand, Feynman diagrams are  
objects whose complexity increases infinitely with the order of perturbation 
theory. But that complexity is not amorphous, it has a structure:
Feynman diagrams can be generated, e.g., by iterating Schwinger-Dyson equations
or via some other equivalent and orderly procedure
(see e.g.\ the construction of perturbation theory in \cite{bog-shir} 
from microcausality condition). To put it shortly,
multiloop diagrams are organised in a recursive fashion. 
For definiteness, here is an example of a causality condition of the kind that
was used by Bogoliubov:
\def\LL{{\cal L}}
\begin{equation}\label{caus}
     T[\LL(x)\LL(y)\LL(z)\LL(u)] = T[\LL(x)\LL(y)] 
     \times T[\LL(z)\LL(u)],  \quad\quad{\rm for\ } x^0,y^0>z^0,u^0,
\end{equation}
which expresses the fact that chronological product of four Lagrangians 
is expressed as a simple product of $T$-products involving lesser number of
Lagrangians, taken in a special order. (Each Lagrangian will correspond, 
in the final respect, to a vertex in a diagram.) 
On the other hand, the elements that participate in the recursion \ref{caus}, 
are singular---their formal nature can be best described by qualifying them as 
distributions---and their products must not be treated formally: products of 
distributions do not, generally speaking, exist (if one insists that they do, 
one runs into UV divergences \cite{bog-shir}). 

Nevertheless, Bogoliubov's finding consists in that  
if all $T$-products of lower order are
known, the totality of such relations 
defines $T[\LL(x)\ldots\LL(u)]$ everywhere except for the
point $x=y=z=u$.

The last step in the definition of $T$-product can be best described 
using the language of distribution theory as a procedure of ``extension
of a functional"
(see below sect.~\ref{s4}). Without entering into detail, we only note that 
at a practical level, such an extension consists in adding to the r.h.s.\ 
of \ref{caus} a counterterm localized at the point $x=y=z=u$.
It can only be a linear combination of derivatives of 
$\delta(x-y)\delta(y-z)\delta(z-u)$. The number of derivatives to be added
is determined by the leading singularity of the r.h.s.\ at the point $x=y=z=u$,
which can be determined by, essentially, power counting, after which 
there are simple ways to fix the coefficients in order to ensure finiteness
of the resulting $T$-product.

The $R$-operation now emerges as a straightforward iteration of 
the same elementary step in situations with increasing number of Lagrangians 
on the l.h.s.: addition of a  counterterm corresponding to the singularity 
at an isolated point. 

The net effect of such a reasoning is, in a profound sense,
``organizational": it allows one to use the inherent 
recursive structures in order to reduce the reasoning to just one 
simple step. It is no secret (see e.g.\ the proof of the Weinberg theorem 
\cite{manoukian:book}) 
that the problems involving multiloop diagrams are reduced, essentially, to
a very cumbersome (if done in a straightforward way) power counting.
A proper ``organization" of the problem 
requires to do power counting only for 
a simple subproblem (isolated point in the above example) 
so that the solution becomes rather obvious.

However,  Bogoliubov in the 50s was not familiar with the 
techniques of the 
modern version of the distribution theory \cite{schwartz}, 
which explains the decision not to try to formalize the 
underlying heuristic derivation 
(such a formalization can be found e.g.\ in \cite{fvt-vvv})
but rather 
to treat the formula for the $R$-operation as given and simply 
``rigorously prove"
it using the simplest available method---reducing the renormalized 
diagram 
to an absolutely convergent integral by resolving all recursions 
in the framework of a parametric representation. Thus the BPHZ method 
was born.

\SUBSECTION{$As$-operation: distributions and ``perfect" expansions}
\label{newpar} 

It turns out \cite{fvt:q82} that the problem of expansions of multiloop
diagrams can be considered in a way similar to Bogoliubov's treatment 
of UV divergences. Since the main body of the present paper
 is devoted to explanation and clarification 
of this fact, only a few general remarks are offered here.

First, the entire set of Feynman diagrams is structured in a 
recursive fashion, even if in more difficult problems (e.g.\ expansions)
it may not be easy to notice the recursion and its relevance 
to the problem at hand.

If the second part of the expansion problem (restoration of expansions
in a global OPE-like form) is to be successful, then the expansions 
of individual diagrams should be done so as to preserve the recursive 
structure. 
The problem gets complicated if properties like gauge invariance are involved
which connect sets of diagrams. Thus, an efficient handling of recursions
is a key to successful organization of any work with multiloop diagrams. 

A scrutiny of the problem of singularities encountered in formal expansions of
integrands from the point of view of the
expansion problem in its entirety---i.e.\ taking into account 
the recursion aspect---reveals that the underlying fundamental 
mathematical problem is that of expansion of products of singular functions 
in the sense of distributions (sect.~\ref{s3}). Its solution involves 
counterterms (similarly to the theory of $R$-operation described above);
each counterterm corresponds, as we will see, to a subgraph, and the
underlying recursive structure of multiloop diagrams is naturally reflected 
in the expansions, allowing simple exponentiation at the second, 
combinatorial stage \cite{II}. 

On the other hand, in order to save effort by avoiding proving 
useless theorems, one must aim at obtaining expansions that 
 run in powers and logarithms of the expansion parameter 
(since we know the analytical structure of expansions; 
otherwise one would have to determine it in the process).

Now we come to the most important point: once the analytical structure of
expansions is determined, a conclusion follows: 
{\em if such an expansion exists, it is unique\/} (see subsect.~\ref{ss4.2}). 
An immediate corrolary is that 
{\em all structural properties of the original collection of diagrams prior 
to expansion are inherited by the expansion in an orderly manner}.
This fact drastically simplifies study of gauge properties of the
resulting expansions.

Thus, the two key ideas of our method are: 
expansions in the sense of distributions (which allows one to make a full
use of the recursive structures of the perturbation series)
and the obligatory requirement of ``perfection" of expansions to be 
obtained. 

One can say that, in the final respect, the two ideas offer a constructive 
heuristic framework to bridge the gap between the lowest level of 
the problem (the underlying power counting, division of integration space into 
subregions etc.) and its highest level (global non-perturbative structure of 
the OPE-like expansions): reasoning using the language of distributions and 
``relentlessly pursuing perfection" in expansions at each step, 
we are guaranteed 
that the resulting formulae will possess all the desirable properties. 

It is worth stressing that within the framework of our techniques 
the problem is represented as an iteration of the same simple step
(construction of counterterms for a singularity localized at an isolated
point), and  
one only has to ensure ``perfection'' of the expansion at this step.
The property of uniqueness will take care of ``perfection'' of the
complete expansion. To appreciate this it is sufficient to recall that
the BPHZ theory (which attempts to deal with the entire
problem without structuring it into simpler steps)  failed to
construct OPE in models with massless particles 
(which problem is closely 
related to that of constructing perfectly factorized expansions)
until the solution was found by other methods 
\cite{fvt:q82}--\cite{IV}.

The recipes of \cite{fvt:ope83}, \cite{alg:83} and the 
multiloop calculations 
performed e.g.\ in \cite{ope3loops} confirm  
that a good organization has its advantages.

It remains to note that neither one of the above two ideas is specific 
to the case of Euclidean asymptotic regimes.

\SUBSECTION{Purpose and plan} 

% This paper:

The aim of the present paper is to explain---at a heuristic level 
comparable to that of the derivation of the $R$-operation in the classic book  
\cite{bog-shir}---both the current understanding of the expansion problem 
and the analytical ideas behind the solution \cite{fvt:q82}--\cite{IV},
presenting the results in a form that proved to be 
most convenient for applications in applied QFT. 
We begin by analyzing 
phenomenological applications in order to arrive at a meaningful 
formulation of the asymptotic expansion problem
and reviewing the results previously known in order to compare them 
with ours. We identify the problem of asymptotic expansions of Feynman 
integrands in the sense of distributions as a central one, and present its 
explicit solution in the form of the \asop. 
In this paper we do not aim at achieving a complete mathematical
formalization but rather concentrate on the key motivations, notions
and ideas. 
We will use the dimensional regularization \cite{dreg} 
(for a review see \cite{collins}) 
as a familiar technical framework to deal with various singularities 
one encounters in this kind of problems.%
\footnote{
The better known rigorous definition of the dimensional regularization 
is in terms of the $\alpha$-parametric representation 
\cite{breit-mais} while our reasoning is essentially based on momentum 
space picture. Formal constructions of dimensional regularization 
in terms of position/momentum representations exist \cite{blekher}, 
\cite{DRinX}, but the original definition of \cite{dreg} 
is the most useful one in applications (which means, by the way, that 
it is this definition that should be a preferred 
subject of investigation; indeed, it is not difficult to adapt it 
for the purposes of formal proofs 
\cite{dregm}, but an in-depth discussion of this point 
goes beyond the scope of the present publication).
}
However, the use of dimensional regularization is by no means 
essential---a regularization-independent 
account of the proofs can be found in \cite{fvt-vvv}--\cite{IV} 
where a fully formalized treatment of 
asymptotic expansion of products of singular functions is presented.

In the companion publications \cite{II} and \cite{III} the \asop\ is 
applied to two problems, respectively: deriving and studying OPE and 
its generalizations in the MS-scheme, and studying UV renormalization 
and related issues. In \cite{II} the combinatorics of 
the transition from expansions for single diagrams to OPE in a global 
form is studied. To this end the so-called \asop\ for integrated 
diagrams is defined (it was first introduced in \cite{IIold}). Its 
combinatorial structure is fully analogous to that of the \rop, which 
fact allows one to obtain exponentiation easily and derive full 
asymptotic expansions of perturbative Green functions in any of the 
so-called Euclidean asymptotic regimes which comprise the familiar 
short-distance OPE, the low-energy effective Lagrangians to all orders 
of inverse masses of heavy particles etc. The derivation automatically 
yields the convenient calculational formulae that have been presented 
earlier \cite{fvt:ope83}, \cite{alg:83} and extensively used in 
applications (see e.g. \cite{ope3loops}). It should be noted that 
the treatment of combinatorics of the ordinary \rop\ given in \cite{II} 
is, probably, one of the simplest in the literature.

In the third paper \cite{III} a new approach to studying UV-divergent 
integrals will be described which is based on the use of the \asop\ of 
the present work, and a new representation for the $R$-operation will be 
given which generalizes the results of sect.~\ref{s7}. (Ref.~\cite{III} 
is a simplified version of the more formal text \cite{gms}.) 
It should be noted that 
one of the major difficulties in studies of asymptotic expansions of 
Feynman diagrams used to be the presence of a subtraction procedure 
for eliminating UV divergences \cite{zimm}. 
The representation \cite{gms} (see also sect.~\ref{s7} below) trivializes 
the problem by reducing it to study of double asymptotic expansions \cite{IV},
which can be accomplished by exactly the same methods as in the case of 
one-parameter expansions. It should be stressed that such a simplification 
mechanism is not limited to Euclidean problem, it is completely general.
An informal description of our treatment of UV divergences 
(presented in full mathematical detail in \cite{gms}) 
will be done in \cite{III}.
We will also explain how the new representation and the \asop\ work together 
to produce the algorithm for calculations of renormalization counterterms 
and renormalization group functions known as the \r*op \cite{r*}. 

The present paper is organized as follows. In the introductory 
sect.~\ref{s1}, we review some phenomenological problems of applied QFT 
falling under the general heading of Euclidean asymptotic expansions, 
and analyze the properties which the corresponding theoretical 
expansions must possess in order that their applications to 
phenomenology make sense. We also review the most important earlier 
results on the subject and explain what improvements and 
generalizations are offered by the new theory. In sect.~\ref{s2} we 
explicitly formulate the Euclidean asymptotic expansion problem 
which we wish to study.

In sects.~\ref{s3}--\ref{s7}, the basic ideas of our method are discussed. 
Heuristic motivations for studying expansions of products of singular 
functions in the sense of the distribution theory (the Master problem) 
are presented in  sect.~\ref{s3}. Sect.~\ref{s4} contains some basic 
background mathematical definitions. The extension principle---our key 
to solving the Master problem---is presented and discussed in 
sect.~\ref{s5}. In sect.~\ref{s6} we demonstrate how the ideas of the 
extension principle work with a simple but representative example. 
Sect.~\ref{s7} deals with expansions of one-loop Feynman integrals and 
introduces some ideas concerning the treatment of UV-divergent graphs, 
which are generalized in \cite{gms} and \cite{III}.

In sects.~\ref{s8}--\ref{s12}, we solve the Master problem in the 
general Euclidean case. First in sect.~\ref{s8} a system of notations 
is developed which allows one to compactly represent and conveniently 
manipulate Feynman integrands, without using integral representations%
---which is important if the benefits of the viewpoint of the 
distribution theory are to be used to the full. Sect.~\ref{s9} 
describes and classifies singularities of the formal Taylor expansion 
of singular functions with respect to a parameter, and the important 
notion of IR-subgraph is introduced. In sect.~\ref{s10} we derive the 
general formulae for the \asop\ for products of singular functions, 
which provides the solution to the Master problem. Explicit 
expressions for the counterterms of the \asop\ are derived in 
sect.~\ref{s11}. Sect.~\ref{s12} contains a detailed discussion of an 
example with a non-trivial pattern of singularities.

\newpage\thispagestyle{myheadings}

\vskip2.3cm
\centerline{\large\bf THE PROBLEM OF EUCLIDEAN}
\vskip3mm
\centerline{\large\bf ASYMPTOTIC EXPANSIONS}

\SECTION{Euclidean expansions in applications}
\label{s1}

Since there are important aspects in which formulation of the
expansion problem within the standard BPHZ theory had been lacking,
and since in publications with emphasis on applications the exact
mathematical nature of the problems is often not understood clearly,
it is appropriate to review at least those phenomenological problems
that are reduced to special cases of asymptotic expansions in
Euclidean regimes.  This will provide motivations for our formulation
of the problem of Euclidean asymptotic expansions, and give us an
opportunity to review the key results on the subject as well as to
explain the improvements which are contained in our results as
compared to those obtained by other authors.  Some important new
notions will also be introduced.

\SUBSECTION{Short-distance operator-product expansions}
\label{ss1.1}

K.~Wilson \cite{wilson:69} considered expansions of 
operator products at short distances of the form:
\begin{equation}\label{(1.1)}
     T [A(x)B(0) ] \approx \sum_i c_i(x) O_i(0),\quad\quad 
     x\rightarrow 0.
\end{equation}
However,  despite the undeniable heuristic value of the point of view 
of position space representation, it is momentum representation picture 
that is immediately relevant for analysis of 
phenomenological problems while the transition to coordinate 
representation somewhat obscures the Euclidean nature of 
the corresponding asymptotic regime. 

In fact, the generic problem 
which is essentially equivalent to short-distance expansions and 
immediately related to phenomenology, is to expand expressions like:
\begin{equation}\label{(1.2)}
       \int dx \, e^{iqx} <T [ A(x)B(0) \prod_i J_i(y_i)]>_0
\end{equation}
as 
$Q^2 = - q^2\rightarrow + \infty.$

It should be stressed that $q$ goes to infinity along Euclidean 
(space-like) directions. The following examples are intended to 
illustrate this point.

($i$) Within the QCD sum rule method \cite{svz} (for a review see 
\cite{SumRules:rev}), when one studies the spectrum of hadrons built 
of light quarks, one considers expansions of the correlator of two 
local operators:
\begin{equation}\label{(1.3)}
       \Pi(Q^2) = i\int\,dx\,e^{iqx}<T[A(x)B(0)]>_0
\end{equation}
at $Q^2$ much greater than $m^2_{u,d,s},$ the masses of light 
quarks.

($ii$) For deep-inelastic lepton-hadron scattering (for a review 
see \cite{rad:rev}) the moments of the observable structure 
functions are directly expressed (see \cite{wandzura}) as integrals of 
the quantity
\begin{equation}\label{(1.4)}
     W(Q^2,2pq_N,p^2_N)=\int\,dx\,e^{iqx} 
     <N|T[A(x)B(0)]|N>.
\end{equation}
One can see from the expressions presented in \cite{wandzura}
 that the behaviour of the $n$-th 
moment at $Q^2 \rightarrow+\infty$  and fixed $n$ is determined 
by the behaviour of $W(Q^2)$ in the deeply Euclidean region
\begin{equation}
     Q^2 \gg \vert 2qp_N \vert \, p^2_N.
\end{equation}
(This does not contradict to the popular ``light-cone'' philosophy 
\cite{dis:book}, where one studies the behaviour of the structure 
functions $F(x,Q^2),$ $x=Q^2/2qp_N,$ at $Q^2 \rightarrow+\infty$  and 
fixed $x,$ which is related to the light-cone expansion of the 
operator product. Indeed, reconstruction of functions from their 
moments is a mathematically ``ill-posed" problem \cite{tikhonov} and 
requires additional information for its solution. This agrees with the 
fact that the short-distance limit is apparently more stringent than 
the light-cone one. However, from the point of view of 
phenomenological usefulness the two approaches are rather 
complementary \cite{MOMvsF}.)

($iii$) Further examples are provided by the problem of 
hadronic formfactors (usually, electromagnetic ones; for a general 
review see \cite{pisaev}) within the QCD sum rules approach 
\cite{svz}. Here one deals with the three-point correlator of the form 
(cf.\ Fig.~1)
\begin{equation}\label{(1.6)}
     \int dx\,dy\,\exp i(q_1x-q_2y)< T \left[ J(x)\,J(y)\,j(0)
                                       \right] >_0,
\end{equation}
where the current $J$ corresponds to the hadron, and $j$ is the 
electromagnetic current. Denote $Q^2  = -q^2_i, Q^2 = - (q_1- q_2)^2.$ 
The case of ``intermediate momentum transfer"~\cite{form:med},
\begin{equation}
     Q^2_i \sim Q^2  \gg m^2_{u,d,\ldots},
\end{equation}
presents an example of a problem with more than one heavy 
momentum. However, this case still belongs to OPE problems ($x$ 
and $y$ tend to 0 simultaneously); the complication due to the 
fact that there are more than two operators to be merged at one 
space-time point is inessential. 

($iv$) The case of ``low momentum transfer" \cite{form:low} is 
described by the asymptotic regime
\begin{equation}
     Q^2_i \gg Q^2  \sim m^2_{u,d\ldots}\,.
\end{equation}
This also is essentially an OPE-like problem: $J(x)$ and $J(y)$ are 
separated by a short distance $x-y\rightarrow0$. Note that here the 
additional spectator current $j$ is present within the same 
$T$-product along with the two operators $J(x)$ and $J(y)$ which are 
to be merged at a point. This fact precludes a straightforward 
substitution of the expansion \ref{(1.1)} into \ref{(1.6)} and gives 
rise to extra terms similar to those that appear in the example ($iii$). 
However, such complications are easily treated within the formalism 
which we will describe.

It is convenient here to point out the difference between the
Euclidean and Minkow\-skian asymptotic regimes. Suppose the momenta
$q_i$ in \ref{(1.6)} are time-like and one considers the so-called
Sudakov asymptotic regime
\begin{equation}
     Q^2 \gg q^2_i  \sim m^2_{u,d\ldots}\,.
\end{equation}
The crucial point is that this does not imply automatically that $q_i$
is of order $m_{u,\ldots}$ componentwise but only that there is a
light-like vector $p_i$, $p_i^2=0$, such that $q_i-p_i\sim
m^2_{u,\ldots}$, while the components of $p_i$ are of order
$\sqrt{Q^2}$. Such asymptotic regimes differ drastically from the
purely Euclidean ones considered in the present paper. Although the
basic ideas of our technique are sufficiently general to make it
applicable to the Minkowskian case, specific implementation should
take into account many new features---e.g.\ non-linearity of manifolds
on which singularities of integrands are localized, etc. One can get a
flavour of what to expect in Minkowskian case from the reviews
\cite{coll-sud}, \cite{korch-rad}.

\SUBSECTION{Technical aspects of OPE. Perfect factorization}
\label{ss1.2}

Within perturbation theory, a first version 
of OPE was obtained by Zimmermann \cite{zimm} 
in the following form:
\begin{equation}\label{(1.9)}
     T[A(x+\xi)B(x)] = \sum_i c_{N,i}(\xi ,m) O_{N,i}(x) + o(\vert\xi \vert^N).
\end{equation}
This result, however, despite the great theoretical significance of \cite{zimm}, 
had a somewhat theoretical rather than practical relation 
to calculations, e.g.\ within pQCD. 
(For justice's sake, however, it should be stressed that the result 
\ref{(1.9)}, which in any case is a theorem, dates back to pre-QCD times.)

First, the above version of OPE relies on momentum subtractions for 
UV renormalization, while all practical calculations beyond one loop 
are always carried out within the MS-scheme introduced by 't Hooft 
\cite{ms} (for a review see \cite{collins}).

Second, for models with massless Lagrangian fields (e.g. for QCD with 
massless gluons) the dependence of the coefficient functions in 
\ref{(1.9)} on $N$ can not be got rid of (this is connected with the 
fact that in \cite{zimm} the MOM-schemes for UV renormalization were 
used and masses in fact played the role of infrared cutoffs)---a 
somewhat awkward property from the viewpoint of practical 
calculations.

Third, the applicability of the OPE to the phenomenology of QCD rests
on the assumption that the coefficient functions are calculable within
perturbation theory. But the coefficient functions in \ref{(1.9)}
depend on the masses $m$ of elementary fields non-analytically and
contain large $\log(m^2 \xi ^2)$ contributions. A renormalization
group analysis \cite{fvt:mass} shows that in pQCD the
renormalization-group resummation of such terms takes one outside the
region of applicability of perturbative methods. Presence of
non-perturbative contributions defeats attempts to extract the
non-perturbative values of matrix elements of the composite operators
by fitting the r.h.s.\ of the above expansion against experimental
data---a procedure typical of phenomenological studies of deeply
inelastic scattering, the QCD sum rules method described above etc.

The notion of {\em perfect factorization}---which had been lacking in, 
and seems to be 
inherently foreign to, the BPHZ theory---was introduced 
in \cite{fvt:ope83}.%
\footnote{
In the context of the theory of $R$-operation perfect factorization
corresponds to renormalization using the so-called massless schemes
among which the MS-scheme  is by far the most useful one.  
\relax}
Its
importance has by now been fully realized.  At the level of
operators/Green functions, perfect factorization means that all the
non-analytic dependence on $m$ in the expansions like \ref{(1.9)} is
localized within the matrix elements of the operators $O_i$ while the
coefficient functions $c_i$ depend on $m$ analytically. It follows
immediately that such an expansion would allow taking the limit $m\to
0$ and, therefore, should be expected to be valid for models with
massless particles. Moreover, since the masses $m$ are no longer
needed as IR regulators, one can expect to get rid of the dependence
on $N$ in \ref{(1.9)}. Furthermore, perfect factorization, implying
that no non-analytic dependences are contained in the OPE
coefficient functions, guarantees that extraction of the
non-perturbative matrix elements from data is a mathematically
meaningful procedure.

At the level of individual diagrams, perfect factorization means that the
asymptotic expansion should explicitly run in powers and logarithms of the
expansion parameter. Such expansions are unique (see below 
subsect.~\ref{ss10.2}) 
which fact is of paramount technical importance; 
for example, it allows one to conclude (cf.\ subsects.~\ref{ss10.2}, \ref{ss11.5}) 
that such asymptotic expansions commute with
multiplication by polynomials and other algebraic operations on 
diagrams---the property that greatly simplifies
combinatorial study of OPE etc.\ in non-scalar models; 
in particular, one need not worry about things like gauge invariance 
of the expansions since such properties are inherited 
by expansions from initial integrals in a well-defined and orderly manner.%
\footnote{
The techniques for studying gauge properties was extended 
to the framework of the
Euclidean $As$-operation in \cite{grishagauge}. In this respect recall 
that gauge invariance plays a central role in studies of 
non-Euclidean asymptotic expansions which represent a major unsolved problem
with important applications to physics of hadrons (see \cite{factoriz}).
\relax} 
Uniqueness of such expansions 
also allows one to exhibit the recursive structure of the problem of 
expansion of Feynman diagrams which can be effectively used in 
derivation of OPE (see below subsects.~\ref{ss10.3}--\ref{ss10.4}).
 
It turned out that the OPE derived directly within the MS-scheme 
\cite{fvt:ope83},%
\footnote{
which result came about as a special case of the general theory of 
EA-expansions developed in \cite{fvt:q82}-\cite{III}, 
\cite{fvt-vvv}-\cite{IV}. Efficiency of the techniques developed 
therein allowed us to obtain very general formulae first made public 
in \cite{IIold}. The results of \cite{IIold} were widely discussed in 
the literature \cite{gorishny}, \cite{smirnov}, \cite{ch:mpi}.
\relax}
\cite{llew}, possess the property of perfect factorization 
so that: the dependence of the 
coefficient functions on masses is analytic (which was earlier 
shown to be a necessary condition of existence of the OPE 
in the MS-scheme \cite{gluing:82});
the expansion is a full asymptotic series; 
it is valid in models with massless particles; 
and due to greatly simplified dependence of coefficient functions on
dimensional parameters, 
there exist algorithms for analytical evaluation of the relevant 
integrals through three loops \cite{ibp}. All these properties are 
interrelated, as is stressed above and 
as will be seen from our derivation. 

More conventional derivations of OPE in the MS-scheme \`a la BPHZ were 
presented by Llewellyn Smith and de Vries \cite{llew} and Gorishny 
\cite{gorishny}. The latter work 
clarified the connection of the results of \cite{fvt:q82}--\cite{IIold} 
with the BPHZ method.%
\footnote{
Still another approach, specifically designed for achieving ultimate 
rigour in treatment of dimensional regularization within the 
techniques of the $\alpha$-parametric representation is sketched in 
\cite{smirnov}. It is unclear, however, how successful the 
latter attempt was because neither that paper nor the publications 
cited therein contain explicit  
treatment of the analytical part of the 
proof (recall in this respect that $\alpha$-parametric integrands for
MS-renormalized diagrams are rather cumbersome 
distributions \cite{MSalphaDistr} so that
advantages of using such a representation are somewhat unclear). 
On the other hand, the combinatorial part of the proof is discussed in the
original publications \cite{IIold}, \cite{II} in a more 
straightforward manner due to recursive economy 
of the formalism of the $As$-operation.

The regularization independent aspects of the original 
derivation have been exhaustively discussed in \cite{fvt-vvv}, 
\cite{gms} and \cite{IV}. Various ways to rigorously treat 
dimensional regularization without parametric representations 
were discussed in \cite{blekher}, \cite{DRinX}, although 
not much practical understanding is added thereby to the heuristic 
treatment of \cite{dreg}. Anyway, the variety of uses of
dimensional regularization in practical calculations is such 
that there is little hope that everything that is being done 
will ever be ``rigorously proved".
In the final respect, mathematics is not about rigour. Mathematics is, 
first and foremost, about calculations.
\relax}

Theoretically speaking, one need not use specifically the MS-scheme to obtain
OPE with the property of perfect factorization. Any of the so-called 
massless schemes (or {\em generalized MS-schemes} \cite{gms}) will do 
\cite{IV}. 

It should also be clearly understood: the most important thing about the 
new results is {\em not\/} the fact that, say expansions were 
proved ``in the MS-scheme", but that such expansions possess the 
property of perfect factorization. This should not be surprising in  view
of the fact that the MS-scheme  has the property of polynomiality 
of its counterterms in masses \cite{collins} which itself is a special case 
of perfect factorization---this point is best understood in the context 
of the definition of generalized MS-schemes in \cite{gms} by way of 
subtraction of UV-dangerous asymptotics from momentum-space
integrands, so that evaluation of those asymptotics is done 
using the same apparatus of the $As$-operation as used in 
obtaining asymptotic expansions of diagrams.  
This similarity is not so much technical as ideological.

\SUBSECTION{Heavy mass expansions}
\label{ss1.3}

Consider a QFT model in which all the fields are divided into two 
sets: light fields collectively denoted as $\varphi$ and heavy fields 
$\Phi.$ Their Lagrangian masses are $m$ and $M,$ respectively. Assume 
that all $M$ are much larger than $m,$ and consider a Green function 
$G$ of light fields only. It depends on $M$ besides other---momentum, 
coordinate etc.---parameters. The problem  is  to study the asymptotic 
expansion of $G$ in inverse  powers of  $M:$
\begin{equation}\label{(1.10)}
     G(\ldots,M) \asy{M}{\infty} ?
\end{equation}

In fact the first two terms in the heavy mass expansion had been known 
for some time. 
The ``decoupling" theorem \cite{decoupl} (for a review and references see 
\cite{collins}, \cite{manoukian:book}) asserts that, up to $O(M^{-1})$ 
corrections, the effect of the virtual presence of heavy particles (in 
phenomenologically meaningful cases) is absorbed into a finite 
renormalization of the Lagrangian of the light world, whose parameters 
become dependent on $M.$ This fixes the form of the leading term of 
the $M\rightarrow\infty$  expansion. The next-to-leading term can be 
described as differing from the initial Green function by one local 
operator insertion. (This result was first obtained in \cite{effL} by 
making use of Wilson's OPE.) 
More formally, let ${\cal L}_{\rm tot}$ be the total Lagrangian of light 
and heavy fields, and let $<T\varphi(x_1)\ldots\varphi(x_n)>_0^{{\cal 
L}_{\rm tot}}$ be a Green function of the light fields in the world 
containing both light and heavy fields. Then the leading and 
next-to-leading contributions to the expansion have the form:
\begin{equation}
     <T\varphi(x_1)\ldots\varphi(x_n)>_0^{{\cal L}_{\rm tot}}
     \asy{M}{\infty} 
     z^n(M) \bigg\{ 
     <T\varphi(x_1)\ldots\varphi(x_n)>_0^{{\cal L}_{\rm 
        tot}^{(0)}}
\end{equation}
\begin{equation}\label{(1.11)}
     + <T\varphi(x_1)\ldots\varphi(x_n)
        \biggr[ \, {g(M)\over M} \int dy\, {\cal L}_{\rm 
        eff}^{(1)}(y) \, \biggl]>_0^{{\cal L}_{\rm tot}^{(0)}}
        \biggr\} + o(M^{-1}),
\end{equation}
where ${\cal L}^{(0)}_{\rm eff}$ 
is obtained from that part of ${\cal 
L}_{\rm tot}$ which describes the light fields, 
by a finite $M$-dependent 
renormalization of its parameters; ${\cal L}^{(1)}_{\rm eff}(y)$ is a 
local operator composed of light fields; $z(M)$ and $g(M)$ are 
functions of $M$ with logarithmic leading behaviour at 
$M\rightarrow\infty$. (The UV renormalization in all cases is 
implicit.)

The notation ${\cal L}^{(1)}_{\rm eff}$ is suggestive, and indeed, 
to all orders in $M^{-1},$
\begin{equation}\label{(1.12)}
     <T\varphi(x_1)\ldots\varphi(x_n)>_0^{{\cal L}_{\rm tot}}
     \asy{M}{\infty} 
          z^n(M) 
     <T\varphi(x_1)\ldots\varphi(x_n)>_0^{{\cal L}_{\rm eff}},
\end{equation}
where
\begin{equation}\label{(1.13)}
     {\cal L}_{\rm eff} = \sum_n M^{-n} g_n(M) 
        {\cal L}^{(n)}_{\rm eff}.
\end{equation}                                            
The derivation of this formula will be given within the MS-scheme, so 
that our results will contain as a by-product a derivation of the 
decoupling theorem and of the formula \ref{(1.11)} directly within the 
MS-scheme, which has been lacking. Moreover, the expansions 
we will obtain possess the property of perfect factorization similar to that
discussed in the preceding subsection, so that
the heavy and light parameters are fully factorized, 
i.e.\ $z(M)$ and $g(M)$ do not depend on $m,$ but only on the heavy 
mass $M$ and, logarithmically, on a renormalization parameter.

The most important phenomenological applications of the heavy mass 
expansion are: the study of effects of heavy quarks on the low-energy 
light quark properties, especially on the parameter $\Lambda_{\rm 
QCD}$ (for a review see \cite{collins}); the low-energy quark 
Lagrangian of the electro-weak interactions \cite{effL} ; evaluation 
of muon contribution to the electron anomalous magnetic moment 
\cite{grinstein}. The latter problem is interesting as an example of 
a problem where a very high precision---and, consequently, 
taking into account many terms in the expansion---is required.

\SUBSECTION{Effects of heavy masses on OPE}
\label{ss1.4}

The QCD sum rules approach to studying the heavy quark bound system 
\cite{svz}, \cite{SumRules:rev} presents an example of a problem where 
a short-distance OPE is modified by presence of a heavy mass. 
Here one should study the 
correlator \ref{(1.3)} but for such an asymptotic regime that the 
squared mass $M^2 $ of the heavy quark were of the same order of 
magnitude as $Q^2 .$ (This point was stressed in \cite{BeylRad}.)

In the context of deep-inelastic lepton-hadron scattering this
asymptotic regime was studied in \cite{wu-ki}.

It follows from our general results that the expansion in this case 
has the form similar to ordinary OPE:
\begin{equation}\label{(1.14)}
     \vbox{\hbox{eq.\ref{(1.3)}}}
       \mathrel{\mathop\simeq_{\mathstrut Q^2\sim M^2\gg m^2_{u,d,s}}}
       \sum_i c_i(M,Q) <O_i(0)>_0,
\end{equation}
but with coefficient functions depending on the heavy mass $M.$ To our
knowledge, an all-order derivation of the expansion \ref{(1.14)} has
heretofore been lacking.

\SUBSECTION{Light-by-light scattering. Linear restrictions on heavy momenta}
\label{ss1.5}

 Consider the problem of 
deeply inelastic light-by-light scattering with both photons deeply 
virtual \cite{DISlight}. Here one studies the four-point correlator of 
hadronic electromagnetic currents with the pattern of external momenta 
as depicted in Fig.~2a, in the asymptotic regime
\begin{equation}
     Q^2_1 \sim Q^2_2 \gg m^2_{u,d,s\ldots} \quad (Q^2_i = q^2_i).
\end{equation}
A new feature in this case is that along with the general momentum 
conservation an additional linear restriction is imposed on the heavy 
momenta (cf.\ Fig.~2b). Motivated by this example, we will allow 
arbitrary restrictions of this kind in our general treatment of 
Euclidean asymptotic expansions.

Note that the Green function to be expanded may not exist at the 
restricted momenta owing to infrared singularities. On the other 
hand, if it does, its expansion can be obtained via our technique in 
the same way as in the non-restricted case. However, the $Q$-dependent 
coefficient functions of the expansion of the non-restricted Green 
function (cf.\ the one corresponding to Fig.~2b) will in general be 
singular at the restricted momenta and have no relation to the 
coefficient functions for the restricted case. In other words, 
restricting heavy momenta and performing expansion are in general 
non-commuting operations.

Allowing restrictions on the heavy momenta has no bearing on our 
formalism for expansion of a single multiloop Feynman 
graph but only on the subsequent combinatorial analysis of the global 
structure of the resulting expansions of Green functions as a whole.

\SUBSECTION{Contact terms}
\label{ss1.6}

In our treatment of Euclidean asymptotic 
expansions we will be considering Green 
functions as distributions with respect to the external momenta.%
\footnote{
The fact that expansions in this case run in powers and 
logs was obtained in \cite{pohl:82}, 
without explicit expressions for the contact terms. 
There are many other 
examples in the literature where distributions and asymptotic expansions 
appeared together---see e.g.\ eq.~(46)  in \cite{ellis}. 
\relax}
Apart from fundamental meta-level arguments 
(finite energy resolution of measuring devices etc.), 
there are more specific motivations for this.

The first one is of a technical nature: we will in any case make an 
essential use of the theory of distributions in order to expand 
integrands of multiloop Feynman graphs so that considering Green 
functions as distributions over external momenta brings uniformity to 
our argument (cf.\ subsect.~\ref{ss3.2} below) 
and is therefore inherently natural.

Another---rather typical---motivation 
comes from the QCD sum rules method \cite{svz}.  
The starting point here is a spectral representation for the 
correlator \ref{(1.3)}, e.g.
\begin{equation}\label{(1.16)}
     \Pi(Q^2 ) = \int^\infty_0 ds \, {\rho(s)\over{s+Q^2}}.
\end{equation}
An important element of the techniques of \cite{svz} is the so-called 
Borelization procedure which is applied to both sides of \ref{(1.16)} 
in order to suppress contributions from large $s$ in the spectral sum 
on the r.h.s.\ of \ref{(1.16)}. It is assumed in \cite{svz} without 
proof that Borelization procedure commutes with the asymptotic 
expansion of $\Pi(Q^2 )$ at $Q^2 \rightarrow+\infty$. However, this 
problem can be avoided (at the cost of some complication of formulae) 
by the following modification of the arguments of \cite{svz}. Consider 
the relation:
\begin{equation}\label{(1.17)}
     F(s/\lambda) = \int d^D\!Q\, \frac{\varphi(Q^2 /\lambda)}{s+Q^2}
           = {\rm const} \int^\infty_0 dx \,x^{D/2} \,
           \frac{\varphi(x/\lambda)}{s+x}.
\end{equation}
One can find---e.g. using tables of integral transforms---pairs of 
$\varphi(Q^2 )$ (related to the Bessel functions $J_\nu(\sqrt{Q^2})$) 
and $F(s)$ (related to the McDonald functions $K_\nu(s)$) such that:

($i$) eq.\ref{(1.17)} is satisfied;

($ii$) $\varphi(Q^2 )$ is smooth in $Q$ everywhere including 
$Q=0$;

($iii$) $\varphi(Q^2 )$ is bounded from above by some negative 
power of $Q^2 $ at $Q^2 \rightarrow\infty;$

($iv$) $F(s)$ decreases at $s\rightarrow\infty$  faster than any 
power $s^{-n}$ (namely, as $\exp(-\sqrt{s})$ times a negative 
power of $s$).

Then using \ref{(1.16)} and \ref{(1.17)} one gets an exact sum rule:
\begin{equation}\label{(1.18)}
     \int d^D\!Q\, \varphi(Q^2 /\lambda) \, \Pi(Q^2) 
     = \int^\infty_0 \!ds\, \rho(s) \,F(s/\lambda).
\end{equation}

The weight function $F$ on the r.h.s.\ provides as good a suppression 
of larger $s$ contributions as the simple exponent used in \cite{svz} 
. At the same time, the l.h.s. is the value of the distribution 
$\Pi(Q^2 )$ on the test function $\varphi(Q^2 /\lambda)$ (the fact 
that $\varphi(Q^2)$ does not decrease at $Q^2\rightarrow+\infty$ as 
fast as is required of the Schwartz's test functions \cite{schwartz} 
is immaterial since we do not have to deal with arbitrary tempered 
distributions but with a specific one, $\Pi(Q^2),$ whose behaviour at 
large $Q^2$ is sufficiently good to make it well-defined on the 
$\varphi(Q^2)$).

The next step in the recipe of \cite{svz} would be to use the expansion 
at $Q^2\rightarrow+\infty$  for $\Pi(Q^2).$ Let us rewrite the l.h.s. 
of \ref{(1.18)} as
\begin{equation}
     \lambda^{D/2} \int d^D\!Q\, \varphi(Q^2)\, \Pi(\lambda Q^2).
\end{equation}
Now it is clear that $\Pi(Q^2)$ should be expanded not as a function 
but as a distribution ``in the sense of distributions". 
Let us explain the meaning of this phrase in our specific case (cf.\ 
subsect.~\ref{ss4.1}).

The asymptotic expansion of $\Pi(Q^2)$ considered as an ordinary 
function of $Q^2$ at $Q^2\rightarrow+\infty$  in the usual sense can 
be written as
\begin{equation}\label{(1.20)}
     \Pi(\lambda Q^2) 
     \asy{\lambda}{\infty}
     \sum_i c_i(\lambda Q^2) \,O_i ,
\end{equation}
where $O_i$ are coefficients independent of $Q^2,$ and $c_i(Q^2)$ are 
ordinary functions, e.g. $c_i(Q^2)=Q^{2(-n)}.$ One can not integrate 
\ref{(1.20)} against a test function because, first, $c_i(Q^2)$ may be 
too singular at $Q\rightarrow0;$ second, there is no guarantee that 
termwise integration of the r.h.s.\ results in a correct asymptotic 
expansion of the integral on the l.h.s.

On the other hand, for an expansion
\begin{equation}\label{(1.21)}
     \Pi(\lambda Q^2) 
     \asy{\lambda}{\infty}
     \sum_{i'} c_{i'}^{\rm distr}(\lambda Q^2)\, O'_{i'} ,
\end{equation}
to hold in the sense of distributions, all $c^{\rm 
distr}_{i'}(Q^2)$ must be well-defined distributions over the whole 
range of $Q,$ and its termwise integration against any test function 
$\psi(Q)$ must generate a correct asymptotic expansion at 
$\lambda\rightarrow\infty$  of the $\lambda$-dependent function
\begin{equation}\label{(1.22)}
     f(\lambda) = \int  d^D \!Q\, \Pi(\lambda Q^2)\, \psi(Q).
\end{equation}
The relation between \ref{(1.20)} and \ref{(1.21)} can be established by 
noticing that both expansions allow integration with test functions 
localized in small neighbourhoods of the points $Q\not=0.$ It follows 
that $c^{\rm distr}_{i'}$ at $Q\not=0$ should be either zero, or it 
should coincide with some $c_i;$ in the latter case $O'_i=O_i.$ 
Therefore,
\begin{equation}\label{(1.23)}
     \sum_{i'} c_{i'}^{\rm distr}(Q) O'_{i'} 
     =
     \sum_i c_i^{\rm R}(Q) O_i 
     +
     \sum_{i'} \Delta c_{i'}(Q) O'_{i'} ,
\end{equation}
where $\Delta c_{i'}(Q)$ are localized at $Q=0,$ which implies that 
they are constructed of $\delta(Q)$ and its derivatives, while $c^{\rm 
R}_i(Q)$ are distributions which by definition coincide with $c_i(Q)$ 
at $Q\not=0.$ For example, if
\begin{equation}\label{(1.24)}
     \Pi(Q) \simeq  {1\over Q^2} \, O_1 + o(Q^{-4}),
\end{equation}
then the expansion in the sense of distributions 
will have the form
\begin{equation}\label{(1.25)}
     \Pi(Q) \simeq  \biggl\{{1\over Q^2}\biggr\}^{\rm R} O_1 
     + \delta(Q) O'_1 + o(Q^{-4}),
\end{equation}
where
\begin{equation}\label{(1.26)}
     \biggl\{{1\over Q^2}\biggr\}^{\rm R}
     =
     \lim_{\epsilon\to0} \; \biggr[\, {1\over Q^2+\epsilon} 
         + Z_\epsilon \delta(Q) \,\biggl],
\end{equation}
with $Z_\epsilon$ chosen so as to render finite the integrals of 
\ref{(1.26)} with arbitrary test functions. Note that the expression 
\ref{(1.25)} is unique---cf.\ below subsect.~\ref{ss4.2}---so that a 
change of $Z_\epsilon$ by a finite constant is compensated by an 
appropriate change of $O'.$ And, of course, one can use any other 
regularization in \ref{(1.26)}.

We see that the expansion in the sense of distributions 
differs from the expansion in the usual sense by $\delta$-functional 
contributions (compare \ref{(1.24)} with \ref{(1.25)} and \ref{(1.26)}). 
Such contributions can be conveniently denoted as {\em contact terms}. Ref. 
\cite{contact} contains an example which demonstrates that such 
contact terms can be numerically significant in applications.

Our formulae for general Euclidean asymptotic expansions 
\cite{IIold}--\cite{II} include 
explicit expressions for the contact terms.

\SECTION{The formal problem of Euclidean expansions}
\label{s2} 

Motivated by the examples considered above, let us now present a 
formal description of the general problem of Euclidean asymptotic 
expansions, following \cite{IIold}. We will stay within the 
framework of perturbative QFT \cite{bog-shir} and assume that the 
MS-scheme \cite{ms} is always used for UV renormalization.

\SUBSECTION{Green functions}
\label{ss2.1}

There are two types of fields collectively denoted as $\varphi$ (light 
fields) and $\Phi$ (heavy fields), with Lagrangian masses $m$ (some of 
which may be zero) and $M.$ Let $h_j(x_j)$ be local monomials built of 
$\varphi,$ and $H_l(y_l)$ built of both $\varphi$ and $\Phi.$ (One 
could allow $h$ to be also built of $\Phi$ but such cases seem to 
never occur in applications.) Denote the full Lagrangian of the model 
as
\begin{equation}\label{(2.1)}
     {\cal L}_{\rm tot}(\varphi,\Phi,m,M).
\end{equation}

Consider the Green function
\begin{equation}\label{(2.2)}
     G(Q,M,k,m,\mu) 
     = \int\Bigl(\prod_i dx_j\Bigr)\Bigl(\prod_l dy_l\Bigr) 
        \exp i \Bigl[\sum_l Q_ly_l + \sum_j k_jx_j \Bigr]
\end{equation}
\begin{equation}
     \times <vac\vert RT\Bigl[ \prod_lH_l(y_l) \prod_jh_j(x_j) 
            \Bigr] \vert vac>^{{\cal L}_{\rm tot}},
\end{equation}
where $R$ is the UV \rop\ and $\mu$ the renormalization parameter; $T$ 
denotes the chronological product. It is assumed that the correlator 
is evaluated within the model described by ${\cal L}_{\rm tot};$ in 
the functional notation:
\begin{equation}\label{(2.3)}
     <vac\vert T[\ldots]\vert vac>^{{\cal L}_{\rm tot}}
     \mathrel {\Longrightarrow} \int d\varphi \, d\Phi\, [\ldots] \exp 
     i\!\int dx\, {\cal L}_{\rm tot}(x).
\end{equation}
The Green function \ref{(2.2)} as defined should be proportional to the 
$\delta$-function which expresses the overall momentum conservation, 
but we will ignore it, just assuming that
\begin{equation}\label{(2.4)}
     \sum_l Q_l + \sum_j k_j = 0.
\end{equation}

The momenta $Q$ and masses $M$ will be called {\em heavy} 
while $k$ and $m$, 
{\em light}. It will be assumed that the ratio of scales of light over 
heavy parameters (denoted as $\kappa $) is vanishing, and $G$ is to be 
expanded with respect to $\kappa .$ More precisely, we wish to expand 
in $\kappa $ the expression
\begin{equation}\label{(2.5)}
     G(Q,M,\kappa k,\kappa m,\mu) \asy{\kappa}{0} ?
\end{equation}

A remark is in order. We have implicitly assumed that the 
renormalization parameter $\mu$ belongs to the heavy parameters. 
However, since we always use the MS-scheme for UV renormalization, 
each Feynman graph contributing to the Green function $G$ is a 
polynomial in $\log\,\mu.$ Therefore, if we take $\mu$ to be 
proportional to $\kappa ,$ the additional dependence on $\kappa $ can 
be easily taken into account. At the level of Green functions, the 
renormalization group can be used to get rid of the $\kappa 
$-dependence of $\mu.$ The net effect is that it is irrelevant 
whether we consider $\mu$ as a heavy or light parameter. Having this 
in view and using the homogeneity of $G$ with respect to its 
dimensional parameters:
\begin{equation}
     G(Q,M,\kappa k,\kappa m,\mu) = \kappa^{\rm const} 
       G(Q/\kappa ,M/\kappa ,k,m,\mu/\kappa ),
\end{equation}
one sees that the problem \ref{(2.5)} is equivalent to studying 
expansions of $G$ in $Q, M\rightarrow\infty,$ which is more 
readily associated with phenomenological problems. The form 
\ref{(2.5)} of the expansion problem is more convenient for our purposes.

\SUBSECTION{Perfect factorization}
\label{ss2.5}

The most important requirement which asymptotic expansions should 
satisfy in the framework of perturbative QFT is that the dependence 
on heavy and light parameters should be fully factorized (see the 
discussion in subsect.~\ref{ss1.2}). Thus, if the expansion has the 
form%
\footnote{
It has become customary to refer to $c_i$ as to ``coefficient functions" 
and to $G_i$ as to ``matrix elements"---the terminology inherited 
from the theory of short-distance OPE.
}
\begin{equation}\label{(2.13)}
     G(Q,M,\kappa k,\kappa m) \asy{\kappa}{0}\sum_i c_i(Q,M) \,
     G_i(\kappa  k,\kappa  m),
\end{equation}
then not only $G_i$ should be independent of $Q$ and $M,$ but also 
$c_i(Q,M)$ of $k$ and $m.$ 

At the level of individual diagrams, the requirement of perfect factorization
means that individual terms on the r.h.s.\ of  \ref{(2.13)} should be 
be proportional to a power of $\kappa$ and be polynomials of 
$\log\kappa$.
This automatically implies that $c_i$
are sums of powers and logs of $Q$ and $M$, while $G_i$, of $k$ and $m$.
In a more general context, one should aim at obtaining expansions 
\ref{(2.13)} in such a form that the scaling properties of, say, 
$c_i(Q,M)$ with respect to $Q$ and $M$ be as simple as possible.%
\footnote{
A priori the form of dependence on the expansion parameter 
may not be known. In more general contexts---e.g.\ 
beyond the framework of perturbation theory---%
the language of dilatation group and its irreducible representations 
might be useful here.  
}

The requirement of perfect factorization is completely universal 
and not limited to the case of Euclidean regimes.

\SUBSECTION{Euclidean regimes}
\label{ss2.2} 

The problem \ref{(2.5)} is yet too general. However, the examples 
considered in sect.~\ref{s1} demonstrate that it will remain 
sufficiently meaningful if we limit ourselves to the special case of 
{\em Euclidean} asymptotic expansions. Namely, we assume that
\begin{equation}\label{(2.7)}
     \left( \sum_{l\in{\cal L}} Q_l \right)^2 < 0,\quad\quad
     {\rm for\ each\ subset\ {\cal L}\ of\ } l.
\end{equation}
The problem \ref{(2.5)} with the restriction \ref{(2.7)} can be called 
{\em Euclidean asymptotic expansion problem}. It 
allows a full and explicit solution \cite{fvt:q82}--\cite{IV}.

The meaning of the restriction \ref{(2.7)} is as follows. Perform the
Wick rotation of all integration momenta in Feynman diagrams into the
Euclidean region. Then only the light external momenta $k$ will remain
non-Euclidean. However, since one performs expansions with respect to
$k$ at $k=0,$ the fact of $k$ being non-Euclidean is inessential from
technical viewpoint. Therefore, from the very beginning one can assume
one works in a purely Euclidean theory. This drastically simplifies the
geometry of IR singularities%
\footnote{
For example, consider the singularity generated 
by the massless propagator $1/p^4.$ 
In Minkowskian region it is smeared over the 
{\em light cone} $p^2=0$ while in Euclidean region it is localized at 
the {\em point} $p=0.$
}
and, in the final respect, allows one to obtain an explicit and complete
solution of the general Euclidean expansion problem. 

Note also that it is for the Euclidean version of the general 
expansion problem that most clean phenomenological predictions can be 
made within QCD where asymptotic freedom allows one to perform 
renormalization-group improved calculations of the $Q$- and 
$M$-dependent coefficient functions of OPE-like expansions.

From now on, we are going to work in a purely Euclidean theory, so that 
the condition \ref{(2.7)} is satisfied automatically.

\SUBSECTION{Linear restrictions on heavy momenta}
\label{ss2.3}

For the heavy momenta on the l.h.s.\ of \ref{(2.2)} the following 
restriction due to momentum conservation holds:
\begin{equation}\label{(2.8)}
     \sum_l Q_l = - \kappa \sum_j k_j  \asy{\kappa}{0} 0,
\end{equation}
i.e. the sum of all heavy momenta is not itself heavy. A less trivial 
example of linear restrictions is provided by the light-by-light 
scattering problem (subsect.~\ref{ss1.5}).

In general, one can allow any number of linear restrictions of the form
\begin{equation}\label{(2.9)}
     \sum_l c_lQ_l = O(\kappa )
\end{equation}
to be imposed on the heavy momenta. In such a case we may assume that 
there exist linearly-independent momentum variables ${\bar Q}_i$ which 
are independent of $\kappa ,$ and each $Q_l$ is expressed as a linear 
combination of $Q_i$ and $k$ with coefficients independent of $\kappa 
:$
\begin{equation}\label{(2.10)}
     Q_l = Q_l(\bar Q,k).
\end{equation}
The restrictions \ref{(2.8)}, \ref{(2.9)} should thus be automatically 
satisfied.

One should take care not to impose restrictions such as would make the 
Green functions to be expanded ill-defined. However, 
even if the restricted Green function develops a singularity at 
$\kappa =0$ but is otherwise well-defined for all $\kappa \not=0,$ 
then the expansion problem still makes sense 
(the singularity will then show up as a contribution proportional 
to a non-positive power of $\kappa$) and can be treated by our methods%
\footnote{
It should be noted that in this case the singular dependence on $\kappa$ 
in the expansion will be localized within the matrix elements 
$G_i$ in \ref{(2.13)} 
while $c_i$ will be insensitive to what value $\kappa$ is set to.
}.

Among all possible linear restrictions, the so-called {\em natural 
restrictions} are of immediate physical interest. Such restrictions 
have the form
\begin{equation}\label{(2.11)}
      \sum_{l\in{ {\cal L}_\alpha}} Q_l = O(\kappa ) 
       \asy{\kappa}{0} 0 ,
\end{equation}
where ${\cal L}_\alpha$ is a subset of the set of all $l,$ and ${\cal 
L}_\alpha$ for different $\alpha$ are pairwise non-intersecting. In 
other words, the external lines corresponding to heavy momenta are 
organized into non-intersecting bunches, and within each such bunch 
the heavy momentum conservation holds separately.

One could also impose similar linear restrictions besides conservation 
on the light momenta $k.$ We will not do this, however, because in all 
interesting cases the final formulae will not be affected thereby 
(unlike the case of restrictions on heavy momenta).

\SUBSECTION{Contact terms}
\label{ss2.4}

As was discussed in subsect.~\ref{ss1.6}, it is natural to consider 
Green functions as distributions over momentum variables; the 
expansions in the sense of the distribution theory are characterized, 
from the pragmatic viewpoint, by presence of contact terms 
proportional to $\delta$-functions of heavy momenta $Q$ in $c_i$ on the r.h.s.
of \ref{(2.13)}.

More specifically, let $F(\bar Q)$ be a test function of the 
independent variables ${\bar Q}_i.$ Then what one has in fact to expand is
the expression
\begin{equation}\label{(2.12)}
     \int \Bigl( \prod_i d\bar Q_i \Bigr) F(\bar Q)\,
      G(Q(\bar Q,\kappa k),M,\kappa k,\kappa m,\mu)  
       \asy{\kappa}{0} \,?
\end{equation}
for arbitrary $F.$ This means, first, that all $c_i$ in \ref{(2.13)} are
distributions and, second, that \ref{(2.13)} retains its asymptotic nature 
after termwise integration against $F(\bar Q)$.

When there is no need to take into account contact 
terms, it is sufficient to consider $Q$ as parameters 
analogous to heavy masses. Then we will say that one deals with the 
simplified expansion problem without contact terms. One should only be 
careful to fix $Q$ at ``non-exceptional" values at which the 
$Q$-dependent factors of the expansion are smooth in $\bar Q.$ 
The exact criteria here are of little practical usefulness.

Note that in most cases of phenomenological interest the Green 
functions are integrable (although singular at some points) functions 
of the light momenta $k.$ Therefore for the sake of simplicity we do 
not introduce test functions for $k;$ it will be sufficient to assume 
that $k$ are fixed at some non-exceptional values. Normally, if the 
Green function and all terms of the asymptotic expansion are 
integrable functions of $k,$ then the expansion allows termwise 
integrations against $k$-dependent test functions; or, to put it 
formally, the operation of asymptotic expansion commutes with such 
integrations.

\vskip0.5cm
To summarize: we work within the perturbative QFT, always use the  
MS-scheme for UV renormalization, consider arbitrary models, and wish to 
obtain full asymptotic expansions for arbitrary Green functions for 
the class of Euclidean asymptotic regimes. We allow linear 
restrictions to be imposed on the heavy momenta of the Green 
functions, and consider and expand the latter in the sense of distributions. 
In the expressions to be obtained, the heavy 
and light parameters must be fully factorized.

%\newpage\thispagestyle{myheadings}\markright{}    
\vskip2.3cm
\centerline{{\large\bf BASIC IDEAS}}

\SECTION{Why expand products of singular functions?}
\label{s3}

The central notion of the theory of \asop\ 
is that of asymptotic expansion in the sense of distributions. 
In this section we present motivations for studying expansions of 
products of singular functions in the sense of distributions.
Although we have assumed to work in Euclidean theories, our 
reasoning here remains valid in the most general Minkowskian 
case as well.

\SUBSECTION{Local study of multiloop integrands}
\label{ss3.1}

We start with 
studying expansions of an arbitrary multiloop Feynman diagram $G$ 
which can be taken to be one-particle irreducible. Let $G(p,\kappa)$ be its 
integrand (prior to UV renormalization) where $p$ is the set of all 
loop momenta, and we have also explicitly shown the dependence on 
the expansion parameter $\kappa.$ The final goal is to obtain an 
expansion in $\kappa$ of the expression
\begin{equation}\label{(3.1)}
     {\bf R} \int dp\, G(p,\kappa),
\end{equation}
where 
          {\bf R}
is the UV 
          \rop,
while integration over $p$ is performed in infinite limits. A scrutiny 
reveals, however, that the starting point should be the expansion not 
of the integral \ref{(3.1)} but of the integrand considered as a 
distribution over $p.$

Indeed, whatever prescription for the \rop\ were chosen, one would not 
be relieved of the necessity to study contributions 
from finite regions of the integration space. Therefore, it 
is natural first to replace the \rop\ by a cut-off at large 
integration momenta. This allows one to strip the expansion problem of 
the collateral complications due to UV renormalization---anyway, 
many diagrams are not UV divergent at all. The cut-off can 
be chosen in the form of a smooth function $H(p/\Lambda)$ equal to zero at all 
sufficiently large $p:$
\begin{equation}\label{(3.2)}
     \int dp \, G(p,\kappa) H(p/\Lambda).
\end{equation}
A straightforward approach to expanding such integrals in $\kappa$ 
would be to first expand their integrands. However, the formal Taylor 
expansion of $G(Q,\kappa)$ in powers of $\kappa$ results, as a rule, 
in non-integrable singularities. For example, for a scalar propagator 
with a light mass $m=O(\kappa)$ one has:
\begin{equation}\label{(3.3)}
     {1\over{p^2+m^2}} = {1\over{p^2}} - m^2 {1\over{p^4}} + \ldots
\end{equation}
The singularities of the formal expansions like those on the r.h.s.\ 
of \ref{(3.3)} will be called {\em Euclidean infrared}---or simply  
IR---{\em singularities}. They correspond to the so-called soft singularities
which appear in studies of IR divergences in QED and QCD, 
while collinear singularities have no analogue in the Euclidean case.

The non-integrable singularities on the r.h.s.\ of \ref{(3.3)} preclude 
substitution of \ref{(3.3)} into \ref{(3.2)}. Emergence of such singularities 
also means that the integrals depend on $\kappa$ non-analytically. 
From the technical point of view, 
it is our major task to develop systematic algorithms 
for treating such singularities in this and more complicated cases and 
to construct correct asymptotic expansions starting from the formal 
series like \ref{(3.3)}. 

The traditional approach to the problem is to split the 
integration region into two parts: a neighbourhood ${\cal O}$ of the 
point $p=0,$ and the complement of ${\cal O}$, 
and then to extract the non-analytic part of the expansion coming 
from the integral over ${\cal O}$ by e.g.\ explicit integrations 
or using special tricks. Although it is the most straightforward way 
to establish the power-and-log nature of the expansion (cf.\ \cite{dslav:73}, 
\cite{french}), it is practically impossible to obtain explicit expressions
that were suitable for restoring the OPE-like combinatorial structures 
out of pieces coming from different diagrams. 
There is, nevertheless, another approach based on the ideas 
and notions of modern mathematics, namely, the theory of distributions 
\cite{schwartz}, which we now proceed to describe.

Our key observation is that the same propagator \ref{(3.3)} can appear 
within different diagrams and even in different places in the same 
diagram, so that it is natural to regard it as a distribution, the 
rest of $G(p,\kappa)$ playing the role of a test function. The same is 
also true of groups of factors in $G(p,\kappa)$ containing more 
than one propagator, and of $G(p,\kappa)$ itself. More precisely, let 
${\cal O}$ be a small region of the integration space. Let $G^{\rm 
sing}(p,\kappa)$ be the product of those factors from $G(p,\kappa)$ 
whose formal expansions are singular within ${\cal O}$. Denote the rest 
of $G$ as $G^{\rm reg}.$ There exist many Feynman diagrams leading to 
the same $G^{\rm sing}$ but different $G^{\rm reg}$'s. Therefore, 
$G^{\rm sing}$ can be regarded as a distribution in $\cal O$, 
while $G^{\rm reg}$ plays the role of a test function. (Such a reasoning 
leads to a very important localization property of the \asop\ considered 
below in subsect.~\ref{ss10.3}.)

One can also observe that besides troublesome propagators like 
\ref{(3.3)}, there are factors---e.g. propagators with a heavy mass, 
\begin{equation}\label{(3.4)}
     {1\over {M^2+(p-\kappa k)^2}},
\end{equation}
or vertex polynomials---whose expansion in $\kappa$ is harmless. 
Denote the product of all such factors in $G$ as $G^{\rm reg}$ and the 
rest of $G$ as $G^{\rm sing}$. Again, $G^{\rm reg}$ may vary for the 
same $G^{\rm sing},$ so that the latter plays the role of a 
distribution and the former of a test function.

Summarizing our observations, we arrive at the conclusion that the 
fundamental technical problem is to learn to expand in $\kappa$ 
the integrals like
\begin{equation}\label{(3.5)}
     \int dp \, \varphi(p) \, G(p,\kappa)
\end{equation}
with arbitrary test functions $\varphi(p)$;
this is exactly what is meant by saying that
one has to obtain an expansion of $G(p,\kappa)$ 
in the sense of distributions \cite{schwartz} 
(for a precise definition see below subsect.~\ref{ss4.1}).

The problem of studying expansions of products of singular functions 
with respect to parameters in the sense of the distribution theory 
will be referred to, in view of its paramount importance within our 
formalism, as {\em the Master problem}. It plays the crucial role in 
our theory, and all mathematical difficulties of analytical nature are 
concentrated in it.

\SUBSECTION{Remarks}
\label{ss3.2}

$\quad$ ($i$) 
If $\varphi$ in \ref{(3.5)} is localized within a small region 
$\cal O$, then one can study local structure of the expansion of $G(p,
\kappa)$ in each $\cal O$. Then an expansion valid for all $p$ can be 
easily obtained using standard techniques of the distribution theory. 
Such a localization trick proves to be a powerful instrument in 
obtaining expansions of products of singular functions---see below 
sect.~\ref{s10} and~\cite{fvt-vvv}.

($ii$) The transition from localized $\varphi$ to $\varphi=1$ requires
an analysis of asymptotics of the corresponding distributions in the
UV region $p\to\infty.$ For example, as will be shown below, studying
UV renormalization of the graph $G$ is equivalent to expansion of
$G(\lambda p,\kappa)$ considered as a distribution in $p\not=0,$ at
$\lambda \to\infty.$ One can show that this is a special case of the
Master problem with a special choice of the expansion parameter.
Studying how asymptotic properties of an expansion are affected by
taking the limit $p\to\infty$ reduces to studying double
$As$-expansions which is done in essentially the same manner as in the
case of expansions with respect to one parameter \cite{IV}.

($iii$) The expansion problem as formulated above requires to consider 
Feynman diagrams as distributions in heavy external momenta. Such a 
formulation is very natural in the context of the Master problem. 
Indeed, if the diagram is integrated over the heavy momenta $Q$ with 
the test function $\chi (Q),$ then instead of the integrals \ref{(3.5)} 
one should expand integrals of the form
\begin{equation}\label{(3.6)}
     \int  dp \, dQ \, \varphi(p)\, \chi (Q) \, G(p,Q,\kappa).
\end{equation}
Combine both sets of momenta into one: $p'=(p,Q),$ consider $G$ as a 
distribution in $p',$ and expand the expression
\begin{equation}\label{(3.7)}
     \int  dp' \, \varphi'(p') \, G(p',\kappa).
\end{equation}
One can see that, essentially, there is nothing new here as compared 
with \ref{(3.5)}. Choosing $\varphi'(p,Q)=\varphi(p)\,\chi (Q),$ we 
return to \ref{(3.5)}.

\SUBSECTION{Analogy with UV renormalization}
\label{ss3.3}

As was pointed out in \cite{fvt:q82} there exists an analogy between 
the problem of UV divergences in Bogoliubov's interpretation 
\cite{bog-shir} and the problem of singularities in expansions like 
\ref{(3.3)}: in both cases the divergences spring up as a 
result of formal manipulations with objects which by their nature are 
distributions. Therefore, Bogoliubov's solution of the UV problem 
provides an insight into how the divergences in expansions 
\ref{(3.3)} should be handled.

Let us reason as follows (for definiteness consider 
\ref{(3.3)}). An important observation is that the expansion \ref{(3.3)} 
(with 4-dimensional $p$)
allows (i.e.\ retains its meaning as an asymptotic expansion after) a 
termwise integration with any test function which is equal to zero in 
any arbitrarily small neighbourhood of the point $p=0.$ Therefore, the 
only way to improve upon the formal expansion \ref{(3.3)} is to add 
terms localized at $p=0$ to the r.h.s. So, it is to be expected 
that an expansion valid in the sense of distributions will look like
\begin{equation}\label{(3.8)}
     {1\over{p^2+m^2}} \asy{m}{0}  {1\over{p^2}} 
     - {m^2\over {p^4}} + c(m) \, \delta(p) + \ldots  ,
\end{equation}
where $c(m)$ is a scalar coefficient. 

The role of the new 
``counterterm" on the r.h.s.\ is two-fold. First, it should provide an 
infinite contribution to counterbalance the divergence due to the 
singularity of $p^{-4}$ and ensure integrability of the r.h.s.\ as a 
whole against arbitrary test functions. Second, it should ensure the 
asymptotic character of the expansion. Note that the dependence of 
$c(m)$ on $m$ can not be trivial, for there is no other place for the 
expected non-analyticity in $m$ to show up.

Thus, heuristically speaking, our problem is to obtain explicit expressions 
for the coefficients $c(m)$ in \ref{(3.8)} and in more complicated 
cases.

\SECTION{Asymptotic expansions of distributions}
\label{s4}

Although asymptotic expansions and distributions did appear 
simultaneously in various contexts in the literature, it seems that 
no systematic study of the corresponding mathematical notions 
has ever been undertaken. 
Below are presented background mathematical definitions and 
some results of a general character. 
Since we are not aiming at attaining 
full mathematical formalization in the present paper, some technical 
details are omitted and others treated at a heuristic level. A 
substantial formal treatment of the subject can be found 
in \cite{fvt-vvv}.

\SUBSECTION{Definition}
\label{ss4.1}

Let $F(p,\kappa)$ be a distribution over $p$ depending on the 
parameter $\kappa.$ We say that the series
\begin{equation}\label{(4.1)}
     F(p,\kappa) \asy{\kappa}{0} \sum_n \kappa^n F_n(p,\kappa),
\end{equation}
represents an asymptotic expansion in the sense of distributions if a 
termwise integration of \ref{(4.1)} with any test function $\varphi(p)$ 
results in a correct asymptotic expansion in the usual sense, i.e.
\begin{equation}\label{(4.2)}
     \int dp\,  F(p,\kappa) \, \varphi(p)      \asy{\kappa}{0}
     \sum_{n\le N}\kappa^n \left[\int  dp \, F_n(p,\kappa) \, 
     \varphi(p)\right] + o(\kappa^N).
\end{equation}
This implies that each $F^n$ is a well-defined distribution over $p.$ 
We say in such cases that the expansion \ref{(4.1)} {\em allows 
integration} with arbitrary test functions.

We will usually have to deal with expansions such that each $F_n$ is a 
polynomial in $\log \kappa :$
\begin{equation}\label{(4.3)}
     F_n(p,\kappa) = \sum_{i=0}^{I_n}\log^i \! \kappa\, F_{n,i}.
\end{equation}
This form of $\kappa$-dependence will be called {\em soft} (see also 
below subsect.~\ref{ss4.3}).

\SUBSECTION{Uniqueness}
\label{ss4.2}

If an expansion in powers and logarithms (like \ref{(4.1)}, \ref{(4.3)})
exists, it is unique.  Indeed, integrating \ref{(4.1)} with test
functions we get asymptotic expansions in the usual sense, for which
the uniqueness is an elementary fact~\cite{olver}. It is important to
understand that there is no point in discussing uniqueness unless the
form of dependence on the expansion parameter is explicitly fixed, as
in
\ref{(4.1)}, \ref{(4.3)}.

The property of uniqueness is extremely important. First, it allows 
one to immediately obtain useful technical results like the 
localization property of the \asop\ for products of singular functions 
(subsect.~\ref{ss10.3}) or its commutativity with multiplications by 
polynomials (see below subsect.~\ref{ss11.5}). Second, it implies that the 
expansions which we will derive cannot, in a sense, be improved upon.

\SUBSECTION{Regularization}
\label{ss4.3}

It is often convenient to introduce a regularization into both sides 
of \ref{(4.1)}. We assume that it is controlled by a parameter 
$\epsilon ,$ and taking it off corresponds to $\epsilon\to0.$ Denote 
the regularized version of the distribution $f$ as $\{f\}^\epsilon,$ 
so that
\begin{equation}\label{(4.4)}
     f(p) = \lim_{\epsilon\to0} \, \{f(p)\}^\epsilon
\end{equation}
on each test function $\varphi.$ Then instead of \ref{(4.1)}, one can 
write
\begin{equation}\label{(4.5)}
     \{F(p,\kappa)\}^\epsilon \asy{\kappa}{0} \sum_n \kappa^n
     \{F_n(p,\kappa)\}^\epsilon.
\end{equation}
We stress that the expression \ref{(4.5)} need not be a true asymptotic 
expansion for $\epsilon\not=0,$ but must only become such upon taking 
the limit $\epsilon\to0.$ Still, the dimensionally regularized 
expansions that we obtain for the standard Feynman integrands are 
asymptotic ones even for $\epsilon \not=0.$ This peculiar fact, 
however, seems to have no practical implications.

Within the dimensional regularization, \ref{(4.3)} should be modified. 
Indeed, the dependence on $\kappa$ takes the form
\begin{equation}\label{(4.6)}
     \{F_n(p,\kappa)\}^\epsilon = \sum_{i=0}^{I_n}
     \kappa^{\epsilon i} \{F'_{n,i}\}^\epsilon,
\end{equation}
which will also be qualified as {\em soft}.

An important fact is that although the expansion \ref{(4.6)} as a whole 
does have a well-defined finite limit at $\epsilon \to 0$ by 
construction, the quantities $\{F'_{n,i}(p)\}^\epsilon,$ in general, 
do not: if one integrates such an object with a test function then the 
resulting expression will contain pole singularities as $\epsilon \to 
0.$ It is, of course, always possible to rearrange the r.h.s.\ in such 
a way as to make each term have a finite limit at $\epsilon \to 0.$ To 
this end it is sufficient to form special linear combinations of the 
braced expressions. However, such a representation is of 
little practical interest, at least as long as one stays within 
the calculational framework of dimensional regularization and
perturbative QFT. This is 
because in applications one has to perform calculations not 
with individual diagrams but with their sums. Summations over full 
sets of relevant diagrams result in considerable cancellations between 
them, especially in gauge models like QCD, so that the efforts spent 
on separate treatment of individual diagrams will be   lost. On 
the other hand, there exist formulae \cite{fvt:ope83}, 
\cite{alg:83} connecting the 
phenomenological quantities that one needs to calculate in the final 
respect, with the corresponding full sums of diagrams. Such formulae 
are compact and, as experience shows \cite{ope3loops}, very 
convenient, which explains our use of expansions in the dimensionally 
regularized form.

\SUBSECTION{$\kappa$-dependent test functions}
\label{ss4.4}

As is clear from the motivations presented in subsect.~\ref{ss3.1}, 
``natural" test functions that emerge after localization of the 
expansion problem, depend on the expansion parameter $\kappa,$ so that 
a typical problem is to expand expressions like
\begin{equation}\label{(4.7)}
     \int dp \, \varphi(p,\kappa) \, F(p,\kappa).
\end{equation}
Strictly speaking, knowing how to expand $F(p,\kappa)$ in $\kappa$ in
the sense of distributions allows one only to expand integrals with
$\varphi$ independent of $\kappa$ (or with trivial---e.g.\
polynomial---dependence).  Nevertheless, under certain conditions the
expansion of \ref{(4.7)} can be obtained by substituting the expansion
\ref{(4.1)} for $F$ and a similar expansion for $\varphi$ and
reordering the terms in the product in ascending powers of $\kappa$.
There are two such conditions.%
\footnote{
Our discussion at this point is simplified in order to avoid somewhat
cumbersome technicalities---the inequalities describing continuity
properties of the distributions involved, the corresponding seminorms
etc. Complete details can be found in \cite{fvt-vvv}.  Suffice it to
mention that the remainder term of expansions of expressions like
\ref{(4.7)} 
is studied using representations of the form 
(for simplicity we consider here
expansions to leading order in $\kappa$)
$\varphi F-\varphi_0 F_0=\varphi_0(F-F_0)+(\varphi-\varphi_0)F$, 
where
subscript 0 denotes the expansion of the corresponding term to
$o(\kappa^0)$. Analyzing the non-trivial second
addendum on the r.h.s.\ one arrives at, loosely speaking, the two
conditions in the main text.
\relax}

($i$) The first one is of little practical significance and is 
included for completeness' sake. Both $F(p,\kappa)$ and all $F_n(p,
\kappa)$ must be distributions in the precise meaning of the word, 
i.e.\ (a) be linear functionals defined on all Schwartz test functions
and (b) possess certain properties of continuity.
However, there is no
practical chance to encounter a functional satisfying (a) but not (b),
as such functionals can only be proved to exist, by making use of the
notorious Axiom of Choice, outside the realm of practical mathematics.
There would be no need to mention (b), but there has not yet been
enough time for the alternative Axiom of Determination due to
Myczelsky and Steinhaus (see e.g. \cite{kanovei}) to purge the
mathematical textbooks of pathological counterexamples.

($ii$) The really important condition imposes restrictions on the 
asymptotic expansion of the test function. Namely, if $\varphi$ is 
expanded as
\begin{equation}\label{(4.8)}
     \varphi(p,\kappa) \simeq \sum_n \kappa^n \, \varphi_n(p,\kappa)
\end{equation}
(where the dependence of $\varphi_n(p,\kappa)$ on $\kappa$ is soft, as in 
\ref{(4.3)}), then the expression
\begin{equation}\label{(4.9)}
     \bigl[\varphi(p,\kappa) - \sum_{n\le N} \kappa^n \, 
     \varphi_n(p,\kappa) \bigr]/\kappa^N
\end{equation}
should tend to zero as $\kappa\to 0$ in the following sense. The 
expression in the square brackets must be non-zero only within a 
compact region $K$ of the integration space for all $\kappa\not=0;$ 
the expression \ref{(4.9)}, as well as its derivatives in $p,$ must 
tend to zero uniformly with respect to $p.$

For the Euclidean expansion 
problem within the standard perturbation theory one has to 
consider only $\kappa$-dependent test functions of the form 
(cf.\ \ref{(6.39)})
\begin{equation}\label{(4.10)}
     \varphi(p,\kappa) = \psi (p) \, f(p,\kappa),
\end{equation}
where $\psi (p)$ is a Schwartz test function, while $f(p,\kappa)$ is a 
rational function of $p$ and $\kappa$ and such that all its 
singularities are localized at points $p$ where $\psi (p)=0,$ for all 
$\kappa.$ It is not difficult to understand that for such $\varphi(p,
\kappa)$ condition ($ii$) is satisfied.

\SECTION{The extension principle}
\label{s5}

Having developed a suitable language, we can turn to the expansion 
problem proper. 
Our purpose now is to offer a very general and abstract 
framework for studying asymptotic expansions of distributions. We will 
introduce some notions and present simple but important 
propositions under a general heading of {\em extension principle} 
which constitutes, essentially, a context in which to work with 
explicit problems. To make it work for a particular problem, specific 
estimates and bounds must be established, which may technically be the 
most cumbersome part of the problem. But the importance of the 
conceptual framework of the extension principle consists in the fact 
that it allows one to guess the structure of the expansions to be 
obtained, shows what kind of estimates are required, and is, 
therefore, a powerful heuristic tool.

\SUBSECTION{Motivations}
\label{ss5.1}

Let $F(p,\kappa)$ be a distribution in $p \in  P$ (where $P$ is a 
Euclidean space) depending on the expansion parameter $\kappa.$ For 
simplicity assume that $F(p,\kappa)$ is an integrable function of $p$ 
for all~$\kappa\not=0.$

A typical situation which we will have to deal with can be described 
as follows. One can obtain (e.g. using the 
Taylor theorem) an approximation%
\footnote{
In this section we consider not full asymptotic 
expansions, but approximations to a given order $o(\kappa^N).$ 
Transition to infinite asymptotic series can also be treated in an 
abstract manner but there is not much wisdom to be gained from it, so 
we postpone the discussion of this point till the next section where 
an example is analyzed in detail.
}
\begin{equation}\label{(5.1)}
     F(p,\kappa) \asy{\kappa}{0} \bar F_N(p,\kappa) + o(\kappa^N),
\end{equation}
where $\bar F_N(p,\kappa)$ is an integrable function for $p \in {\cal 
O}$ where $\cal O$ is an open region in $P,$ and has non-integrable 
singularities at the points from $P\backslash {\cal O};$ the notation%
\footnote{
The more familiar notation $O(\kappa^N)$ is often used in theoretical 
physics to denote terms behaving as $\kappa^N$, possibly with 
logarithmic corrections. Note that a priori we can say nothing about 
the behaviour of the remainder of the full expansion we wish to 
construct except that it should vanish faster then the last term 
retained. That is why the $o$-notation is preferable to the 
$O$-notation in our case.
}
$o(\kappa^N)$ stands for the terms that vanish faster than $\kappa^N$ 
as $\kappa \to  0.$ Suppose that the expansion \ref{(5.1)} is such that 
it is valid pointwise for each fixed $p \in  {\cal O},$ and allows 
termwise integration with arbitrary test functions localized within 
$\cal O$. It is required to obtain an approximation for $F(p,\kappa)$ 
to the same order $o(\kappa^N)$ valid in the sense of distribution 
theory on the entire $P.$

Let us look at the problem from a different angle. Let $L$ be the 
vector space of all test functions $\varphi(p),$ $p \in  P,$ and $L_N$ 
the set of the test functions $\varphi$ localized within $\cal O$. 
$L_N$ forms a vector subspace in $L$. The distribution $F(p,\kappa)$ 
is, by definition, a linear functional on $L,$ which we will denote as 
$F_\kappa.$ On the other hand, $\bar F_N(p,\kappa)$ generates a linear 
functional $\bar F_{\kappa,N}$ defined on $L_N,$ and the restriction 
of $F_\kappa$ to $L_N$ can be approximated as
\begin{equation}\label{(5.2)}
     \left[F_\kappa\right]_{L_N} \asy{\kappa}{0} 
     {\bar F}_{\kappa,N} + o(\kappa^N),
\end{equation}
which is to be understood so that \ref{(5.2)} should be valid in the 
usual sense if one evaluates both sides of \ref{(5.2)} on arbitrary 
vectors $\varphi \in  L_N.$ The problem is to construct an 
approximation
\begin{equation}\label{(5.3)}
     F_\kappa  \asy{\kappa}{0} F_{\kappa,N} + o(\kappa^N),
\end{equation}
which would be valid on the entire $L.$ Therefore, one can say that 
the problem consists in extending the approximating functional defined 
on the subspace onto the entire space in such a way as to preserve its 
approximation properties. Its solution is provided by the so-called 
{\em extensions principle} \cite{fvt:q82}.%
\footnote{
Our extension principle belongs to the Hahn-Banach type results. 
The classical Hahn-Banach theorem (see any textbook on functional analysis) 
considers extension of functionals 
preserving the property of being bounded by a seminorm. In our case,
the extensions should preserve the approximation property.
}

\SUBSECTION{The extension principle}
\label{ss5.2}

Let $L$ be a linear space and $F_\kappa$ a linear functional on it 
which depends on the expansion parameter $\kappa.$ Assume there exists 
a subspace $L_N \subset L$ and a functional $\bar F_{\kappa,N}$ 
defined on $L_N,$ such that $\bar F_{\kappa,N}$ approximates 
$F_\kappa$ to order $o(\kappa^N)$ on $L_N.$ More precisely, for each 
$\varphi \in L_N$
\begin{equation}\label{(5.4)}
     \mathopen< F_\kappa - {\bar F}_{\kappa,N} , \bar \varphi \mathclose> = o(\kappa^N).
\end{equation}
We wish to construct a functional $F_{\kappa,N}$ defined and 
approximating $F_\kappa$ to order $o(\kappa^N)$ on the entire $L,$ and 
coinciding with $\bar F_{\kappa,N}$ on $L$.

To this end, let $\varphi_\alpha$ be a set of vectors transverse to 
$L_N$ and such that each $\varphi \in  L$ can be uniquely expanded as
\begin{equation}\label{(5.5)}
     \varphi = \sum k_\alpha \varphi_\alpha + {\bar \varphi} ,
\end{equation}
where $\varphi \in L_N.$ It is sufficient to define $F_{\kappa,N}$ on 
$\varphi_\alpha,$ because on $\bar \varphi$ its value is known. Now, the 
desired definition is:
\begin{equation}\label{(5.6)}
     \mathopen< F_{\kappa,N} , \varphi_\alpha \mathclose> = \mathopen< F_\kappa , \varphi_\alpha \mathclose> 
     + o(\kappa^N),
\end{equation}
where we have indicated that the definition allows an arbitrariness 
of order $o(\kappa^N),$ which should not be surprising.

It is not difficult to check that $F_{\kappa,N}$ thus defined is 
unique within accuracy $o(\kappa^N).$

\SUBSECTION{Counterterms and consistency conditions}
\label{ss5.3}

The above construction can be represented as the following two-step 
procedure. Suppose it is possible to find an extension $\bar F^{\rm 
R}_{\kappa,N}$ of $\bar F_{\kappa,N}$ from $L_N$ onto the entire $L,$ 
which does not necessarily approximate $F_\kappa$ on $L$ but has, 
perhaps, some other nice properties. Let $\delta_\beta$ be a set of 
functionals on $L$ such that
\begin{equation}\label{(5.7)}
     \mathopen< \delta_{\beta} , \varphi \mathclose> = 0,  \quad  {\rm for\ all\  }\varphi \in  L_N,
\end{equation}
and
\begin{equation}\label{(5.8)}
     \mathopen< \delta_\beta , \varphi_\alpha \mathclose> = \delta_{\alpha,\beta} \quad   
     {\rm(the\ Kronecker\ symbol).}
\end{equation}
(Instead of \ref{(5.8)}, one can simply require that $\delta_\beta$ be a 
maximal linearly independent set of functionals satisfying 
\ref{(5.7)}.) Then $F_{\kappa,N}$ should be representable as
\begin{equation}\label{(5.9)}
     F_{\kappa,N} = {\bar F}_{\kappa,N}^{\rm R} 
     + \sum_\alpha c^{\rm R}_{\alpha} (\kappa) \delta_\alpha ,
\end{equation}
where $c^{\rm R}_\alpha(\kappa)$ are unknown coefficients. $F_{\kappa,N}$ thus 
defined reduces to $\bar F_{\kappa,N}$ on each $\bar \varphi \in  L_N,$ 
so owing to \ref{(5.4)} it is sufficient to require that \ref{(5.9)} 
approximate $F_\kappa$ to $o(\kappa^N)$ on each $\varphi_\alpha:$
\begin{equation}\label{(5.10)}
     \mathopen< [ F_\kappa - F_{\kappa,N} ] , \varphi_\alpha \mathclose> = o(\kappa^N).
\end{equation}
This {\em consistency condition} \cite{fvt:q82}, \cite{I} allows one 
to find the unknown coefficients in \ref{(5.9)}:
\begin{equation}\label{(5.11)}
     c^{\rm R}_{\alpha} (\kappa) 
     = 
     \mathopen< [ F_\kappa - {\bar F}_{\kappa,N}^{\rm R} ] , \varphi_\alpha \mathclose> 
     + o(\kappa^N).
\end{equation}
It is not difficult to check the equivalence of this construction and 
that of subsect.~\ref{ss5.2}. Note that the recipe \ref{(5.9)} is more 
in the spirit of the distribution theory, for one can find a correct 
form of the required extension \ref{(5.9)}, up to some coefficients, 
practically without reference to test functions that only make their 
appearance in the consistency conditions \ref{(5.10)}.

\SUBSECTION{Remark}
\label{ss5.4}

Assume that $\varphi_\alpha$ in \ref{(5.10)} depend on a parameter 
$\Lambda$ without violating \ref{(5.8)}: 
$\varphi_\alpha=\varphi_\alpha^\Lambda,$ and assume that there exists 
a finite limit:
\begin{equation}\label{(5.12)}
     \lim_{\Lambda\to\infty} \mathopen< \left[ F_\kappa 
     - {\bar F}_{\kappa,N}^{\rm R} \right] , \varphi_\alpha^\Lambda \mathclose>
\end{equation}
even if $\varphi_\alpha^\Lambda$ itself does not converge to any
vector in $L.$ Now, if taking the limit $\Lambda \to \infty$ does not
take one outside the $o(\kappa^N)$ uncertainty, then instead of
\ref{(5.11)} one may take:
\begin{equation}\label{(5.14)}
     c^{\rm R}_{\alpha} (\kappa) 
     = \lim_{\Lambda\to\infty} \mathopen<\left[ F_\kappa 
                              - {\bar F}_{\kappa,N}^{\rm R} \right] , 
           \varphi_\alpha^\Lambda  \mathclose> + o(\kappa^N).
\end{equation}
The option  to take such limits proves to be very important---see 
below subsect.~\ref{ss6.5}.

\SUBSECTION{Regularized form of the extension principle}
\label{ss5.5}

There is a third way to represent the extended approximating 
functional $F_{\kappa,N},$ which will be widely used in what follows. 
Define a {\em regularization} of $\bar F_{\kappa,N},$ i.e.\ a 
functional $\bar F_{\kappa,N}^\epsilon$ defined on the entire $L$
which depends on an additional {\em regularization parameter\/} $\epsilon$ 
and such that  
$\bar F_{\kappa,N}^\epsilon \to \bar F_{\kappa,N}$ on $L_N$ as  $\epsilon \to 0,$ 
i.e.
\begin{equation}\label{(5.15)}            
     \lim_{\epsilon\to0} \mathopen<\left[ {\bar F}_{\kappa,N}^{\epsilon}
     - {\bar F}_{\kappa,N} \right] , {\bar \varphi}  \mathclose> = 0,
     \quad {\rm for\ each\ \ } {\bar \varphi} \in L_N.
\end{equation}
It is easy to check that the definition \ref{(5.9)} can now be 
rewritten as
\begin{equation}\label{(5.16)}
     F_{\kappa,N} = \lim_{\epsilon\to0}\left[ { {\bar F}_{\kappa,N}^{\epsilon} 
     + \sum_\alpha c_\alpha^\epsilon(\kappa) \delta_\alpha  } \right]
     + o(\kappa^N),
\end{equation}
where
\begin{equation}\label{(5.17)}
     c_\alpha^\epsilon(\kappa) = \mathopen< \left [ F_\kappa 
     - {\bar F}_{\kappa,N}^{\epsilon}\right ] , 
     \varphi_\alpha \mathclose> + o(\kappa^N).
\end{equation}
Here one can also use parameterized sequences of $\varphi_\alpha$ as in \ref{(5.14)}.

\SUBSECTION{Summary}
\label{ss5.6}

The procedure of constructing an approximating functional now is as 
follows. One has the linear functional $F_\kappa$ on $L$ which is to 
be expanded in $\kappa,$ and one is given another functional $\bar 
F_{\kappa,N}$ which approximates $F_\kappa$ to a given accuracy on 
many but not all vectors from $L.$ The key point is to identify the 
maximal subspace $L_N$ on which $\bar F_{\kappa,N}$ approximates 
$F_\kappa.$ After this is done, the rest consists essentially in 
adding to $\bar F_{\kappa,N}$ a linear combination of functionals 
$\delta_\alpha$ which vanish on $L_N,$ with coefficients determined by 
the consistency conditions.

In the context of subsect.~\ref{ss5.1}, 
the fact that $\delta_\alpha$ must be 
zero on $L_N$ can be interpreted as that the corresponding 
distributions are localized in the complement $P\backslash{\cal O}$ of 
the open set $\cal O$.

Let us now turn to an example.

\SECTION{Example: expansion of \protect{$(p^2 + \kappa^2)^{-1}.$}}
\label{s6}

The abstract considerations of the previous section provide us with a 
general approach to constructing expansions of distributions. Now we 
wish to show how it works in practice. There are also some 
important details---e.g. the transition from approximations to a given 
order to full asymptotic expansions---which can be more conveniently 
explained using an example.

\SUBSECTION{Formal expansion}
\label{ss6.1}

Let $P$ be a 4-dimensional Euclidean space; its elements will be 
denoted as $p.$ Consider the scalar propagator
\begin{equation}\label{(6.1)} 
     F(p,\kappa) = {1\over {p^2+\kappa^2}}.
\end{equation}
We wish to obtain its expansion for $\kappa\to 0$ valid in  the  sense 
of  the distribution theory.

First consider the formal expansion to order $o(\kappa^2):$
\begin{equation}\label{(6.2)}
     {1\over {p^2+\kappa^2}} = {1\over p^2}
     - \kappa^2 {1\over {p^4}} + o(\kappa^2).
\end{equation}

It holds pointwise for any $p \not= 0.$ Moreover, it allows 
integration with any test function which is equal to zero in some 
small neighbourhood of $p = 0.$ But most important is that although it 
does not allow integrations with arbitrary test functions owing to the 
non-integrable singularity in the $O(\kappa^2)$ term on the r.h.s., 
still it allows integration 
with test functions $\bar \varphi$ which only 
satisfy the condition $\bar \varphi(0) = 0,$ apparently because then 
$\bar \varphi(p) \sim p,$ $p\to 0,$ and the singularity is effectively 
suppressed.

\SUBSECTION{Approximation properties of the formal expansion}
\label{ss6.2}

More precisely, one has to prove that
\begin{equation}\label{(6.3)}
     \int d^4p \,  \bar \varphi(p) 
     \left[ {1\over {p^2+\kappa^2}} 
     - {1\over {p^2}} + \kappa^2 {1\over {p^4}}\right] = o(\kappa^2).
\end{equation}
The simplest way
to do this is as follows. One splits the integral as
\begin{equation}\label{(6.4)}
     \int = \int_{|p|<c\kappa} + \int_{|p|>c\kappa}.
\end{equation}
For $\vert p\vert < c\kappa ,$ one rescales $p\to \kappa p,$ uses the 
scaling properties of the bracketed expression and the fact that $\bar 
\varphi(p) \sim p,$ $p\to 0,$ and finds that
\begin{equation}\label{(6.5)}
      \int_{|p|<c\kappa} \sim O(\kappa^3) = o(\kappa^2).
\end{equation}
For $\vert p\vert > c\kappa ,$ one uses the Taylor theorem to estimate 
the bracketed expression by an expression similar to the first 
discarded term in \ref{(6.2)}, namely, ${\rm const} \times \kappa^4/p^6,$ 
estimates $\bar \varphi(p)$ by 
${\rm const} \times \vert p\vert$ and integrates 
over $c\kappa < \vert p\vert < M < + \infty$  where $M$ depends on 
$\bar \varphi.$ The result is
\begin{equation}\label{(6.6)}
      \int_{|p|>c\kappa} \sim O(\kappa^4 \log\kappa) = o(\kappa^2).
\end{equation}
The desired result \ref{(6.3)} follows immediately from \ref{(6.4)}--\ref{(6.6)}.

We have discussed the proof in detail because the same 
steps are performed in the most general case as well.

Now we are going first to implement the recipe of subsect.~\ref{ss5.3}. 
After this is done, we will cast the results into a regularized form 
corresponding to subsect.~\ref{ss5.4}, using the dimensional 
regularization.

\SUBSECTION{Intermediate operation $\tilde{\bf R}$}
\label{ss6.3}

Before proceeding further, let us construct a distribution 
${\tilde{\bf R}}[p^{-4}]$ defined on all test functions and such that 
the equality
\begin{equation}\label{(6.7)}
     {1\over p^4} = {\tilde{\bf R}} \left[{1\over p^4}\right]
\end{equation}
is valid on the test functions $\bar \varphi$ defined above. An explicit 
expression for such a distribution can be found, if one fixes any one 
test function $\psi (p)$ such that $\psi (0) = 1,$ and integrates both 
sides of \ref{(6.7)} with $\varphi(p)-\varphi(0)\psi (p).$ Then one 
finds:
\begin{equation}\label{(6.8)}
     \int d^4p\, {\tilde{\bf R}} \left[{1\over p^4}\right] \,\varphi(p)
      = \int d^4p \left\{ {1\over p^4} \left[ \varphi(p) 
                                    - \varphi(0) \psi (p)\right] 
      + c \delta (p) \varphi(p)\right\} ,
\end{equation}
where
\begin{equation}\label{(6.9)}
     c = \int d^4p \, \, {\tilde{\bf R}} 
     \left[{1\over p^4} \right] \psi (p)
\end{equation}
is a single constant to be fixed---which can be done arbitrarily---in 
order that ${\tilde{\bf R}}[p^{-4}]$ be fully defined. Introducing a 
regularization into the r.h.s.\ of \ref{(6.8)}, one gets a 
representation of ${\tilde{\bf R}}[p^{-4}]$ ``in terms of infinite 
counterterms":
\begin{equation}\label{(6.10)}
     {\tilde{\bf R}} \left[{1\over p^4} \right] 
     = \lim_{\epsilon\to0} 
     \left[ \left\{ {1\over p^4} \right\}^\epsilon 
     + Z_\epsilon \delta (p) \right],
\end{equation}
where $\{\ldots\}^\epsilon$ denotes the regularization, e.g.
\begin{equation}\label{(6.11)}
     \left\{ {1\over p^4} \right\}^\epsilon  
     = {1\over{p^4+\epsilon^4}}, \quad 
     {\rm or} \quad  
     \left\{ {1\over p^4} \right\}^\epsilon  
     =      \Theta(\vert p\vert >\epsilon ) {1\over p^4} \quad  {\rm etc.},
\end{equation}
where we have introduced a convenient function
\begin{equation}\label{(6.12)}
     \Theta(x) \equiv 1, {\ \rm if\ the\ logical\ expression\ } 
                       x {\rm\ is\ ``true",}
\end{equation}
\begin{equation}
                    0, {\ \rm otherwise.}  \hbox{\kern32mm}
\end{equation}
It might be helpful to note that the distribution ${\tilde{\bf 
R}}[p^{-4}]$ defined above is similar to the so-called 
``+"-distribution,
\begin{equation}\label{+distr }
     \int_0^1 \varphi(x) \left[{1\over x}\right]_+  dx 
     \equiv \int_0^1 \frac{ \left[\varphi (x)-\varphi (0)\right] }{x} \, dx ,
\end{equation}
which is widely used in the QCD parton model calculations.

\SUBSECTION{Consistency condition}
\label{ss6.4}

According to the recipe of subsect.~\ref{ss5.3}, we should first 
redefine the r.h.s.\ of \ref{(6.2)} in such a way as to make it 
well-defined on all test functions. To this end we replace $p^{-4}$ in 
\ref{(6.2)} by ${\tilde{\bf R}}[p^{-4}]$ without changing values of the 
r.h.s.\ on $\bar \varphi,$ and obtain an expression
\begin{equation}\label{(6.13)}
     {1\over p^2+\kappa^2} = {1\over p^2} - \kappa^2 \,{\tilde{\bf R}} 
                         \left[{1\over p^4} \right] + o(\kappa^2),
\end{equation}
which holds, as \ref{(6.2)}, on $\bar \varphi,$ but unlike \ref{(6.2)}, the 
r.h.s.\ is well-defined on all $\varphi(p).$ 

The only distribution that 
can be added to the r.h.s.\ of \ref{(6.13)} without violating the 
asymptotic character of \ref{(6.13)} on $\bar \varphi$ is $\delta (p).$
The immediate conclusion is that the expansion of $(p^2 + \kappa^2)^{-1}$ to 
order $o(\kappa^2)$ in the sense of the distribution theory can only 
have the form
\begin{equation}\label{(6.14)}
     {1\over p^2+\kappa^2} 
     = {1\over p^2} 
     - \kappa^2 \, {\tilde{\bf R}} \left[ {1\over p^4} \right] 
     + c_0^{\rm R} (\kappa) \delta (p) 
     + o(\kappa^2),
\end{equation}
and is defined up to a single numeric-valued (non-analytic) function 
of $\kappa, c_0^{\rm R}(\kappa).$

In order to determine $c_0^{\rm R}(\kappa),$ recall the 
prescription of subsect.~\ref{ss5.3}. One notes that any test 
function $\varphi(p)$ can be represented as
\begin{equation}\label{(6.15)}
     \varphi(p) = \varphi(0) \psi (p) + \bar \varphi(p),
\end{equation}
where $\psi (p)$ has already been defined and $\bar \varphi(0) = 0.$ On 
such $\bar \varphi$ \ref{(6.14)} degenerates into \ref{(6.2)} and, 
therefore, is valid. Hence, in order that \ref{(6.14)} be valid on 
all test functions, one only has to ensure that it holds on $\psi 
(p).$ Let us take this as a definition of $c_0^{\rm R}(\kappa)$ (a 
``consistency condition", cf.\ \ref{(5.10)}). Then, integrating both 
sides of \ref{(6.14)} with $\psi (p),$ one gets:
\begin{equation}\label{(6.16)} 
     c_0^{\rm R}(\kappa) = \int d^4p \, \psi (p) \left\{
         {1\over p^2+\kappa^2} - {1\over p^2} 
        + \kappa^2 {\tilde{\bf R}} \left[ {1\over p^4} \right] 
          \right\}
        + o(\kappa^2).
\end{equation}

Note that $c_0^{\rm R}(\kappa)$ is determined at this stage only up to 
$o(\kappa ^2).$ This agrees with the fact that $\psi$  has been chosen 
arbitrarily, because a change in $\psi$  preserving the condition 
$\psi (0)=1$ will change $c_0^{\rm R}(\kappa )$ by a term of order 
$o(\kappa ^2).$

\SUBSECTION{Dependence of counterterms on $\kappa$}
\label{ss6.5}

Although eqs.\ref{(6.14)}, \ref{(6.16)} provide a correct expansion of the 
propagator in the sense of the distribution theory to $o(\kappa^2),$ 
one further important refinement is possible in the spirit of 
subsect.~\ref{ss5.4}. It is natural to try to find a $\psi (p)$ such 
that would simplify \ref{(6.16)} as much as possible. To this end, 
notice that the square-bracketed expression in \ref{(6.16)} is bounded 
by ${\rm const} \times  \kappa^4/p^6$ at $p\to \infty .$ Therefore, $c_0^{\rm 
R}(\kappa)$ will be changed by a finite $o(\kappa^2)$ contribution if 
one simply takes $\psi (p)\equiv1$. More precisely, let $\psi (p) = 
\Phi (p/\Lambda )$ and let $\Lambda \to \infty .$ The resulting 
expression for $c_0^{\rm R}(\kappa)$ is:
\begin{equation}\label{(6.17)}
     c_0^{\rm R}(\kappa) = \int \, d^4p \left\{
       {1\over p^2+\kappa^2} - {1\over p^2} 
     + \kappa^2 {\tilde{\bf R}} \left[ {1\over p^4} \right] \right\}.
\end{equation}

The distribution ${\tilde{\bf R}}[p^{-4}]$ has a simple scaling 
behaviour:
\begin{equation}\label{(6.18)}
     F(\lambda p) = \lambda ^{-4} \left[ F(p) + {\rm const} \times \log\lambda 
     \, \delta (p) \right]
\end{equation}
(which can be easily deduced from the definition \ref{(6.8)}). 
Therefore, replacing $p\to \kappa p$ in \ref{(6.17)} and using 
\ref{(6.18)}, one gets:
\begin{equation}\label{(6.19)}
     c_0^{\rm R}(\kappa) = \kappa^2 \left[ K_1 + K_2\log\kappa \right],
\end{equation}
where $K_1$ and $K_2$ are constants independent of $\kappa.$ The fact 
of a simple dependence of $c_0^{\rm R}(\kappa)$ as defined in 
\ref{(6.17)} on $\kappa$ singles out this definition from the family 
\ref{(6.16)}.

\SUBSECTION{The remainder of the expansion}
\label{ss6.6}

Another way to check the validity of the expansion \ref{(6.14)} with 
$c_0^{\rm R}(\kappa)$ given by \ref{(6.17)}, is to notice that the 
remainder can be represented as:
\begin{equation}\label{(6.20)}
     \int\ d^4p \, \varphi(p) \left\{ 
         {1\over p^2+\kappa^2} 
       - {1\over p^2} 
       + \kappa^2 {\tilde{\bf R}} \left[ {1\over p^4} \right] 
       - c_0^{\rm R}(\kappa) \delta (p) 
     \right\} 
\end{equation}
\begin{equation}
     = \int  d^4p \, \left[ \varphi(p)-\varphi(0) \right] 
     \left\{ 
        {1\over p^2+\kappa^2} 
      - {1\over p^2} 
      + \kappa^2 {1\over p^4}
     \right\} .
\end{equation}
(By the way, this representation verifies that the r.h.s.\ of \ref{(6.14)} 
is independent of the arbitrary constant in the definition  of 
${\tilde{\bf R}}[p^{-4}]$ \ref{(6.8)}.) 

To check that the r.h.s.\ is 
$o(\kappa^2)$ one splits the integration region into two subregions: 
$\vert p\vert  < K\kappa$ and $\vert p\vert  > K\kappa$ with $K$ 
appropriately chosen. For the first subregion one represents 
$\varphi(p)-\varphi(0)$ as $\sum p_i \tilde\varphi_i(p)$ where all 
$\tilde\varphi_i(p)$ are smooth everywhere including $p=0$ 
\cite{schwartz:analyse}, and then performs the scaling $p\to \kappa 
p.$ In the second region, one bounds the first factor in the integrand 
on the r.h.s.\ by a constant, and uses the Taylor theorem to 
estimate the second factor. Then one easily obtains the desired 
estimate. (For more details see \cite{fvt-vvv}.)

\SUBSECTION{IR-counterterms as UV-renormalized integrals}
\label{ss6.7}

Let us discuss the expression \ref{(6.17)}. Introduce again the cut-off 
$\Phi (p/\Lambda )$ explicitly into the integrand of \ref{(6.17)}. Then 
one can rewrite \ref{(6.17)} as:
\begin{equation}\label{(6.21)}
     c_0^{\rm R}(\kappa) = \lim_{\Lambda\to\infty} 
     \left \{
         \int_{|p|<\Lambda} d^4p \, {1\over p^2+\kappa^2} 
     - K'_1 \Lambda ^2 
     - \kappa^2 \left( K'_2 + K'_3 \log\Lambda  \right)
    \right \},
\end{equation}
and the limit is finite by construction.

The first term on the r.h.s.\ is the vacuum average of the operator 
$T[\varphi^2(x)]$ (without normal ordering), while the rest play the 
role of its UV-counterterms.

We take note of this fact for two reasons. First, the coefficients of 
IR-counterterms analogous to $c_0^{\rm R}$ in more complicated cases 
will also admit simple interpretation as contributions to Green 
functions with local operator insertions. Second, reversing the course 
of thought, one can view \ref{(6.17)} as a peculiar%
\footnote{
or natural---depending on one's prejudices.
}
representation of 
an UV-renormalized one-loop Feynman graph in terms of subtraction of 
asymptotics of the integrand at large integration momenta. Such a 
representation will play an important role in our further study of 
EA-expansions, effectively reducing the case of UV-divergent diagrams 
to that of UV-convergent ones (cf.\ below subsect.~\ref{ss7.3} and \cite{gms}).

\SUBSECTION{Dimensionally regularized form}
\label{ss6.8}

The form \ref{(6.14)}, \ref{(6.17)} for the expansion is not the only one
possible, and, definitely, not the most convenient one. If one uses a 
regularization, then, taking into account \ref{(6.17)}, one has the 
following expression instead of \ref{(6.14)} and \ref{(6.17)}:
\begin{equation}\label{(6.22)}
     {1\over p^2+\kappa^2} = \lim_{\epsilon\to0} 
     \left[ 
         {1\over p^2} 
       - \kappa^2 \left\{ {1\over p^4} \right\}^\epsilon 
       + c_0^{\epsilon} (\kappa) \delta (p)  + o(\kappa^2),
     \right]
\end{equation}
where
\begin{equation}\label{(6.23)}
     c_0^{\epsilon} (\kappa) 
     = \int  d^4p\, \left[ 
          {1\over p^2+\kappa^2} 
        - {1\over p^2} 
        + \kappa^2 \left\{ {1\over p^4} \right\}^\epsilon 
     \right].
\end{equation}

Much simpler expressions emerge within the dimensional regularization. 
(Note that in this case one regularizes also the expression to be 
expanded as well as those terms of the expansion which actually do not 
need to be regularized, but this can do no harm. For simplicity of 
notation we will omit the limit signs.) Thus, instead of 
\ref{(6.22)}--\ref{(6.23)} one gets:
\begin{equation}\label{(6.24)}
     {1\over p^2+\kappa^2} 
     = {1\over p^2} 
     - \kappa^2 {1\over p^4} 
     + c_0(\kappa) \delta (p) 
     + o(\kappa^2),
\end{equation}
where
\begin{equation}\label{(6.25)}
     c_0(\kappa) = \int           d^Dp \, \left[ 
         {1\over p^2+\kappa^2} 
       - {1\over p^2} 
       + \kappa^2 {1\over p^4} 
     \right]
     = \int d^Dp \, {1\over p^2+\kappa^2} 
\end{equation}
(we take $\epsilon  = (4-D)/2$). The terms on the r.h.s.\ vanished 
owing to the fact that the dimensional regularization preserves the 
formal scale invariance, so that e.g.
\begin{equation}\label{(6.26)}
     \int           d^Dp \, {1\over p^{2\alpha}} 
       \;\;\Longrightarrow\;\; 
    {p\to \lambda p}  
       \;\;\Longrightarrow\;\; 
    \lambda ^{D-2\alpha} \int           d^Dp \, {1\over p^{2\alpha}} = 0.
\end{equation}
Another consequence of this fact is:
\begin{equation}\label{(6.27)}
     c_0(\kappa) = \left( \kappa^2 \right)^{D/2-1} c_0(1) ,
\end{equation}
i.e.\ the dependence on $\kappa$ takes the form discussed in 
subsect.~\ref{ss4.3}.

Of the three forms of the expansion (\ref{(6.14)} and \ref{(6.17)}; 
\ref{(6.22)} and \ref{(6.23)}; \ref{(6.24)} and \ref{(6.25)}), the most 
important are: the first one since it does not use artificial 
regularizations and contains only explicitly convergent terms, and the 
third one which owing to its simplicity is most convenient in 
applications. Note, however, that the representation in 
terms of dimensional regularization \ref{(6.24)}, \ref{(6.25)} is not 
devoid of some mysticism: indeed, the IR divergence of the term 
$p^{-4}$ on the r.h.s.\ of \ref{(6.24)} is compensated by a divergent 
part of $c_0(\kappa)$ which, judging from the r.h.s.\ of \ref{(6.25)}, 
has an UV origin. However, one should firmly remember that that is due 
to the fact that the dimensional 
regularization nullifies certain types of integrals 
(cf.\ \ref{(6.26)}) and  
formally preserves scale invariance which establishes a connection 
between IR and 
UV contributions. Therefore, it becomes possible for an expression to 
contain both types of divergences balanced so as to render the whole 
finite. In more conventional regularizations there occur only 
compensations of divergences of the same nature (cf.\ \cite{IV}).

\SUBSECTION{Complete expansion}
\label{ss6.9}

Now let us turn to higher terms in the expansion \ref{(6.14)}. To 
derive them, instead of following the recipe of the extension 
principle, we will use a trick motivated by the representation of the 
remainder of the expansion in the form of \ref{(6.20)}.%
\footnote{
An essentially similar trick was used in \cite{gorishny}. 
}
 The 
advantage of it is that it provides a short-cut for the reader 
familiar with the dimensional regularization to arrive at the final 
formulae. However, this trick cannot be extended to more complicated 
situations of non-Euclidean asymptotic regimes. 

Consider the expression
\begin{equation}\label{(6.28)}
     \int d^4p A(p)B(p) = o(\kappa^{2N}),
\end{equation}
where
\begin{equation}\label{(6.29')}
     A(p) = {1\over p^2+\kappa^2} 
          - \sum_{n=0}^N \left( -\kappa^2 \right)^n 
               {1\over p^{2(n+1)}}, 
\end{equation}

\begin{equation}\label{(6.29")}  
     B(p) = \varphi(p) - {\rm\bf T}_{\omega_N} \varphi(p),
\end{equation}

\begin{equation}\label{(6.30)}
     \omega_N = 2(N-1),
\end{equation}
and
\begin{equation}\label{(6.31)}
     {\rm\bf T}_{\omega_N} \varphi(p) 
     \equiv \sum_{n=0}^{\omega_N} 
         {p^n\over n!} \varphi^{(n)} (0)
\end{equation}
is a partial Taylor series. The expression \ref{(6.28)} will 
turn out to be the remainder of an expansion to order 
$o(\kappa^{2N}).$

The fact that the r.h.s.\ of \ref{(6.28)} is indeed $o(\kappa^{2N})$ can 
be explained as follows. $B(p)$ is $O(p^{\omega_N+1})$ at $p\to 0$ and 
$O(p^{\omega_N})$ at $p\to \infty ,$ and is independent of $\kappa. $ 
$A(p)$ is $o(\kappa^{2N})$ for $p\not=0,$ has a non-integrable 
singularity $O(p^{-2(N+1)})$ at $p\to 0,$ and is bounded by 
$O(\kappa^{2(N+1)}/p^{2(N+2)})$ at $p\to \infty .$ One can see that 
the non-integrable singularity of $A(p)$ at $p\to 0$ is  
suppressed by $B(p),$ while the growth of $B(p)$ at $p\to \infty$  is 
not rapid enough to spoil integrability of $A(p)$ at infinity. A more 
formal proof proceeds along the lines of subsect.~\ref{ss6.6}. 

There are two remarks worth making here. First, the interplay of 
asymptotics in \ref{(6.28)} is only possible owing to the uniformity of 
$(p^2+\kappa^2)^{-1}$ with respect to simultaneous scaling in $p$ and 
$\kappa.$ Second, the expression on the l.h.s.\ of \ref{(6.28)} contains 
no cut-offs, and is therefore devoid of arbitrariness.

To represent the final result in a more convenient form, introduce a 
full basis of homogeneous polynomials ${\cal P}_\alpha (p),$ such that
\begin{equation}\label{(6.32)}
     {\cal P}_\alpha (\lambda p) 
     = \lambda ^{|\alpha|} {\cal P}_\alpha (p),
\end{equation}
and the dual set of derivatives of the $\delta$-function :
\begin{equation}\label{(6.33)}
     \int  d^Dp \, {\cal P}_\alpha (p) \delta_\beta (p) 
     = \delta _{\alpha,\beta}
\end{equation}
(cf.\ \ref{(5.8)}). Then the Taylor expansion can be represented as:
\begin{equation}\label{(6.34)}
     {\rm\bf T}_{\omega} \varphi(p) 
     = \sum_{|\alpha|\le\omega} {\cal P}_\alpha (p)\, 
           \left[ 
               \int  dp'\, \delta _\alpha (p') \varphi(p')
           \right].
\end{equation}

For the example under consideration, one can take
\begin{equation}\label{(6.35)}
     {\cal P}_\alpha (p) \to  p^{2\alpha}, \quad \quad
     \delta _\alpha (p) \to \partial^{2a} \delta (p),  a=0,1,2\ldots,
\end{equation}
owing to the $O(D)$-invariance.

Finally, introduce dimensional regularization into the 
(convergent) expression on the l.h.s.\ of \ref{(6.28)}. Then integrate 
it using \ref{(6.26)}. Simple transformations using \ref{(6.34)} lead to 
the following final result, wherein $\int d^Dp\, \varphi(p)$ has been 
omitted from both sides:
\begin{equation}\label{(6.36)}
     {1\over p^2+\kappa^2} 
     = \sum_{n=0}^N \left( -\kappa^2 \right)^n 
         {1\over p^{2(n+1)}} 
       + \sum_{|\alpha|\le\omega_N} 
             c_\alpha (\kappa) \delta_\alpha (p) 
       + o(\kappa^{2N}), 
\end{equation}
where
\begin{equation}\label{(6.37)}
     c_\alpha (\kappa) 
     = \int d^Dp \: {\cal P}_\alpha (p) \, 
          {1\over p^2+\kappa^2} 
     = \kappa ^{D-2+|\alpha|}\, c_\alpha (1),
\end{equation}
and $\omega_N$ is given by \ref{(6.30)}. One can easily find from 
\ref{(6.37)}, which order in $\kappa$ each counterterm belongs to, and 
one can see that the additional dependence on $\kappa$ in each order 
in $\kappa$ is soft in the sense of subsect.~\ref{ss4.3}.

To avoid misunderstanding, we stress once more that the expansion 
\ref{(6.36)} is to be understood in the following sense (cf.\ the 
remarks in subsect.~\ref{ss4.3}). One should first integrate both sides 
termwise against a test function, and take off the regularization 
(i.e.\ take the limit $D\to 4$). The resulting expression will 
represent a correct expansion of a $\kappa$-dependent function in 
powers and logarithms of $\kappa$ to order $o(\kappa^{2N}).$ 

Concerning cancellations of the IR divergences in the first sum on the 
r.h.s.\ of \ref{(6.36)} by the apparently UV divergent expressions in 
\ref{(6.37)}, see the end of subsect.~\ref{ss6.8}.

Note also that \ref{(6.36)} is a true infinite asymptotic expansion of 
$(p^2+\kappa^2)^{-1}$ in the sense of distributions, and 
after the regularization is taken off, one only has powers and 
logarithms of $\kappa$ on the r.h.s.\ of \ref{(6.36)}, and such an 
expansion is unique (cf.\ subsect.~\ref{ss4.2}).

\SUBSECTION{Several $\kappa$-dependent factors}
\label{ss6.10}

Let us consider an example with several factors, such that the 
singularities of their expansions are located at different points. For 
example, consider the following integrand corresponding to a 
self-energy contribution:
\begin{equation}\label{(6.38)}
     {1\over \left(p^2+\kappa^2\right)} \times
     {1\over \left[(p-Q)^2+\kappa^2 \right]}.
\end{equation}
We regard the external momentum $Q$ as a heavy parameter and fix it at 
a non-zero value, which corresponds to the simplified version of the 
expansion problem without contact terms (subsect.~\ref{ss2.4}). The final 
recipe for expanding \ref{(6.38)} in the sense of distributions 
will be to simply multiply the expansions for the two factors. 
But the reasoning behind it is more instructive than the result itself 
(cf.\ a similar reasoning in a more complicated case below in 
subsects.~\ref{(12.2)} and \ref{(12.3)}).

The key observation here is that the expansion of the second factor is 
regular around the singular point of the first factor; more 
precisely, if $\varphi_1$ is such that it is equal to zero around the 
singular part of the second factor, then the $\kappa$-dependent test 
function
\begin{equation}\label{(6.39)}
     \bar \varphi_1(p,\kappa) = \varphi_1(p)\,
         {1\over { (p-Q)^2+\kappa^2 } }
\end{equation}
satisfies the condition ($ii$) of subsect.~\ref{ss4.4}. Therefore, one 
can formally expand $\bar \varphi_1$ and use \ref{(6.36)} for 
$(p^2+\kappa^2)^{-1},$ and we conclude that for $p\not=Q,$ the 
expansion of \ref{(6.38)} is given by the Taylor expansion of the 
second factor times the full expansion of the first factor. Swapping 
the factors and considering test functions $\varphi_2$ which are 
identically zero around the singular point of the first factor, we 
arrive at a similar conclusion. The final observation is that since 
the singular points of the two factors are separated, any $\varphi(p)$ 
can be represented as $\varphi_1(p)+\varphi_2(p),$ whence the desired 
result can be easily deduced. (Another way to see it is to notice that products 
of $\delta$-functions from expansions of the two products result in zeros.)

\vskip 5mm

To summarize, we have demonstrated with a simple example how the ideas 
of the extension principle discussed in the preceding section work in 
practice. We have also demonstrated a short-cut way to derive our 
final result, the expansion \ref{(6.36)}, represented in the 
dimensionally regularized form. It can be represented in an explicitly 
finite form with the regularization taken off (for more on this see 
\cite{fvt-vvv}), but the form \ref{(6.36)} is more convenient for our 
purposes.

\SECTION{Applications to one-loop integrals}
\label{s7}

The results of the previous sections can now be applied to 
one-loop integrals. We first consider the UV-convergent case, and then 
show how the case of UV-divergent MS-renormalized integrals can be 
reduced, essentially, to the convergent case. The most important 
non-technical element of our reasoning is a new definition of the 
MS-scheme in terms of subtractions of UV-asymptotic part directly from 
the integrand, prior to any momentum integrations (subsect.~\ref{ss7.4} and \cite{gms}).

\SUBSECTION{UV-convergent one-loop integral}
\label{ss7.1}

Consider the following UV-convergent one-loop integral:
\begin{equation}\label{(7.1)}
     \int d^4p \, {1\over p^2+\kappa^2} \times
     {1\over \left( p^2+M^2 \right)^2 }. 
\end{equation}
One would expect (we discuss this in more detail below) that although 
the ``test function" $(p^2+M^2)^{-2}$ is not a true Schwartz test 
function, its decrease at $p\to \infty$  is sufficiently rapid to 
allow one to substitute the expansion \ref{(6.36)} for 
$(p^2+\kappa^2)^{-1}$ into \ref{(7.1)} in order to get the asymptotic 
expansion for $\kappa\to 0$ of the integral \ref{(7.1)} as a whole. 
Doing this and integrating the $\delta$-functions, one obtains:
\begin{equation}\label{(7.2)}
     \vbox{\hbox{eq.\ref{(7.1)}}} 
     = \lim_{D\to4} \biggl\{ \int           d^D\!p \, {1\over p^2} \times
       {1\over \left( p^2+M^2 \right)^2 } 
     - \int           d^D\!p \,\,{\kappa^2\over p^4} \times 
                      {1\over \left( p^2+M^2 \right)^2 } 
\end{equation}
\begin{equation}
     + \int d^Dp \, {1\over p^2+\kappa^2} \times
                      {1\over M^4} 
     \biggr\}
     + o(\kappa^2).
\end{equation}
The validity of \ref{(7.2)} can be easily checked by explicit integrations.

In order to make connection with what follows, let us briefly discuss 
why the rapid decrease of the ``test function" $(p^2+ M^2)^{-2}$ in 
\ref{(7.1)} allows one to substitute \ref{(6.36)} directly into 
\ref{(7.1)}. By definition, integrations over infinite regions involve 
a limiting procedure. Let us make it explicit in \ref{(7.1)}. Take a 
smooth function $\Phi(p)$ such that
\begin{equation}\label{(7.3)}
     \Phi(p) = 1,   {\rm\ \ for\ \ }\vert p\vert  < 1,
\end{equation}
\begin{equation}
     \quad\quad\quad \,\,0,   {\rm\ \ for\ \ } \vert p\vert  > 2.
\end{equation}
We can introduce the cut-off $\Phi(p/\Lambda )$ into \ref{(7.1)}:
\begin{equation}\label{(7.4)}
     \hbox{eq.\ref{(7.1)}} 
     = \lim_{\Lambda\to\infty}  \int d^4p \, 
         {1\over p^2+\kappa^2} \left\{ 
           {\Phi(p/\Lambda)\over \left( p^2+M^2 \right)^2 } 
         \right\} .
\end{equation}

Now it is fully correct to use \ref{(6.36)} in order to expand the 
expression under the limit sign to order, say, $o(\kappa^2).$ Then one 
has to check that ($i$) the limit $\Lambda \to \infty$  exists for 
each term of the expansion of $(p^2+\kappa^2)^{-1}$ and ($ii$) the 
remainder which was $o(\kappa^2)$ prior to taking limit will be 
$o(\kappa^2)$ after it. While ($i$) is obvious, to check ($ii$) one 
can recall the representation \ref{(6.28)} of the remainder term. The 
desired result easily follows therefrom; moreover, the proof here 
closely repeats the reasoning of the paragraph after \ref{(6.3)}.

The conclusion is that for absolutely convergent 1-loop integrals the 
asymptotic expansions are normally 
obtained by substituting the expansions in 
the sense of the distributions into the integrand, and then 
performing termwise integrations over the loop momentum.

\SUBSECTION{Separation of UV and IR divergences}
\label{ss7.2}

A remark is in order here. The expansion \ref{(7.2)} contains two terms 
that diverge as $D\to 4:$ the second integral is logarithmically 
divergent for $p\to 0,$ and the third one, for $p\to \infty .$ However, 
by construction, the expansion as a whole is finite as $D\to 4.$ 
Therefore, the divergences must cancel. But the peculiar feature of 
\ref{(7.2)} is that one divergence occurs in a $\kappa$-dependent 
integral, while the other one in an integral depending on $M$ but not 
on $\kappa.$ This is not satisfactory because in the final respect in 
applications one has to deal with {\em finite} functions of $\kappa$ and 
$M.$

As we have already noted, all our expansions can be represented in an 
explicitly convergent form which makes no use of the dimensional 
regularization, but this requires a special techniques (see 
\cite{fvt-vvv}). However, it is still possible to rearrange \ref{(7.2)} 
in terms of $\kappa$- and $M$-dependent contributions which are 
separately finite, by using only notions known to the practitioners of 
perturbative QFT who are familiar with the techniques of UV 
renormalization within the MS-scheme. The price to be paid for this is 
that IR divergences will be canceled by UV-counterterms, but we have 
already explained in subsect.~\ref{ss6.4} why such things can happen.

Look at the third integral on the r.h.s.\ of \ref{(7.2)}. It is an UV 
divergent 1-loop Feynman integral which can be made finite by adding 
an UV-counterterm which can be evaluated e.g. in the MS-scheme. Such 
UV-counterterms are polynomials in masses and external momenta, and in 
our case can be taken in the form:
\begin{equation}\label{(7.5)}
     \kappa^2 \,{ {\rm const}\over D-4}\, \mu^{D-4},
\end{equation}
where $\mu$  is introduced to preserve dimensionality (cf.\ \ref{(7.9)} 
below). Now, using \ref{(7.5)}, the expansion \ref{(7.2)} can be 
identically rewritten as:
\begin{equation}\label{(7.6)}
     \hbox{eq.\ref{(7.1)}} 
     = \lim_{D-4} \Biggl\{ 
         \left[ 
             \int\, d^D\!p \, {1\over p^2} \times
                           {1\over \left( p^2+M^2 \right)^2 } 
         \right]
\end{equation}
\begin{equation}
      - \kappa^2 \left[ \int           d^Dp \, 
            {1\over p^4} \times {1\over \left( p^2+M^2 \right)^2 } 
          + { {\rm const} \times \mu^{D-4} \over M^4(D-4)} 
        \right]
\end{equation}
\begin{equation}
      + {1\over M^4} \left[ \int           d^Dp \, 
                 {1\over p^2+\kappa^2} 
           + {\kappa^2 {\rm const} \times \mu^{D-4} \over D-4 }
         \right]
      \Biggr\} + o(\kappa^2).
\end{equation}
Since the first and third square-bracketed terms as well as the 
expression as a whole are finite in the limit $D\to 4,$ so finite must 
be the second term too, which is not difficult to verify by explicit 
calculations.

Note that by rescaling the integration variables, one can exhibit the 
dependence of various terms in \ref{(7.6)} on $\kappa$ and $M:$
\begin{equation}\label{(7.7)}
     \hbox{eq.\ref{(7.6)}} 
     = {1\over M^4} \int           d^4p \, 
                  {1\over p^2} \times
                  { 1\over \left( p^2+1 \right)^2 } 
\end{equation}
\begin{equation}
     - {\kappa^2\over M^4} \lim_{D\to4} \Biggl\{ 
          \left[ 
              \int           d^Dp \, 
              {1\over p^4} \times
              { 1\over \left( p^2+1 \right)^2 } 
            + {{\rm const} \times (\mu/M)^{D-4}\over D-4 } 
          \right]
\end{equation}
\begin{equation}
     - \left[ \int           d^Dp \, 
           {1\over p^2+1} 
         + {{\rm const} \times  \left( \mu/\kappa \right)^{D-4} \over D-4 }
       \right] \Biggr\} 
     + o(\kappa^2).
\end{equation}
We stress once more that the numerical connection of the UV 
counterterm in the third term and the counterterm in the second term, 
whose role is to cancel the IR divergence---this connection can be 
traced back to the fact that the dimensional regularization nullifies 
certain integrals containing both IR and UV divergences 
simultaneously. One could pursue the topic further and analyze such 
cancellations more explicitly, but as the relevant formulae are quite 
complicated while the results perfectly useless, we stop here.

To conclude, we have demonstrated a trick for recasting expansions 
that emerge from our methods into an explicitly convergent form. The 
trick employs specific properties of UV renormalization of Feynman 
integrals, and can be generalized to the 
multiloop case where it takes the form of an inversion of the 
\rop\ \cite{IIold}--\cite{II}.

\SUBSECTION{UV divergent one-loop integrals}
\label{ss7.3}

Let us now turn to UV divergent one-loop integrals, e.g.
\begin{equation}\label{(7.8)}
     \int \, d^4p {1\over p^2+\kappa^2} \times
                  {1\over  p^2+M^2 } .
\end{equation}
Its MS-renormalized version which we wish to expand in $\kappa,$ is:
\begin{equation}\label{(7.9)}
     \lim_{D\to4} \left[ \int           d^Dp \, 
                         {1\over p^2+\kappa^2} \times
                         {1\over p^2+M^2 } 
                       - {z\over D-4} \, \mu^{D-4}
                  \right] ,
\end{equation}
where $z$ is a numerical constant and $\mu$  is 't Hooft's unit of 
mass which plays the role of the subtraction parameter in the 
MS-scheme.

Our purpose is to show that the expansion \ref{(6.24)} (and the more 
general expansion \ref{(6.36)}) can be directly substituted into the 
integral in \ref{(7.9)} in order to get the asymptotic expansion for 
\ref{(7.9)}. (Correctness of this recipe can also be checked by 
explicit calculations.) The motivation comes from the fact that the 
divergent contribution to the integral in \ref{(7.9)} is generated by 
the asymptotics of the integrand at $p\to \infty ,$ while replacing 
$(p^2+\kappa^2)^{-1}$ by its expansion leaves that asymptotics intact. 
These heuristic arguments can be formalized in the following way. 

The crucial step is to redefine the MS-subtraction in \ref{(7.9)} in a 
manner independent of dimensional regularization (because the 
expression \ref{(7.9)} does not allow one to apply the reasoning like 
that in subsect.~\ref{ss7.1}). To this end recall (subsect.~\ref{ss6.7}) 
that a subtraction of large-$p$ asymptotics from the integrand is 
equivalent to UV renormalization. One has:
\begin{equation}\label{(7.10)}
     {1\over p^2+\kappa^2} \times {1\over p^2+M^2} 
     \vtop{ \hbox{ \quad\quad $\simeq$ }
            \hbox{ \quad $p \to \infty$ }
            \hbox{ $\left( \kappa,M \to 0 \right)$ }
     }
     {1\over p^4} ,
\end{equation}
so that subtracting $p^{-4}$ from the integrand in \ref{(7.9)} would 
produce an UV convergent integral. However, $p^{-4}$ is ill-defined at 
$p = 0.$ Let us modify it at $p = 0$ so as to ensure its 
integrability:
\begin{equation}\label{(7.11)}
     \left[ {1\over p^4} \right] ^{\rm R} 
     = {1\over p^4} 
     + {z'\over D-4} \, \mu^{D-4} \, \delta (p)
\end{equation}
(cf.\ \ref{(6.10)}). $z'$ in \ref{(7.11)} is a numeric constant, and the 
mass-like parameter $\mu$ is introduced to 
preserve dimensionality.

Now, the expression
\begin{equation}\label{(7.12)}
     \int           d^D\!p \, \left\{ {1\over p^2+\kappa^2} \times
                            {1\over p^2+M^2 } 
                          - \left[ {1\over p^4} \right] ^{\rm R} 
                    \right\}
\end{equation}
is finite at $D = 4$ both at $p\to 0$ and $p\to \infty .$ Owing to 
\ref{(6.26)}, it can be represented as:
\begin{equation}\label{(7.13)}
     \int           d^D\! p \, {1\over p^2+\kappa^2} \times
                    {1\over p^2+M^2 } 
     - {z'\over D-4} \mu^{D-4} . 
\end{equation}
Recall that the constant $z$ in \ref{(7.9)} was chosen solely from the 
requirement of finiteness of \ref{(7.9)}. Hence, comparing 
\ref{(7.9)} and \ref{(7.13)} we see that $
     z = z',
$
and
$
     \hbox{eq.\ref{(7.9)} $=$ eq.\ref{(7.12)}}.
$
Finally, recall that the distribution \ref{(7.11)} can be represented 
as \ref{(6.8)} with a suitably chosen $c,$ and no regularization is 
used in \ref{(6.8)}. Therefore:
\begin{equation}\label{(7.16)}
     \hbox{eq.\ref{(7.9)}} 
     = \int           d^4p \, \left\{ 
            {1\over p^2+\kappa^2} \times
            {1\over p^2+M^2 } 
          - \left[ {1\over p^4} \right] ^{\rm R}
       \right\}.
\end{equation}

So, we have represented the MS-renormalized integral \ref{(7.9)} in the 
form of an integral over $p$ of the expression which differs from the 
original integrand \ref{(7.8)} by a contribution independent of 
$\kappa.$

Further reasoning presents no problems, as the representation 
\ref{(7.16)} allows one to use the arguments of subsect.~\ref{ss7.1} 
with only inessential modifications in order to obtain the expansion 
in $\kappa.$ We leave it to an interested reader to check that the 
final result is such as if we had substituted the expansion 
\ref{(6.36)} directly into \ref{(7.9)}.

The method for recasting the expansion into an explicitly convergent 
form described in subsect.~\ref{ss7.2} is also applicable here.

\SUBSECTION{Generalized MS-schemes}
\label{ss7.4}

Now consider the following definition of a class of subtractions 
schemes. (It is shown elsewhere that this definition works in the 
case of multiloop Feynman integrals \cite{gms}, \cite{III}.) Let 
$F(p)$ be the integrand of a 1-loop Feynman diagram. Evaluate its 
asymptotic expansion for $p\to \infty$ and discard the terms that are 
integrable at large $p.$ Denote the result as $F^{\rm as}(p).$ The 
terms in $F^{\rm as}(p)$ that generate logarithmic divergences at 
$p\to \infty ,$ also have a logarithmic singularity at $p=0,$ while 
the rest are integrable at that point. Redefine the logarithmically 
divergent terms at $p=0$ similarly to \ref{(6.8)}, \ref{(6.10)} to make 
them integrable at all finite $p.$ Denote the result as $[F^{\rm 
as}(p)]^{\rm R}.$ (Note that the redefinition should only affect 
powers of $p$ and commute with other dimensional parameters; in 
particular, the constant $c$ in \ref{(6.8)} should be independent of 
the masses and external momenta of the diagram.) The renormalized 
integral is now defined~as
\begin{equation}\label{(7.17)}
     \int           d^4p \, \left\{ F(p) 
                          - \left[F^{\rm as}(p)\right]^{\rm R}
                     \right\} .
\end{equation}
It is easy to see (cf.\ the reasoning in subsect.~\ref{ss7.3}) that the 
class of subtraction schemes thus defined includes the MS-scheme as a 
special case. Note also that the UV-counterterms in these schemes are 
always polynomials in masses and external momenta of the diagram.

To take an example, consider the integral that has already been 
encountered above in subsect.~\ref{ss6.7}:
\begin{equation}\label{(7.18)}
     \int           d^4p \, {1\over p^2+\kappa^2} ,
\end{equation}
then 
\begin{equation}\label{(7.18.1)}
     F(p)= {1\over p^2+\kappa^2}, 
   \quad
     F^{\rm as}(p) 
     = {1\over p^2} 
     - \kappa^2 {1\over p^4} ,
\end{equation}
and
\begin{equation}\label{(7.20)}
     \left[ F^{\rm as}(p)\right]^{\rm R} 
     = {1\over p^2} 
     - \kappa^2 \left[ {1\over p^4} \right] ^{\rm R} .
\end{equation}
The renormalized integral is
\begin{equation}\label{(7.21)}
     \int           d^4p \, \left\{ {1\over p^2+\kappa^2} 
                          - {1\over p^2} 
                          + \kappa^2 \left[ {1\over p^4} 
                                     \right] ^{\rm R} 
                    \right\}.
\end{equation}
Introducing the dimensional regularization into \ref{(7.21)}, using 
\ref{(7.11)} and \ref{(6.26)}, we arrive at an expression,
\begin{equation}% [(7.22)]
     \int           d^Dp \, {1\over p^2+\kappa^2} 
     + \kappa^2 {z'\over D-4} \mu^{D-4}, 
\end{equation}
where $z' = c_0 + (D-4)c_1 + \ldots$ and $c_i$ are arbitrary finite 
constants. If $c_i = 0,$ $i \ge 1,$ one recovers the standard 
MS-scheme.

\SUBSECTION{Summary}
\label{ss7.5}

The representation of the MS-renormalization as subtractions from the 
bare integrand of those terms from its asymptotics at large loop 
momentum which are responsible for UV divergences, allows one to 
reduce the problem of expansion of 1-loop integrals to the UV 
convergent case. And the reasoning of the above sections shows that 
the correct expansion of the renormalized integrals can be obtained by 
direct substitution of the corresponding expansion in the sense of 
distributions in place of the $\kappa$-dependent terms, and 
rearranging the resulting expression along the lines of 
subsect.~\ref{ss7.2}, in order to obtain the expansion in an explicitly 
convergent form.

In order to generalize this recipe to the multiloop case we need, 
first, to learn to expand more complicated products of singular 
functions (which is done in the present paper), second, to 
generalize the above treatment of UV-divergent integrals to the 
multiloop case \cite{gms}, \cite{III}, and third, to 
generalize the rearrangement trick of subsect.~\ref{ss7.2} 
\cite{IIold}--\cite{II}.

\vskip2cm

\thispagestyle{myheadings}\markright{} 
%\vskip2.3cm
\centerline{{\large\bf THE MASTER EXPANSION}}

\SECTION{Notations for products of singular functions}
\label{s8}

Admittedly, multiloop Feynman diagrams are cumbersome objects. 
Nevertheless, the 
analytical aspects of the problem, say, of UV renormalization or the 
Euclidean expansion problem are by no means 
as complicated as one can imagine regarding 
them from the point of view 
of the traditional techniques 
prevailing in rigorous studies of perturbative QFT like e.g.\ the 
$\alpha$-representation which completely destroys 
the essential multiplicative structures of 
Feynman integrands. However, to reveal the 
underlying simplicity and help one to ignore the plethora of 
irrelevant detail, a careful choice of notations is crucial. Below we 
describe such a system of notations (discussed in a more formalized 
manner in \cite{fvt-vvv}) which does not make use of parametric 
representations and applies equally well to both scalar and non-scalar 
theories.

\SUBSECTION{Graphs and products}
\label{ss8.1}

Let $G$ be a Feynman graph. Its loop momenta $p_1, p_2, \ldots, p_l$ 
are always Euclidean vectors of 4 dimensions. (For studying toy models 
and examples it is sometimes convenient to consider the 2-dimensional 
case.) The loop momenta are combined into a single vector of $4 \times 
l$ dimensions:
\begin{equation}\label{(8.1)}
     p = \left(p_1,p_2\ldots p_l \right).
\end{equation}
The variable $p$ may also incorporate the heavy external momenta---see 
remark ($iii$), subsect.~\ref{ss3.2}. In such a case
\begin{equation}\label{(8.2)}
     p = \left(p_1,p_2\ldots p_l , {\bar Q}_1 \ldots \right),
\end{equation}
where ${\bar Q}$ are defined in subsect.~\ref{ss2.3}.

The integrand of the graph $G$ in momentum representation is denoted 
as $G(p).$ Dependence on other parameters can be shown with additional 
arguments after $p.$ $G(p)$ is a product of factors corresponding to 
both lines and vertices. The factors are numerated by the label $g.$ 
The set of values of the label is denoted also as $G.$ The use of the 
same symbol for three different but related entities should not be 
misleading. Similarly, it should not be misleading if we denote as 
$g(p)$ that factor in the product $G(p),$ which corresponds to the 
label $g.$ So,
\begin{equation}\label{(8.3)}
     G(p) = \prod_{g\in G} g(p).
\end{equation} 
We will also have to study subproducts of $G.$ Let $\gamma \subset G.$ 
Then
\begin{equation}\label{(8.4)}
     \gamma (p) \equiv  \prod_{g\in \gamma} g(p)
\end{equation}
and also
\begin{equation}\label{(8.5)} 
     G \backslash \gamma (p) 
     \equiv
     \prod_{g\in G\backslash\gamma} g(p) = G(p)/\gamma (p).
\end{equation}
where $G \backslash \gamma$ is the standard notation for the 
difference of sets.

\SUBSECTION{Conventions for non-scalar factors}
\label{ss8.2}

We allow each function $g(p)$  to carry sub- (or super-) scripts: Lorentzian, 
${\rm SU}(n)$ etc. Then the product \ref{(8.3)} may contain implicit 
contractions. If, for example, the product \ref{(8.4)} carries a pair 
of subscripts, which are contracted in \ref{(8.3)}, then this 
contraction is implied in \ref{(8.4)} as well.

Now, let $\Gamma'$ and $\Gamma^{\prime\prime}$ be non-intersecting 
subproducts from $G.$ In $G$ there may exist such a pair of contracted 
subscripts that one of them belongs to $\Gamma',$ while the other one 
to $\Gamma^{\prime\prime}.$ Then the product
\begin{equation}\label{(8.6)}
     \Gamma'(p) \, \Gamma^{\prime\prime}(p)
\end{equation}
implies contraction of such a pair. Furthermore, we will build new 
expressions from $G$ by replacing some subproducts $\gamma$  by 
certain expressions, say, $Z_\gamma,$ so that $Z_\gamma$ will carry 
exactly the same subscripts as $\gamma.$ Then the products like
\begin{equation}\label{(8.7)} 
     \Gamma' \, Z_{\Gamma^{\prime\prime}} ,\quad \quad   
     Z_{\Gamma'} \, Z_{\Gamma^{\prime\prime}} \quad\quad {\rm etc}.
\end{equation}
imply the same contractions as \ref{(8.3)}.

The implicit presence of such contractions will have no complicating 
effect on our reasoning as compared with the scalar case.

\SUBSECTION{Momentum dependence}
\label{ss8.3}

Let us discuss the dependence of factors on $p.$ One should 
distinguish factors of two types: those corresponding to lines and to 
vertices of Feynman graphs.

Let $g$ correspond to a line. The momentum flowing through this line 
is a linear combination of various momentum variables of our problem:
\begin{equation}\label{(8.8)}
     \left( 
        \sum_{i=1\ldots l} c_{g,i} p_i + Q_g 
     \right) 
     + k_g ,
\end{equation}
where $Q_g$ and $k_g$ are linear combinations of heavy and light 
external momenta from the sets $Q$ and $k.$ Then $g(p)$ is a 
propagator depending on \ref{(8.8)} and a mass. The mass may be light, 
i.e.\ equal to zero or proportional to $\kappa,$ or heavy, i.e.\  
non-zero and independent of $\kappa.$ Since $k_g$ in \ref{(8.8)} is 
proportional to $\kappa$ and will in fact play the role of an 
expansion parameter, it is convenient to introduce a notation for the 
non-vanishing part of \ref{(8.8)}:
\begin{equation}\label{(8.9)} 
     L_g(p) \equiv  {\rm the\ parenthesised\ expression\ in\
                         \vtop{\hbox{\ref{(8.8)}}} 
                    }.
\end{equation}
Now one can write:
\begin{equation}\label{(8.10)}
     g(p) = F_g( L_g(p)+k_g , \kappa ).
\end{equation}
In fact our arguments are true for a wide class of functions 
$F_g(L,\kappa)$ including any standard perturbation theory propagators 
raised to integer powers and multiplied by polynomials of $L.$

\SUBSECTION{Assumptions on the properties of factors}
\label{ss8.4}

Let us describe those properties of $F_g(L,\kappa)$ that are essential 
for our theory. (For a formalized description see \cite{fvt-vvv}.) One 
should consider two cases:

($i$) The function $F_g(L,\kappa)$ at $L = \kappa = 0$ is regular 
(which corresponds to propagators with heavy masses). Then for 
$\kappa\to 0$, $F_g$ is expanded into a series of functions which are 
smooth in $L,$ and the expansion is uniform in $L$ in any bounded 
region and allows arbitrary termwise differentiations in $L.$

($ii$) The function $F_g(L,\kappa)$ at $L = \kappa = 0$ is singular 
(which corresponds to propagators with a light mass). Here we assume 
that $F_g(L,\kappa)$ has the following properties:

(a) scaling:
\begin{equation}
     F_g(\lambda L,\lambda \kappa) = \lambda^{d_g} F_g(L,\kappa);
\end{equation}

(b) asymptotic expansion at each $L \not= 0:$
\begin{equation}
     F_g(L,\kappa) = \sum_n \kappa^n F_{g,n}(L),  \quad L \not= 0;
\end{equation}

(c) all $F_{g,n}(L)$ are smooth at $L \not= 0$ and have simple scaling 
properties:
\begin{equation}
     F_{g,n}(\lambda L) = \lambda^{d_g-n} F_{g,n}(L);
\end{equation}

(d) regularity properties: the expansion in (b) allows arbitrary 
termwise differentiations in $L,$ and the remainder of the 
expansion to order $O(\kappa^N )$ has an upper bound of the form of 
the modulo of the first discarded term: $\kappa^{N+1}|L|^{d_g-N-1}$ 
(cf.\ property (c); also note that property (d) ensures that the 
$\kappa$-dependent test functions that arise at intermediate steps of 
our reasoning will possess the properties described in 
subsect.~\ref{ss4.4}).

In case $i/ii$ we will say that the factor is of {\em regular/singular 
type}, respectively. 

It should be stressed that the described properties are sufficient for 
our purposes and comprise the problems within the standard 
perturbation theory. But our methods will work in a more general 
context, e.g.\ if the singular functions have logarithmic or power 
behaviour with a non-integer exponent.

\SUBSECTION{Factors corresponding to vertices}
\label{ss8.5}

Now, let $g$ correspond to a vertex of the graph. One can assume that 
$g(p)$ is a uniform polynomial of momenta entering the vertex. There 
are three ways to include such factors into a common framework with 
lines. First, all vertex polynomials can be merged with test 
functions. Second, each vertex polynomial can be split so that each 
part may be included into one of the propagators attached to the 
vertex (such modifications of propagators are allowed by our 
definitions---see subsect.~\ref{ss8.4}). The third way---which we 
prefer---consists in generalizing the notations introduced for lines: 
it is sufficient to allow $L_g$ to consist of several independent 
momentum components like \ref{(8.8)}, i.e.:
\begin{equation}\label{(8.11)}
     L_g = \left( 
              \sum_{i=1\ldots l}c_{g,v,i} p_i + Q_{g,v} 
           \right)_{v=1,\ldots} 
\end{equation}
For convenience, we assume that, by definition, the vertex functions 
belong to the singular type (see subsect.~\ref{ss8.4} above).

Equivalence of expansions obtained in all three cases follows from the 
uniqueness of the final result (subsect.~\ref{ss4.2}). More explicitly, 
the equivalence can be checked using the property of the \asop\ to 
commute with multiplications of the expression to be expanded by 
polynomials of $p$ (see below subsect.~\ref{ss11.5}).

\SUBSECTION{Summary}
\label{ss8.6}

Let us summarize the entire scheme. The product \ref{(8.3)} is defined 
on the space of the multicomponent variable \ref{(8.11)}, and consists 
of factors of the form \ref{(8.10)}. The factors need not be scalar 
functions---the conventions of subsect.~\ref{ss8.2} are operative. The 
first argument in \ref{(8.10)} is a set of momentum variables of the 
form \ref{(8.11)}. The functions $F_g$ satisfy the conditions described 
in subsect.~\ref{ss8.4}. Finally, the factors may be of two 
types---singular or regular (see the end of subsect.~\ref{ss8.4}).

\SECTION{Formal expansions and IR-subgraphs}
\label{s9}

An important element of our technique is classification 
of singularities of the 
formal (Taylor) expansions in $\kappa$ of the products \ref{(8.3)}. The 
key notion here is that of IR-subgraph, introduced in 
subsect.~\ref{ss9.2}. Note that there are similarities between 
our Euclidean space  
classification of IR-singular points and the Minkowski space 
reasoning of Libby and Sterman \cite{libby}.

\SUBSECTION{Some notations}
\label{ss9.1}

Denote the operation of formal expansion in $\kappa$ as ${\rm\bf 
T}_\kappa .$ The expansion ${\rm\bf T}_\kappa  g$ is the same as the 
one described in property (b), subsect.~\ref{ss8.4}, up to a 
replacement of the momentum variable. Then
\begin{equation}\label{(9.1)}
     {\rm\bf T}_\kappa  g(p,\kappa) 
     = \prod_{g\in G} {\rm\bf T}_\kappa  g(p,\kappa) ,
\end{equation}
and similarly for each subset $\gamma \subset G.$ The r.h.s.\ of 
\ref{(9.1)} should be understood as a simple infinite series in integer 
powers of $\kappa,$ obtained by formal multiplication of the series 
with reordering of terms.

It is convenient to denote the operation of partial expansion up to 
terms $O(\kappa^N )$ as ${\rm\bf T}_{\kappa,N}.$ We will also omit the 
subscript indicating the expansion parameter 
as ${\rm\bf T}_\kappa \to {\rm\bf T},$ ${\rm\bf T}_{\kappa,
N}\to {\rm\bf T}_N,$ if misunderstanding is excluded.

\SUBSECTION{Singularities of the formal expansion}
\label{ss9.2}

Let us describe the structure of singularities of the formal expansion 
\ref{(9.1)}. Each factor of the singular type generates singularities 
at those points where $L_g(p)=0.$ We call the set of such points {\em 
singular plane of\/} $g$ and denote it as
\begin{equation}\label{(9.2)}
     \pi_g = \left\{ p \,\vert\,  L_g(p)=0 \right\} .
\end{equation}
Now consider an arbitrary subset $\gamma\subset G$ such that all its 
elements are of singular type. Define the {\em singular plane of the 
subset\/} $\gamma:$
\begin{equation}\label{(9.3)}
     \pi_\gamma = \bigcap_{g\in\gamma} \pi_g .
\end{equation}
On this plane, all factors $g\in\gamma$ are singular simultaneously.

We will consider only such subsets $\gamma,$ to which  there 
correspond non-empty singular planes. 

An important point is that the 
same singular plane may correspond to different subsets. For example, 
in Fig.~3 the pairs of propagators with momenta $p_1$ and $p_2;$ $p_1$ 
and $p_1 + p_2$ generate the same singular plane described by the 
equations $p_1 = p_2 = 0.$ 
Further, there is one largest among the subsets with the same singular 
plane. This follows from the fact that if $\pi_\gamma = \pi_{\gamma'}$ 
then $\pi_\gamma = \pi_{ \gamma \cup \gamma' }.$ In Fig.~3 such a 
subproduct consists of all three propagators. Subsets, to which new 
factors cannot be added without reducing their singular plane, will be 
described as {\em complete} and will be called ({\em complete}) {\em 
IR-subgraphs} or, if confusion is excluded, simply subgraphs.

\SUBSECTION{Diagrammatic interpretation of IR-subgraphs}
\label{ss9.3}

Take a Feynman graph and evaluate momenta for each line. To check if a 
given set of singular lines $\gamma$ is complete, perform the following 
test. Set to zero all the momenta flowing through the lines of 
$\gamma,$ and all light external momenta. Using momentum conservation 
at vertices, reevaluate the momenta for the rest of the lines. If 
there is a line of singular type not belonging to $\gamma,$ whose 
momentum will vanish after reevaluation, then the set $\gamma$ is not 
complete. If there are no such lines, then $\gamma$ is an IR-subgraph. For 
example, in Fig.~4a the pair of vertical lines forms an IR-subgraph.

Note that the same subset $\gamma$ in the same diagram $G$ may be or 
be not an IR-subgraph, depending on whether the full Euclidean 
expansion problem is 
considered or its simplified version without contact terms (see 
subsect.~\ref{ss2.4}). For example, for the graph in Fig.~5 (masses in 
both lines are light) the entire graph will be its own IR-subgraph if 
the graph is considered as a distribution over $Q,$ so that $Q$ is 
included into integration momenta; the corresponding singular plane is 
described by the equations $p = Q = 0.$ In the simplified version one 
considers integration only over the loop momentum $p$, while $Q$ is 
fixed at a non-exceptional value (this means here simply that $Q \not= 
0$); in this case the pair of lines cannot be an IR-subgraph because 
its singular plane is empty. 

The property of being IR-subgraph also depends on the additional 
restrictions specified for the heavy momenta in the  
particular problem. For example, in Fig.~4a which differs from Fig.~4b 
by a linear restriction $Q_1 = Q_2,$ the pair of vertical lines forms 
an IR-subgraph, while this is not so for Fig.~4b.

When should a vertex be included into an IR-subgraph? Recall that any 
vertex is a singular-type factor by definition (cf.\ 
subsect.~\ref{ss8.5}), and its singularities are localized at the 
points where all the momenta from the set $L_g$ are equal to zero. 
Therefore, in the most general case the recipe is as follows: if all 
the lines incident to this vertex belong to the subgraph, then the 
vertex is to be included, too. However, in the simplified version of 
the expansion problem, an IR-subgraph may not contain a vertex with a 
non-zero total heavy external momentum entering into it.

Starting from the basic analytical definition, one can always 
enumerate IR-sub\-graphs in any specific case. However, we do not 
attempt translating the definition of the IR-subgraph into the 
graph-theoretic language: the result would be cumbersome and 
practically useless. For the purposes of doing the combinatorics of 
factorization in \cite{II}, we will only need the properties of 
complements of IR-subgraphs that can be easily derived from the above 
definition (cf.\ subsects.~\ref{(4.2)}, \ref{(5.1)} and \ref{(5.4)} in 
\cite{II}).

\SUBSECTION{Why IR-subgraphs?}
\label{ss9.4}

The reason for considering only IR-subgraphs among all subproducts of 
$G(p)$ is as follows. If $\gamma$ is a subgraph, then singularities of 
the formal expansion
\begin{equation}\label{(9.4)}
     {\rm\bf T}_\kappa  G(p) 
     = \left[ {\rm\bf T}_\kappa \gamma (p) \right] \times 
       \left[ {\rm\bf T}_\kappa G \backslash \gamma (p) \right]
\end{equation}
near almost any point at the plane $\pi_\gamma$ (except for the 
intersections of $\pi_\gamma$ with singular planes of factors not 
belonging to $\gamma$ ), are determined solely by $\gamma .$

Vice versa, let $p_0$ be any point from the space of $p$. Select from 
the graph the smallest subproduct $\gamma$  such that ${\rm\bf 
T}_\kappa  G \backslash \gamma (p)$ have no singularities at $p = 
p_0.$ Then $\gamma$ will be an IR-subgraph.

\SUBSECTION{Hierarchy of IR-subgraphs}
\label{ss9.5}

The set of all IR-subgraphs of a given graph has a natural ordering. 
Namely, $\gamma \le \Gamma$ if and only if $\gamma$ belongs to 
$\Gamma.$ Note that if $\gamma \le \Gamma,$ then $\pi_\gamma \supset 
\pi_\Gamma .$ Correspondingly, one can define {\em maximal} and {\em 
minimal} subgraphs. Note that there may be several maximal subgraphs; 
their singular planes do not intersect.

It is convenient to introduce the notion of {\em immediate precedence\/} 
for subgraphs. We say that a subgraph $\gamma$ immediately precedes 
the subgraph $\Gamma ,$ and write $\gamma  \triangleleft  \Gamma,$ \ if 
\ $\gamma  < \Gamma$ and there are no subgraphs between $\gamma$ and 
$\Gamma,$ i.e. no subgraphs $\Gamma'$ such that $\gamma  < \Gamma' < 
\Gamma .$ We describe the subgraph which immediately precedes a 
maximal one as {\em submaximal}.

\SUBSECTION{Proper variables of subgraphs}
\label{ss9.6}

Let $\Gamma$ be a subgraph of the graph $G.$ In general, $\Gamma(p,
\kappa)$ is independent of some components of $p$. Indeed, it is easy 
to check that $\Gamma(p,\kappa)$ is always invariant with respect to 
shifts in $p$ parallel to $\pi_\Gamma.$ Split $p$ into two parts: 
transverse and longitudinal with respect to $\pi_\Gamma :$ $p = 
(p_\Gamma,p_\Gamma^{\rm L})$, so that the singular plane $\pi_\Gamma$ 
is described by the equation $p_\Gamma = 0.$ Then $\Gamma(p,\kappa)$ 
depends only on $p_\Gamma.$ 
The variables $p_\Gamma$ will be referred to as the set 
of {\em proper variables\/} of $\Gamma.$

Although the choice of proper variables is not unique, transition to 
another set of proper variables is equivalent to a change of 
coordinates in the space of $p_\Gamma.$ Here invariant coordinateless 
formulations are possible and useful---for more on this see 
\cite{fvt-vvv}.

The proper variables of an IR-subgraph can be easily determined using 
its graphical representation. Thus, for the subgraph in Fig.~3 the 
proper variables are the set of momenta $p_1$ and $p_2$ (or $p_1$ and 
$p'_2 = p_1 + p_2$ etc.).

Consider the case when the IR-subgraph has loops (see e.g.\ Fig.~6). The 
loop momenta belong to the proper variables of the subgraph. Denote 
the set of all loop momenta as $p_\gamma^{\rm int},$ and the rest of 
the proper variables as $p_\gamma^{\rm ext}.$ So,
\begin{equation}\label{(9.5)}
     p_\gamma = \left( p_\gamma^{\rm int}, p_\gamma^{\rm ext} \right).
\end{equation}

To conclude this section, note that any IR-subgraph can be considered 
as a graph in its own right. Then the loop momenta of the graph should 
comprise all proper variables of the IR-subgraph. Graphically, this 
corresponds to merging all the ``external" vertices of the IR-subgraph 
into one. (In Fig.~6, the external vertices are $a$ and $b.$) The 
resulting graph possesses an additional property: it is its own (and 
the only maximal) IR-subgraph. In that case the singular plane of the 
entire graph is reduced to a point which can be assumed to be the zero 
point. It is clear that for each subgraph one can consider the problem 
of its asymptotic expansion. Moreover, we will see in 
subsect.~\ref{ss10.3} that construction of the \asop\ for the graph is 
logically preceded by its construction for the subgraphs. It is worth 
noting that the \asop\ for a subgraph is formulated entirely in terms 
of its proper variables.

\SUBSECTION{Quantitative characteristics}
\label{ss9.7}

Let us introduce some quantitative characteristics for description of 
singularities of IR-sub\-graphs. For each IR-subgraph $\Gamma$ define:
\begin{equation}\label{(9.6)}
     d_\Gamma = \sum_{g\in\Gamma} d_g ,
\end{equation}
which describes simultaneous scaling in $\kappa$ and the proper 
variables:
\begin{equation}\label{(9.7)}
     \Gamma(\lambda p,\lambda \kappa) 
     = \lambda ^{d_\Gamma} \Gamma(p_\Gamma,\kappa)
\end{equation}
(cf.\ property (a), subsect.~\ref{ss8.4});
\begin{equation}\label{(9.8)}
     {\rm dim}\, p_\Gamma
\end{equation}
is the dimension of the space of proper variables of the 
subgraph;
\begin{equation}\label{(9.9)}
     \omega _\Gamma = - d_\Gamma
     - {\rm dim}\, p_\Gamma
\end{equation}
is the {\em singularity index}. Its meaning is as follows. If for all 
$\Gamma \le G$ one has $\omega _\Gamma+ N < 0$ (strict inequality!) 
then the formal expansion of $G$ to order $\kappa^N$ does not contain 
non-integrable singularities and is a correct asymptotic expansion in 
the sense of distributions.

\SECTION{\kern-1pt\asop\ for products of singular functions}
\label{s10}

In this section, following the recipe of the extension principle 
(sect.~5), we establish formulae for the expansion of products of 
singular functions in the sense of distributions in the form 
of the so-called \asop\ whose structure is similar to that of the 
Bogoliubov \rop. The derivation presented bears a resemblance to, 
and was partially inspired by the 
analysis of the \rop\ in \cite{bog-shir}. Explicit expressions for 
counterterms of the \asop\ (which, unlike the \rop, is determined 
uniquely) will be obtained in the next section.

\SUBSECTION{General remarks and notations}
\label{ss10.1}

We are going to derive the recipe of the \asop, the instrument for 
evaluating expansions in the form of infinite asymptotic series for 
any product (graph) $G(p)$ of the described type. We will use the 
notation:
\begin{equation}\label{(10.1)}
     G(p,\kappa) \asy{\kappa}{0} {\rm\bf As}\, G(p,\kappa) 
     = \sum_n \kappa^n G_n(p,\kappa).
\end{equation}
Here each $G_n$ should be a distribution in $p$, and the dependence on 
$\kappa$ should be soft (see the definition in subsect.~\ref{ss4.1}). 
The expansion \ref{(10.1)} should be valid in the sense 
of distributions (see subsect.~\ref{ss4.1}). This means that for each 
test function $\varphi$  the following estimate should hold:
\begin{equation}\label{(10.2)}
     \int  dp\,  \varphi (p) 
     \left[ G(p,\kappa) 
          - {\rm\bf As}_N G(p,\kappa) 
     \right] 
     = o(\kappa^N ).
\end{equation}
The subscript on the \asop\ means that in \ref{(10.2)} all the terms of 
order $o(\kappa^N )$ are discarded, i.e.
\begin{equation}\label{(10.3)}
     {\rm\bf As}_N G = \sum_{n\le N} \kappa^n G_n .
\end{equation}
Existence of such expansions is either obvious or non-trivial, 
depending on the point of view. It is 
rather obvious if one recalls the results of
\cite{dslav:73} wherefrom one concludes that
expansion of $G$ integrated with a test function 
runs in powers and logarithms of the expansion parameter 
and takes into account that such an expansion should be linear in 
test functions. On the other hand, from the point of view of
the technique of \cite{zimm} existence of such expansions is not 
at all trivial because the construction of \cite{zimm} 
corresponds to an expansion similar to \ref{(10.3)} for each $N$ but 
with $G_n$ depending on $N$, too.

\SUBSECTION{Uniqueness of the expansion}
\label{ss10.2}

From subsect.~\ref{ss4.2} it follows that if an expansion of the form 
\ref{(10.1)} exists, then it is unique (provided one always expands in 
powers and logarithms of the small parameter). Uniqueness of the 
\asop\ will allow us in subsect.~\ref{ss10.3} to determine the necessary 
conditions which it must satisfy. This will give us a sufficient 
number of hints to construct it explicitly.

\SUBSECTION{Localization property of the \asop}
\label{ss10.3}

Let us study the local structure of the \asop\ (cf.\ remark ($i$) in 
subsect.~\ref{ss3.2}). First of all note that the formal expansion
\begin{equation}\label{(10.4)}
     G(p,\kappa) \sim {\rm\bf T} G(p,\kappa) ,\quad\quad\kappa\to 0,
\end{equation}
allows integrations with test functions $\varphi$ whose support does 
not intersect singular planes of the formal expansion:
\begin{equation}\label{(10.5)}
     {\rm supp} \, \varphi  \bigcap \biggl( \bigcup_\gamma \pi_\gamma 
                                  \biggr) = \emptyset .
\end{equation}
This follows from the properties of the functions of which $G$ is 
built (cf.\ subsect.~\ref{ss8.4}). The r.h.s.\ of \ref{(10.4)} defines a 
functional which is an asymptotic expansion of the functional on the 
l.h.s., on the subspace of test functions satisfying \ref{(10.5)}. From 
uniqueness it follows that on test functions satisfying \ref{(10.5)} 
the \asop\ should coincide with the operation of the formal expansion 
{\rm\bf T}:
\begin{equation}\label{(10.6)}
     {\rm\bf As} \, G(p,\kappa) = {\rm\bf T} G(p,\kappa), 
     \quad\quad p \notin \bigcup \pi_\gamma .
\end{equation}

Further, consider any small region ${\cal O},$ and let $\Gamma$ be a 
subgraph such that ${\rm\bf T} G \backslash \Gamma(p)$ has no 
singularities within ${\cal O}$ (cf.\ subsect.~\ref{ss3.1}). Take an 
arbitrary $\varphi$  such that ${\rm supp}\, \varphi  \subset {\cal O}.$ 
Then
\begin{equation}\label{(10.7)}
     \int  dp \, G(p,\kappa) \varphi (p) 
     = \int  \Gamma(p,\kappa) \varphi ^{\prime\prime}(p,\kappa) ,
\end{equation}
where
\begin{equation}\label{(10.8)}
     \varphi ^{\prime\prime}(p,\kappa) 
     = G \backslash \Gamma(p,\kappa) \varphi (p)
\end{equation}
is a test function with ${\rm supp}\, \varphi  \subset  {\rm supp}\, 
\varphi ^{\prime\prime},$ but depending on $\kappa.$

Recall (subsect.~\ref{ss9.6}) that $\Gamma$ is independent of that 
component of $p$ which is longitudinal with respect to $\pi_\Gamma .$ 
Therefore we may perform integration of $\varphi ^{\prime\prime}$ over 
the longitudinal components on the r.h.s.\ of \ref{(10.7)}. So, we 
arrive at the problem of expansion for the subgraph in terms of its 
proper variables, and with test functions depending on the expansion 
parameter. The assumptions made in subsect.~\ref{ss8.4} ensure that the 
expansion procedure described in subsect.~\ref{ss4.4} will be 
applicable here.

So, assume that the problem of expanding $\Gamma(p,\kappa)$ is solved, 
i.e. that the action of the \asop\ on $\Gamma$ is already known. Then 
the expansion of the r.h.s.\ of \ref{(10.7)} can be represented as
\begin{equation}\label{(10.9)}
     \int dp\,\left[ {\rm\bf As}\Gamma(p,\kappa) \right] \times
               \left[ {\rm\bf T}\varphi^{\prime\prime}(p,\kappa) \right]
     = \int dp\, \left[ {\rm\bf As}\, \Gamma(p,\kappa) \right] \times
               \left[{\rm\bf T} G \backslash \Gamma(p,\kappa) \right] 
               \varphi (p).
\end{equation}
On the other hand, the expansion of the r.h.s.\ of \ref{(10.7)} should be 
given by 
\begin{equation}
     \int  dp \, {\rm\bf As} \, G(p,\kappa) \varphi (p).
\end{equation}
Then again from uniqueness of the \asop\ we conclude that, given the 
relation described above between the region ${\cal O}$ and the 
subgraph $\Gamma,$ on all test functions localized in ${\cal O}$ the 
following {\em localization property} must be valid:
\begin{equation}\label{(10.10)}
     {\rm\bf As} \, G(p) 
     = [{\rm\bf T} G \backslash \Gamma(p)] \times [{\rm\bf As}\, \Gamma(p) ], 
     \quad\quad p \in  {\cal O}.
\end{equation}
This is the most important structural property of the \asop. 
It exhibits the recursive structure of the expansion problem 
considered on the entire collection of Feynman diagrams.

\SUBSECTION{Structure of the \asop}
\label{ss10.4}

Let us represent the \asop\ in terms of counterterms localized at 
singular points of the formal expansion, in analogy with the 
well-known expression of the Bogoliubov \rop\ in terms of quasilocal 
counterterms \cite{bog-shir}. ${\rm\bf As} \, G$ may differ from ${\rm\bf 
T} G$ only by corrections localized on singular planes of the formal 
expansion. Consequently, one can take the following ansatz for the 
full expansion:
\begin{equation}\label{(10.11)}
     {\rm\bf As} \, G = {\rm\bf T} G 
     + \sum_\gamma \left( {\rm\bf E} \gamma \right) K_{G,\gamma} ,
\end{equation}
where summation runs over all IR-subgraphs, while the distribution 
$({\rm\bf E} \gamma )$ is localized on $\pi_\gamma .$ Let us determine 
$K_{G,\gamma}$ from \ref{(10.10)}, expanding the \asop\ in accordance with 
\ref{(10.11)}:
\begin{equation}\label{(10.12)}
     {\rm\bf T} G + \sum_{\gamma\le G} 
         \left( {\rm\bf E} \gamma \right) K_{G,\gamma}
     = \left( {\rm\bf T} G\backslash\Gamma \right) 
        \bigl({\rm\bf T} \Gamma 
             + \sum_{\gamma\le\Gamma} 
               \left( {\rm\bf E} \gamma \right) K_{\Gamma,\gamma} 
        \bigr) .
\end{equation}
We get the conditions:
\begin{equation}\label{(10.13)}
     K_{G,\gamma} = K_{\Gamma,\gamma}\,
     ({\rm\bf T} G\backslash\Gamma) .
\end{equation}
Without loss of generality we may assume that for each $\Gamma:$
\begin{equation}\label{(10.14)}
     K_{\Gamma,\Gamma} = 1 .
\end{equation}
(Indeed, within the problem for the subgraph $\Gamma$ considered in 
terms of its proper variables, the distribution $({\rm\bf E} \Gamma)$ 
is localized at the origin, so that $K_{\Gamma,\Gamma}$ must be a 
constant independent of momentum variables. And such a constant can 
always be included into $({\rm\bf E} \Gamma).$)

Setting $\gamma  = \Gamma$ in \ref{(10.13)} and using \ref{(10.14)}, we 
get:
\begin{equation}\label{(10.15)}
     K_{G,\Gamma} = {\rm\bf T} G \backslash \Gamma
\end{equation}
for each subgraph $\Gamma \le  G.$ We see that the dependence on 
$\Gamma$ disappears from the r.h.s.\ of \ref{(10.13)}, as expected. 
Finally:
\begin{equation}\label{(10.16)}
     {\rm\bf As} \, G = {\rm\bf T} G 
                   + \sum_{\gamma\le G} \left( 
                          {\rm\bf E} \gamma
                      \right) 
                      \left( 
                        {\rm\bf T} G\backslash \gamma 
                      \right) .
\end{equation}
The expressions $({\rm\bf E} \gamma )$ will be referred to as {\em 
counterterms for subgraphs} $\gamma .$

For \ref{(10.1)} to be true, it is necessary that the counterterm had 
the form of an infinite series
\begin{equation}\label{(10.17)}
     \left( {\rm\bf E} \gamma \right) 
     = \sum_n \kappa^n \left( {\rm\bf E} \gamma \right)_n ,
\end{equation}
where $({\rm\bf E} \gamma )_n$ depend softly on $\kappa.$

\SUBSECTION{Summary}
\label{ss10.5}

So, if the \asop\ \ref{(10.1)}--\ref{(10.3)} exists, it is unique and, in 
accordance with the extension principle, must have the form 
\ref{(10.16)}--\ref{(10.17)}. However, the analysis of 
subsect.~\ref{ss10.3} indicates that the \asop\ can be naturally 
defined by \ref{(10.16)} using the induction based on the natural order 
among IR-subgraphs (see also below sect.~\ref{s11}). Then if the \asop\ 
has been constructed for all $\gamma  < \Gamma,$ it only remains to 
determine the counterterms $({\rm\bf E} \Gamma)$ possessing all the 
required properties. Note that the reasoning of sect.~\ref{s6} is in 
fact applicable to any minimal IR-subgraphs, which provides a correct 
starting point for the induction. The general formulae for the 
counterterms derived in the next section will include the case of 
minimal subgraphs as a simple special case.

\SECTION{Counterterms of the \asop}
\label{s11}

\SUBSECTION{Structure of the {\bf As$'$}-operation}
\label{ss11.1}

Suppose that the existence of the \asop\ has been established for all 
$\gamma < \Gamma,$ and explicit expressions for $({\rm\bf E} \gamma )$ 
of the form of \ref{(10.17)} have been found. From the reasoning of the 
preceding section it follows that the expansion for $\Gamma$ valid on 
test functions which are equal to zero around $p = 0$ has the form:
\begin{equation}\label{(11.1)}
     \Gamma( p_\Gamma ,\kappa ) 
     \asy{\kappa}{0} 
     {\rm\bf As}' \Gamma\left( p_\Gamma ,\kappa \right),
     \quad\quad    p_\Gamma \not= 0 ,
\end{equation}
where ${\rm\bf As}'$ is the \asop\ without the last counterterm:
\begin{equation}\label{(11.2)}
     {\rm\bf As}' \Gamma 
     = {\rm\bf T} \Gamma 
     + \sum_{\gamma<\Gamma} \left({\rm\bf E} \gamma \right) 
       \left( {\rm\bf T} \Gamma \backslash \gamma \right) ,
\end{equation}
so that
\begin{equation}\label{(11.3)}
     {\rm\bf As}\, \Gamma = {\rm\bf As}' \Gamma + ({\rm\bf E} \Gamma) .
\end{equation}
Indeed, let $H$ run over all submaximal subgraphs of $\Gamma,$ i.e. $H 
\triangleleft \Gamma$ (see subsect.~\ref{ss9.5}). Then pairwise 
intersections of the singular planes $\pi_H$  consist only of the 
point $p_\Gamma = 0.$ Therefore, if $\varphi (p_\Gamma) = 0$ around 
$p_\Gamma = 0$ then $\varphi$  can be represented as
\begin{equation}\label{(11.4)}
     \varphi  = \sum_{H\triangleleft\Gamma} \varphi _H ,
\end{equation}
where $\varphi _H (p_\Gamma)\equiv0$ in a neighbourhood of $\pi_{H'}$  
for any $H' \not= H .$ Then
\begin{equation}\label{(11.5)}
     < \Gamma \varphi  >\quad 
     = \sum_{H\triangleleft\Gamma}  < \Gamma \varphi _H  >
     = \sum_{H\triangleleft\Gamma} 
     < H, \Gamma \backslash H \varphi _H > 
\end{equation}
\begin{equation}
    \asy{\kappa}{0} 
     \sum_{H\triangleleft\Gamma} 
     < {\rm\bf As}\, H, {\rm\bf T} \Gamma \backslash H \varphi _H  > 
\end{equation}
(cf.\ the reasoning in subsect.~\ref{ss10.3}, especially \ref{(10.9)}). 
On the other hand, the operation ${\rm\bf As}'$ as defined in 
\ref{(11.2)} inherits the localization property analogous to 
\ref{(10.10)}:
\begin{equation}\label{(11.6)}
     < {\rm\bf As}' \Gamma, \varphi _H  > 
     = < {\rm\bf As}\, H, {\rm\bf T} \Gamma \backslash H \varphi _H  > ,
\end{equation}
so that the r.h.s.\ of \ref{(11.5)} is equal to
\begin{equation}\label{(11.7)}
     \sum_{H\triangleleft\Gamma} 
      < {\rm\bf As}' \Gamma, \varphi _H > 
      = < {\rm\bf As}' \Gamma, \varphi > ,
\end{equation}
whence follows \ref{(11.1)}.

\SUBSECTION{Approximation properties of the {\bf As$'$}-operation}
\label{ss11.2}

Our inductive assumptions imply that all counterterms $({\rm\bf E} 
\gamma )$ for $\gamma  < \Gamma$ have the form \ref{(10.17)}. Then 
${\rm\bf As}' \Gamma$ can be represented as an expansion in powers of 
$\kappa:$
\begin{equation}\label{(11.8)}
     {\rm\bf As}' \Gamma(p_\Gamma,\kappa) 
     = \sum_n \kappa^n \Gamma'_n(p_\Gamma,\kappa) ,
\end{equation}
where $\Gamma'_n(p_\Gamma,\kappa)$ can softly depend on $\kappa.$ 
Besides, the functions $\Gamma'_n$ possess a number of natural 
properties following from our assumptions. First, they inherit the 
scaling property \ref{(9.7)}:
\begin{equation}\label{(11.9)}
     \Gamma'_n( \lambda p_\Gamma, \lambda \kappa ) 
     = \lambda^{d_g-n} \Gamma'_n(p_\Gamma,\kappa) .
\end{equation}
Second, from \ref{(11.9)} and the soft dependence on $\kappa,$ one 
immediately gets the scaling property in the momentum argument:
\begin{equation}\label{(11.10)}
     \Gamma'_n( \lambda p_\Gamma, \kappa ) = \lambda ^{d_g-n}
     \left( \Gamma'_n(p_\Gamma,\kappa) 
          + {\rm soft\ corrections}
     \right).
\end{equation}
To put it differently, the dependence of the expression 
$\lambda^{n-d_g} \Gamma'_n(\lambda p_\Gamma,\kappa)$ on $\lambda$ is 
soft.

Note the important parallel between what we have here and what we had 
in subsect.~\ref{ss6.1}: the expansion \ref{(11.8)}, similarly to \ref{(6.2)}, 
contains terms progressively more singular at $p_\Gamma \sim 0$ as 
$n\to\infty,$ and both expansions are valid in the sense of the 
distribution theory for $p_\Gamma \not= 0.$ We can push the analogy 
even further. Consider the expansion \ref{(11.8)} truncated at the 
terms of order $O(\kappa^N ):$
\begin{equation}\label{(11.11)}
     \Gamma = {\rm\bf As}'_N \Gamma + o(\kappa^N ) ,
     \quad\quad p_\Gamma \not= 0 .
\end{equation}
(The use of the subscript $N$ here is analogous to \ref{(10.3)}.) From 
\ref{(11.10)} it follows that the r.h.s.\ of \ref{(11.11)} becomes 
non-integrable at zero for $\omega_\Gamma+N\ge0$ ($\omega_\Gamma$ is 
defined in \ref{(9.9)}). However if the test function has a zero of 
order $\omega _\Gamma+N+1$ at $p_\Gamma=0,$ then the r.h.s.\ of 
\ref{(11.11)} is well defined and, moreover, on such test functions the 
estimate $o(\kappa^N)$ of the expansion \ref{(11.11)} is valid.

The last property is fully natural. Its proof is essentially not 
difficult, is entirely based on  power counting and has the same 
structure as that in subsect.~\ref{ss6.2}. One considers a neighborhood 
of the origin of a radius $O(\kappa)$ wherein the uniformity property 
\ref{(11.9)} is used for explicitly extracting the factor 
$O(\kappa^{N+1})$ from the integral of the remainder term of the 
expansion over this neighbourhood. Then one uses the properties of the 
\asop\ on subgraphs (the inductive assumption) to estimate the 
integral over the rest of the space. This last step is somewhat 
cumbersome though fully straightforward. A detailed discussion can be 
found in \cite{fvt-vvv}. (See also sect.~\ref{s12} below.)

\SUBSECTION{Expressions for counterterms}
\label{ss11.3}

The above results allow one to apply here the same reasoning as in the 
example in sect.~\ref{s6}. From the extension principle it follows that 
the expansion \ref{(11.11)} can be extended to all $p_\Gamma$ by 
addition of counterterms localized at $p_\Gamma = 0:$
\begin{equation}\label{(11.12)}
     \Gamma = {\rm\bf As}'_N \Gamma 
     + \left({\rm\bf E}_N \Gamma\right) + o(\kappa^N ) .
\end{equation}
$({\rm\bf E}_N\Gamma)$ may only contain derivatives of the 
$\delta$-function of order not higher than $\omega _\Gamma+ N:$
\begin{equation}\label{(11.13)}
     \left( {\rm\bf E}_N \Gamma \right) \left( p_\Gamma \right) 
     = \sum_{|\alpha|\le\omega_\Gamma+N} c_{\Gamma,\alpha} (\kappa) 
       \delta _{\Gamma,\alpha} (p_\Gamma)
\end{equation}
(here $\delta _{\Gamma,\alpha} (p_\Gamma)$ is a full basis of 
$\delta$-functions and their derivatives localized at $p_\Gamma = 0;$ 
the order of derivatives is denoted as $|\alpha|$). To find explicit 
expressions of the coefficients $c_{\Gamma,\alpha}$ (recall that in 
general they depend on $N,$ but it will turn out possible to choose 
them independent of $N),$ we use the trick of subsect.~\ref{ss6.9}.

Consider the expression
\begin{equation}\label{(11.14)}
     \int  dp_\Gamma A(p_\Gamma) B(p_\Gamma) 
     = o(\kappa^N ) ,
\end{equation}
where
\begin{equation}\label{(11.15)}
     \!\!\! A = \Gamma - {\rm\bf As}'_N \Gamma,
\end{equation}
\begin{equation}\label{(11.16)}
     B = \varphi - {\rm\bf T}_{\omega_\Gamma+N} \varphi,
\end{equation}
with $\omega_\Gamma$ as defined in \ref{(9.9)}. The fact that the  
l.h.s.\ of \ref{(11.14)} is indeed $o(\kappa^N)$ is proved exactly as in 
subsect.~\ref{ss6.9}; one only has to note that owing to the scaling 
properties \ref{(9.7)}, \ref{(11.9)} one has:
\begin{equation}\label{(11.17)}
     \left( I - {\rm\bf As}'_N \right) \Gamma(\lambda p_\Gamma,\kappa) 
     = \lambda^{d_\Gamma} \left( I - {\rm\bf As}'_N \right) 
       \Gamma( p_\Gamma, {\kappa/\lambda} )
\end{equation}
\begin{equation}
     \asy{\lambda}{\infty} \lambda^{d_\Gamma} 
     \times o\left( \left( { \kappa / \lambda } \right)^N  \right), 
     \quad\quad p_\Gamma \not= 0.
\end{equation}
Therefore, the integral converges at $p_\Gamma\to \infty$ by power 
counting. (Note, however, that this is not an ordinary absolutely 
convergent integral because the term ${\rm\bf As}'_N \Gamma$ is a 
distribution. A more accurate definition of the integral in 
\ref{(11.14)} involves introducing a cut-off 
$\Phi(p_\Gamma/\Lambda)$ with 
any $\Phi$ defined as in \ref{(7.3)} and taking the limit 
$\Lambda\to\infty.$ The resulting value is independent of the exact 
form of $\Phi.$ Also note that the integral \ref{(11.14)} can be 
transformed into an absolutely convergent integral of ordinary 
functions by invoking the explicit expressions for the operation 
${\rm\bf As}'$ and integrating out the $\delta$-functions.)

Introduce the dimensional regularization into the l.h.s.\ of 
\ref{(11.14)}, multiply $A$ and $B,$ and perform termwise integrations 
using the fact that the dimensionally regularized integration 
nullifies integrands with simple scaling properties---cf.\ \ref{(6.26)}. 
Finally, we arrive at the following analogue of \ref{(6.37)}:
\begin{equation}\label{(11.18)}
     c_{\Gamma,\alpha} (\kappa) 
     = \int  dp_\Gamma \, {\cal P}_{\Gamma,\alpha} (p_\Gamma) 
       \,\Gamma(p_\Gamma,\kappa) ,
\end{equation}
where ${\cal P}_{\Gamma,\alpha}$ is the basis of polynomials of the 
variable $p_\Gamma,$ which is dual to the basis of $\delta$-functions 
$\delta_{\Gamma,\alpha} (p_\Gamma)$---cf.\ \ref{(6.33)}. The 
counterterms $({\rm\bf E} \Gamma)_n$ from \ref{(10.17)} now are as 
follows:
\begin{equation}\label{(11.19)}
     \left( {\rm\bf E} \Gamma \right)_n 
     = \sum_{|\alpha|=\omega_\Gamma+n} c_{\Gamma,\alpha} (\kappa) 
       \, \delta _{\Gamma,\alpha} .
\end{equation}

Using the scaling properties, we get:
\begin{equation}\label{(11.20)}
     c_{\Gamma,\alpha} (\kappa) 
     = \kappa^{(|\alpha|-\omega_\Gamma+\ldots)}
       c_{\Gamma,\alpha} (1) ,
\end{equation}
where the dots denote a correction proportional to $D - 4,$ the 
deviation of the complex parameter of dimensionality from the 
canonical integer value. The integer part of the exponent in 
\ref{(11.20)} shows which order in $\kappa$ this term belongs to. One 
can also see that the dependence on $\kappa$ in the expansion ${\rm\bf 
As} \Gamma$ is as discussed in subsect.~\ref{ss4.3}.

The formulae \ref{(10.16)}, \ref{(10.17)}, \ref{(11.19)} and \ref{(11.18)} 
present the full solution of the Master problem in the Euclidean case.

\SUBSECTION{Remarks}
\label{ss11.4}

$\quad$
($i$) It should be stressed that although the above formulae are 
represented in a form which is specific to dimensional 
regularization, there are no serious obstacles (except for necessity 
to introduce a large number of new notations) for writing analogous 
formulae in other regularizations (e.g.\ using cut-offs). But this 
falls outside the scope of the present paper (see, however, 
\cite{fvt-vvv}).

($ii$) Concerning the mathematical aspects of our derivation, the full 
details of the proof that are regularization independent are presented 
in \cite{fvt-vvv}. As to the use of the dimensional regularization in 
the preceding section, it can be justified along the following lines. 
First, the definition of dimensionally regularized integrals of 
arbitrary smooth test functions over momentum variables was presented 
in \cite{blekher}. Second, as noted after \ref{(11.17)}, the integral 
in \ref{(11.14)} can be represented as an absolutely convergent 
integral of ordinary smooth functions. Therefore, one only has to 
extend the results of  \cite{blekher} to a wider class of functions, 
which seems to present no difficulties. 
This issue will, hopefully, be addressed in a future 
publication \cite{dregm}. 
Another approach would be to check the final formulae for 
expansions of Feynman diagrams using the $\alpha$-parametric integral 
representation. But the attempts to do so \cite{smirnov} show that a 
full proof would be a straightforward, cumbersome and unilluminating 
excercise.

\SUBSECTION{Commutativity with multiplication by polynomials}
\label{ss11.5}

In conclusion of the discussion of the Master problem, a remark is in 
order. We have proved existence of the \asop\ and its uniqueness 
(subsect.~\ref{ss4.2}). It follows that it commutes with multiplication 
of $\Gamma$ by an arbitrary polynomial of momenta---indeed, the 
polynomial can either be included into the distribution to be expanded
($\Gamma$) or, equivalently, it can be considered to be 
a part of the test function;
the final result cannot depend on which option is chosen.
This means that the operation 
${\rm\bf E}$ which generates the counterterm for a subgraph, commutes 
with multiplication by polynomials, too. This can also be checked with 
the help of the explicit representations obtained above.

The described property of the \asop\ is useful in 
situations when one needs to transform expansions of Feynman diagrams 
to a convenient form.

\SECTION{Example of an {\it As}-expansion}
\label{s12}

\SUBSECTION{The expression to be expanded and its singularities}
\label{ss12.1}

Let the integration variable $p$ consist of two 4-dimensional 
components:
\begin{equation}\label{(12.1)}
     p = ( p_1 , p_2 ).
\end{equation}
The product to be expanded is (cf.\ Fig.~7a):
\begin{equation}\label{(12.2)}
     G(p,\kappa) = g_1(p,\kappa) \,g_2(p,\kappa) \,g_3(p,\kappa),
\end{equation}
where 
\begin{equation}\label{(12.3)}
     g_1(p,\kappa) = {1\over p_1^2+\kappa^2} ,
\end{equation}
\begin{equation}
     g_2(p,\kappa) = {1\over p_2^2+\kappa^2} ,
\end{equation}
\begin{equation}
     \quad\quad\;\;\,\;\kern2mm
      g_3(p,\kappa) = {1\over (p_1-p_2)^2+\kappa^2} .
\end{equation}
The expansion of an isolated propagator in the sense of the 
distribution theory has been obtained and studied in sect.~\ref{s6}.

Each propagator in \ref{(12.2)} generates, upon formal Taylor expansion 
in $\kappa,$ singularities at the points where the corresponding 
momentum variable vanishes. Thus, the singular planes for each factor 
are:
\begin{equation}\label{(12.4a)}
     \pi_1 = \{ p \mid p_1 = 0 \},
\end{equation}
\begin{equation}\label{(12.4b)}
     \pi_2 = \{ p \mid p_2 = 0 \},
\end{equation}
\begin{equation}\label{(12.4c)}
     \quad\quad\, \pi_3 = \{ p \mid p_1 - p_2 = 0 \},
\end{equation}
i.e. $\pi_1$ consists of all $p$ such that $p_1 = 0,$ etc.

The set of all IR-subgraphs of $G$ comprises three subgraphs 
$\gamma_i,$ $i = 1,2,3,$ each consisting of one factor, and $G$ 
itself. The two-factor subsets do not satisfy the completeness 
condition and therefore are not IR-subgraphs.

\SUBSECTION{Fixing singularities corresponding to subgraphs}
\label{ss12.2}

Now we note that in the space of $p$ {\em without\/} the point $p=0,$ 
the planes $\pi_i$ do not intersect, so that a reasoning similar to 
that of subsects.~\ref{(6.10)} and \ref{(11.1)} can be used. Indeed, 
consider a test function $\varphi_1(p)$ which is non-zero within the 
region ${\cal O}$ shown in Fig.~7b and therefore is zero in 
neighbourhoods of both $\pi_2$ and $\pi_3.$ Then
\begin{equation}\label{(12.5)}
     \int  dp_1\,  dp_2\,  G(p,\kappa) \, \varphi _1(p) 
     \equiv 
     \int  dp_1\, g_1(p,\kappa) \, {\tilde\varphi}_1(p_1,\kappa),
\end{equation}
where
\begin{equation}\label{(12.6)}
     \tilde\varphi_1(p_1,\kappa) 
     = \int  dp_2\,  G \backslash g_1(p,\kappa) \, \varphi_1(p) 
     = \int  dp_2\,  
        g_2(p,\kappa)\,  g_3(p,\kappa)\,  \varphi _1(p) .
\end{equation}

The expansion of the expression \ref{(12.5)} can be obtained (cf.\ 
subsect.~\ref{ss10.3}; also note that we present the formulae in 
dimensionally regularized form) by Taylor-expanding 
$\tilde\varphi_1(p_1,\kappa)$ (which does not give rise to any 
singularities in $p_1$) and using \ref{(6.36)} to expand the propagator 
$g_1$ (cf.\ \ref{(10.10)}). Denote:
\begin{equation}\label{(12.7)}
     \left[ {\rm\bf E} g_1 \right] (p,\kappa) 
     = \sum_\alpha c_\alpha (\kappa) \, \delta _\alpha (p_1)
\end{equation}
with $c_\alpha (\kappa)$ given by \ref{(6.37)}, where $\delta_\alpha$ 
and ${\cal P}_\alpha$ have been introduced in \ref{(6.32)}, \ref{(6.33)} 
(it is convenient to assume that $\delta_0 = \delta$ and ${\cal P}  = 
1$). Then one can rewrite the obtained result as follows:
\begin{equation}\label{(12.8)}
     G(p,\kappa) \asy{\kappa}{0} 
     \left( {\rm\bf T} g_1(p,\kappa) \, 
                       g_2(p,\kappa) \,
              g_3(p,\kappa)
     \right) 
     + \left( {\rm\bf T} g_2(p,\kappa) \, g_3(p,\kappa) \right) 
       \left[ {\rm\bf E} g_1 \right] (p,\kappa),
\end{equation}
which is valid on the test functions $\varphi_1$ described above. 

Analogous expansions can be obtained for the cases of test functions 
$\varphi_2$ and $\varphi_3$ that are in the same relation to the 
singular planes $\pi_2$ and $\pi_3,$ respectively, as $\varphi_1$ is 
to $\pi_1.$

%\SUBSECTION{}
%\label{ss12.3}

Consider a test function $\varphi_0(p)$ which is identically zero in 
some neighbourhood of the point $p = 0.$ It can always be represented 
as
\begin{equation}\label{(12.9)}
     \varphi_0(p) = \varphi_1(p) + \varphi_2(p) + \varphi_3(p),
\end{equation}
where all $\varphi_i$ are as described above. Therefore, the three 
expansions that are valid on each $\varphi_i$ can be glued together 
into an expansion that is valid on any $\varphi_0:$
\begin{equation}\label{(12.10)}
     G(p,\kappa) \asy{\kappa}{0} 
     \left( {\rm\bf T} g_1(p,\kappa)\, g_2(p,\kappa)\, g_3(p,\kappa)
     \right) 
     + \left( {\rm\bf T} g_2(p,\kappa)\, g_3(p,\kappa) \right)\, 
       \left[ {\rm\bf E} g_1 \right] (p,\kappa)
\end{equation}
\begin{equation}
     + \left( {\rm\bf T} g_1(p,\kappa)\, g_3(p,\kappa) \right)\, 
       \left[ {\rm\bf E} g_2 \right](p,\kappa)
     + \left( {\rm\bf T} g_1(p,\kappa)\, g_2(p,\kappa) \right)\, 
       \left[ {\rm\bf E} g_3 \right] (p,\kappa)
\end{equation}
\begin{equation}
     \equiv {\rm\bf As}' G(p,\kappa) .\kern9cm
\end{equation}
This agrees with \ref{(11.1)}--\ref{(11.2)}.

Consider the expansion \ref{(12.10)} to order $o(\kappa^2):$
\begin{equation}\label{(12.11)}
     {\rm\bf As}'_2 G(p,\kappa) = 
     {1\over p_1^2 p_2^2 (p_1-p_2)^2} 
\kern5cm\end{equation}
\begin{equation}
     {}- \kappa^2 \left[ {1\over p_1^4 p_2^2 (p_1-p_2)^2} 
                     + {1\over p_1^2 p_2^4 (p_1-p_2)^2} 
                     + {1\over p_1^2 p_2^2 (p_1-p_2)^4} 
                \right]
\end{equation}
\begin{equation}
     {}+ c_0 (\kappa) \left[ \delta (p_1) {1\over p_2^2 (p_1-p_2)^2} 
     + {1\over p_1^2 } \delta (p_2) {1\over (p_1-p_2)^2 } 
     + {1\over p_1^2 p_2^2 } \delta (p_1-p_2) \right].
\end{equation}
Recall that 
\begin{equation}\label{(12.12)}
     c_0(\kappa) = \int  d^Dq\,  (q^2 + \kappa^2)^{-1} 
     = \kappa^{2(D-4)}\, c_0(1) .
\end{equation}

One sees that the $O(\kappa^2)$ terms possess a logarithmic 
singularity at $p = 0.$ Therefore, the r.h.s.\ of \ref{(12.11)} is 
well-defined on test functions $\varphi (p)$ such that $\bar \varphi 
(0) = 0.$ Moreover, on such test functions the asymptotic character of 
the expansion is preserved which means (to order $o(\kappa^2)$) that:
\begin{equation}\label{(12.13)}
     \int  d^{2D}p\,  \bar \varphi (p) 
     \left[ G(p,\kappa) - {\rm\bf As}'_2 G(p,\kappa) 
     \right] 
     = o(\kappa^2).
\end{equation}
This should be compared with \ref{(6.3)}. An explicit proof of 
\ref{(12.13)} proceeds along the same lines as that of \ref{(6.3)}. A 
new element here is that one should consider cone regions in the 
integration space, as follows. (There is a limited resemblance between 
such cone regions and the Hepp sectors used in the theory of 
$\alpha$-representation. However, the Hepp sectors correspond to a 
complete resolution of the recursive structure of singularities, while 
in our case only one-level descent---form the complete graph to its 
largest subgraphs---is performed.)

\SUBSECTION{Approximation properties of As$'$}
\label{ss12.4}

Let $\theta _1(p)$ be such that $\theta _1(p) \equiv 1$ around each 
non-zero point of $\pi_1,$ $\theta _1(p) \equiv 0$ around non-zero 
points of $\pi_2$ and $\pi_3,$ and $\theta _1(\lambda p) = \theta 
_1(p)$ for each $\lambda,$ $p \not= 0.$ Consider the following 
contribution to \ref{(12.13)}:
\begin{equation}\label{(12.14)}
     \int  d^{2D}\! p \,\theta _1(p) \bar \varphi (p) 
     \left[ G(p,\kappa) - {\rm\bf As}'_2 G(p,\kappa) 
     \right].
\end{equation}
One splits the integration region into two parts: $|p_1| > \kappa$ and 
$|p_1| < \kappa$ (cf.\ Fig.~7c and subsect.~\ref{ss6.2}). In the region 
$|p_1| < \kappa,$ one rescales $p\to \kappa p$ then takes into account 
the scaling properties of $G$ (which are not violated by the \asop) 
and the fact that $\bar\varphi (p) \sim p$ near $p = 0,$ and finds 
that
\begin{equation}\label{(12.15)}
     \int_{|p_1|<\kappa} d^{2D}p\, \theta _1(p)\, \bar \varphi (p) 
     \left[ G(p,\kappa)
          - {\rm\bf As}'_2 G(p,\kappa) 
     \right] = O(\kappa^3).
\end{equation}
In the region $|p_1| > \kappa$, it is convenient to reorder the 
integrand of \ref{(12.14)} as follows:
\begin{equation}\label{(12.16)}
     \int_{|p_1|>\kappa} d^{2D}p\, \theta _1(p)\, \bar \varphi (p) 
     \Biggl\{ 
         G(p,\kappa) - { 1\over p_1^2 p_2^2 (p_1-p_2)^2 } 
\end{equation}
\begin{equation}
     + \kappa^2 \left[
            {1\over p_1^4 p_2^2 (p_1-p_2)^2 } 
          + {1\over p_1^2 p_2^4 (p_1-p_2)^2 } 
          + {1\over p_1^2 p_2^2 (p_1-p_2)^4 } 
       \right]
      \Biggr\} 
\end{equation}
\begin{equation}
     = \int_{|p_1|>\kappa} d^{2D}p\, \theta _1(p)\, \bar \varphi (p) 
     \times \Biggl\{ \left[ {1\over p_1^2+\kappa^2} 
                          - {1\over p_1^2} 
                          + \kappa^2 {1\over p_1^4} 
                    \right] 
                    {1\over p_2^2+\kappa^2} 
                    {1\over (p_1-p_2)^2+\kappa^2 }
\end{equation}
\begin{equation}
                  + \left[ {1\over p_1^2} \right] 
                    \biggl[{1\over p_2^2+\kappa^2} 
                           {1\over (p_1-p_2)^2+\kappa^2} 
                         - { 1\over p_2^2 (p_1-p_2)^2 } 
                         + \kappa^2 \left( 
                                       {1\over p_2^4 (p_1-p_2)^2 } 
                                     + {1\over p_2^2 (p_1-p_2)^4 } 
                                    \right) 
                    \biggr] 
\end{equation}
\begin{equation}
     + \left[ - \kappa^2 {1\over p_1^4} \right] 
       \left[ {1\over p_2^2+\kappa^2} 
              {1\over (p_1-p_2)^2+\kappa^2} 
            - {1\over p_2^2 (p_1-p_2)^2 } 
       \right] 
           \Biggr\}.
\end{equation}
Consider e.g.\ the first term contributing to the r.h.s.:
\begin{equation}\label{(12.17)}
     \int_{|p_1|>\kappa} d^{2D}p\, \theta _1(p)\, \bar \varphi (p) 
\end{equation}
\begin{equation}
     \times \left[ {1\over p_1^2+\kappa^2} 
                 - {1\over p_1^2} 
                 + \kappa^2 {1\over p_1^4} 
                 - c_0(\kappa) \delta (p_1)
            \right] 
         {1\over p_2^2+\kappa^2} \; {1\over (p_1-p_2)^2+\kappa^2} 
\end{equation}
\begin{equation}
     = \int_{|p_1|>\kappa} d^Dp_1\, \bar{\bar\varphi}(p_1,\kappa) \, 
       \left[ {1\over p_1^2+\kappa^2} 
            - {1\over p_1^2} 
            + \kappa^2 {1\over p_1^4} 
       \right] ,
\end{equation}
where
\begin{equation}\label{(12.18)}
     \bar {\bar \varphi} (p_1,\kappa) = 
     \int  d^Dp_2 
     { \theta_1\, \bar\varphi
         \over 
       (p_2+\kappa^2)
       ((p_1-p_2)^2+\kappa^2)
     }. 
\end{equation}
The square-bracketed term on the r.h.s.\ of \ref{(12.17)} is bounded by 
${\rm const}\times\kappa^4/|p_1|^6.$ To estimate $\bar \varphi,$ one 
should take into account that the integration over $p_2$ in 
\ref{(12.18)} runs over the region $|p_2| < k |p_1|,$ and that $\bar 
\varphi$ can be represented as $p_1\tilde\varphi_1+p_2\tilde\varphi_2,
$ where both $\tilde\varphi_i$ are smooth functions 
\cite{schwartz:analyse}. Rescaling the integration variable as $p_2\to 
|p_1|\cdot p_2,$ one gets:
\begin{equation}\label{(12.19)}
     \bar {\bar \varphi} (p_1,\kappa) = |p_1|^{D-4+1} 
     \int_{|p_2|<k} \, d^Dp_2 \,
     {\theta_1(\hat p_1,p_2)\, 
      \bar \varphi_1(p_1,|p_1|\cdot p_2)
          \over
      (p_2^2+\kappa^2/|p_1|^2)
      ((\hat p_1-p_2)^2+\kappa^2/|p_1|^2 )
     } .
\end{equation}
Since $\bar \varphi_1(p_1,|p_1|\cdot p_2)$ can be bounded by ${\rm 
const}\times |p_1|,$ one sees that $\bar\varphi(p_1,\kappa)$ is bounded 
by ${\rm const}\times|p_1|.$ (It is not difficult to understand that 
since the expansion \ref{(12.10)} should be asymptotic only at $D=4,$ 
one may check estimates like \ref{(12.13)} only at $D=4.$) One arrives 
at the conclusion that the expression \ref{(12.17)} is $o(\kappa^2),$ 
as expected. The remaining terms in \ref{(12.16)} are considered in the 
same way and, finally,  one obtains \ref{(12.13)}.

\SUBSECTION{The last counterterm}
\label{ss12.5}

The result of the preceding subsection (the estimate \ref{(12.13)}) 
allows one to repeat the reasoning of subsect.~\ref{ss11.3}. 

One finally obtains:
\begin{equation}\label{(12.20)}
     G(p,\kappa) \asy{\kappa}{0} 
     {\rm\bf As}_2 G(p,\kappa) + o(\kappa^2),
\end{equation}
where 
\begin{equation}\label{(12.21)}
     {\rm\bf As}_2 G(p,\kappa) = {\rm\bf As}'_2 G(p,\kappa) 
     + c_{G,0} (\kappa) \delta (p)
\end{equation}
($\delta(p)\equiv\delta(p_1)\delta(p_2))$ and
\begin{equation}\label{(12.22)}
     c_{G,0} (\kappa) = \int  d^{2D}p\, G(p,\kappa) 
     \equiv  \int  d^Dp_1\, d^Dp_2\, G(p_1,p_2,\kappa).
\end{equation}
Examples of higher-order counterterms are:
\begin{equation}\label{(12.23)}
     c_{G,2,0} (\kappa) \, \partial^2\delta (p_1) \delta (p_2),
\end{equation}
where
\begin{equation}\label{(12.24)}
     c_{G,2,0} (\kappa) = \int  d^Dp_1\, d^Dp_2 \, p_1^2\, 
     G(p_1,p_2,\kappa),
\end{equation}
and 
\begin{equation}\label{(12.25)}
     c_{G,0,2} (\kappa) \, \partial_\mu\delta (p_1) 
                        \partial^\mu\delta(p_2),
\end{equation}
where
\begin{equation}\label{(12.26)}
     c_{G,0,2} (\kappa) = \int  d^Dp_1\, d^Dp_2 \, (p_1\cdot p_2) 
                                                    G(p_1,p_2,\kappa),
\end{equation}
etc. Concerning diagrammatic interpretation of \ref{(12.22)}, 
\ref{(12.24)} and \ref{(12.26)}, see Fig.~7d.

This completes our discussion of the example.

The essential point to be stressed here is that the proof involves 
nothing but the power counting. The same remains true in the most 
general case as well, which justifies the conclusions of 
subsect.~\ref{ss11.2}. For a more formal treatment of this point see 
\cite{fvt-vvv}.

Note that in practice there is no real need to perform the detailed
analysis as above. It is simply sufficient to enumerate singularities
of the formal expansion, determine (by power counting) what
counterterms are needed, and to find counterterms by formal
integration of the ansatz for the \asop\ with suitable polynomials to
project out the coefficients.
{}

{}

\noindent{\sc Acknowledgments}.

I would like to express my gratitude to 
A.~V.~Radyush\-kin and A.~A.~Vla\-di\-mi\-rov for 
the interest in, and support of this work since its early stages.  I  
am  indebted  to  G.~B.~Pi\-vo\-va\-rov, V.~V.~Vla\-sov and 
A.~N.~Kuz\-ne\-tsov for collaboration on putting  the  theory of 
asymptotic expansions of distributions on  a  rigorous  foundation, 
and to A.~A.~Pivovarov for a discussion of the sum rules method. 
I would also like to  thank  N.~Ata\-ki\-shiev,  C.~Burgess, 
G.~Efim\-ov,  M.~Fu\-ku\-gi\-ta, 
E.~Re\-mid\-di, D.~Ro\-ba\-schik,  D.~Shir\-kov,  A.~Slav\-nov,  
D.~Slav\-nov,  and   S.~Zla\-tev   for encouragement and stimulating 
discussions.

The main part of this paper  was  written  during  my  stay  at  Jilin 
University  (Chang\-chun,  China),  and  it  is  my  pleasure  to  thank 
Professor Wu Shishu and the staff of the Physics Department 
for their warm hospitality. 
I am  grateful to  A.~P.~Contogouris  for  hospitality  at the 
McGill University (Montreal, Canada)
and to W.~A.~Bardeen and R.~K.~Ellis for  their kind hospitality  at 
the Theory division of Fermilab---where the manuscript was completed.

Unfortunately, Sergei Gorishny, who contributed so much to the  theory 
of Euclidean asymptotic expansions, 
will never read these lines. The only thing I can do 
to appreciate his  generous  help  and  advice  is  to  dedicate  this 
publication to his memory.

Finally, my thanks are due to J.~C.~Collins 
for a careful reading of the manuscript and 
stimulating discussions concerning extension 
of  the  described  methods 
to non-Euclidean asymptotic regimes; 
and to A.~V.~Radyushkin whose support at various stages 
of the work was crucial.

\newpage\thispagestyle{myheadings}\markright{}

\newpage
\thispagestyle{myheadings}\markright{}
\centerline{\bf Figure captions}

{}

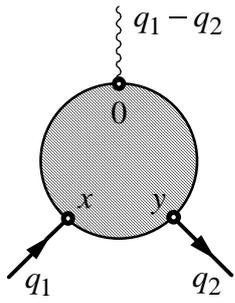
Fig.1. Kinematics of three-point functions.

Fig.2. (a) Kinematics of light-by-light scattering. It differs from 
the general 4-point function (b) by the restriction $q_1 + q_3 = 0.$

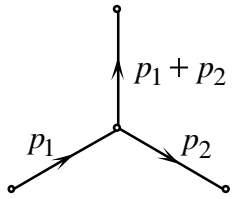
Fig.3. Three light-mass propagators connected to a common vertex form 
a complete IR-subgraph. 

Fig.4. The property of being an IR-subgraph depends on restrictions 
imposed on heavy external momenta: in (a) the pair of vertical lines 
constitutes an IR subgraph, while in (b) it does not.

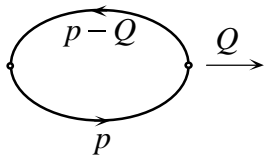
Fig.5. This graph is its own IR-subgraph if the contact terms 
localized at $Q=0$ are to be taken into account; it is not, if $Q$ is 
just fixed at a non-zero value.

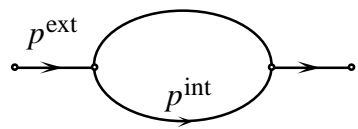
Fig.6. A simple example of the division of the proper variables of 
IR-subgraphs into ``internal" components corresponding to loops and the 
`` external" components.

Fig.7. (a) A graphical representation of the product \ref{(12.2)}. (b) 
The test function $\varphi_1$ is non-zero within the region enclosed 
by the dashed line. (c) Geometry of integrations in \ref{(12.17)} and 
\ref{(12.18)}. (d) A graphical interpretation of the expressions 
\ref{(12.22)}, \ref{(12.24)} and \ref{(12.26)}.

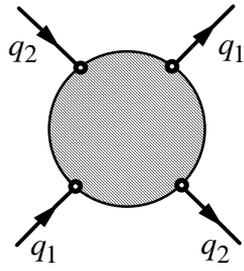
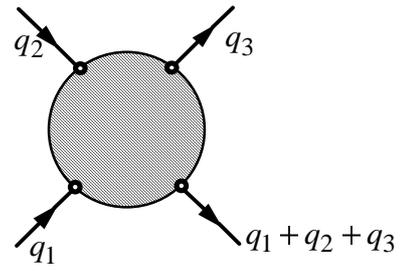
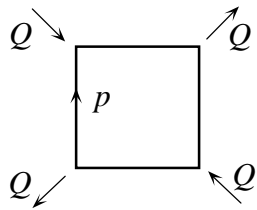
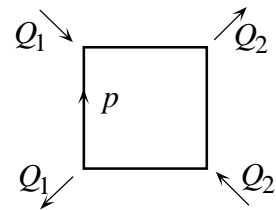
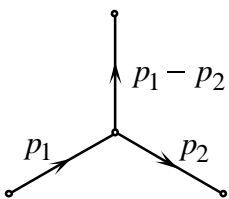
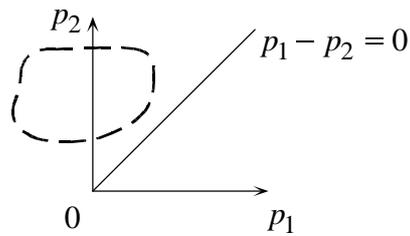
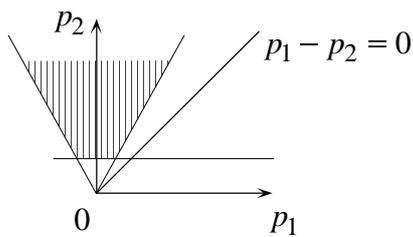
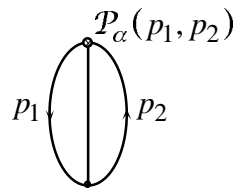
\end{document}